\documentclass[sigconf,nonacm]{acmart}
\setcopyright{none}
\renewcommand\footnotetextcopyrightpermission[1]{}

\usepackage{graphicx}
\usepackage{subcaption}
\usepackage{booktabs}
\usepackage{array}
\usepackage{makecell}
\usepackage{multirow}
\usepackage{changepage}
\usepackage[dvipsnames,table]{xcolor}
\definecolor{mygrey}{gray}{0.4}
\usepackage{amsmath}
\usepackage{mathtools}
\usepackage{amsthm}
\usepackage{bbm}

\usepackage{algorithm}
\usepackage{algorithmic}

\usepackage{microtype}
\usepackage{pifont}
\usepackage{enumitem}
\usepackage[T1]{fontenc}

\usepackage[most]{tcolorbox}
\usepackage{listings}
\usepackage{fontawesome5}
\usepackage{svg}

\tcbuselibrary{listings,breakable,skins}

\newcommand{\drop}[1]{%
  {\scriptsize\textcolor{red}{($\downarrow$#1)}}%
}

\AtBeginDocument{%
  }

\setcopyright{acmlicensed}
\copyrightyear{2018}
\acmYear{2018}
\acmDOI{XXXXXXX.XXXXXXX}
\acmConference[Conference acronym 'XX]{Make sure to enter the correct
  conference title from your rights confirmation email}{June 03--05,
  2018}{Woodstock, NY}
\acmISBN{978-1-4503-XXXX-X/2018/06}

\begin{document}

\title{A-MapReduce: Executing Wide Search via Agentic MapReduce}

\author{Mingju Chen}
\authornote{These authors contributed equally to this work.}
\affiliation{%
  \institution{Beijing Advanced Innovation Center for Future Blockchain and Privacy Computing, School of Artificial Intelligence, Beihang University}
  \city{Beijing}
  \country{China}
}
\email{}

\author{Guibin Zhang}
\authornotemark[1]
\affiliation{%
  \institution{National University of Singapore}
  \city{Singapore}
  \country{Singapore}
}
\email{}

\author{Heng Chang}
\authornote{Lead contributor.}
\affiliation{%
  \institution{Tsinghua University}
  \city{Beijing}
  \country{China}
}
\email{}

\author{Xiao Wang}
\affiliation{%
  \institution{School of Software, Beihang University}
  \city{Beijing}
  \country{China}
}
\email{}

\author{Shiji Zhou}
\authornote{Corresponding author.}
\affiliation{%
  \institution{Beijing Advanced Innovation Center for Future Blockchain and Privacy Computing, School of Artificial Intelligence, Beihang University}
  \city{Beijing}
  \country{China}
}
\email{zhoushiji25@buaa.edu.cn}
\renewcommand{\shortauthors}{Chen et al.}

\begin{abstract}
  Contemporary large language model (LLM)-based multi-agent systems exhibit systematic advantages in \textit{deep research} tasks, which emphasize iterative, vertically structured information seeking. However, when confronted with \textit{wide search} tasks characterized by large-scale, breadth-oriented retrieval, existing agentic frameworks, primarily designed around sequential, vertically structured reasoning, remain stuck in expansive search objectives and inefficient long-horizon execution. To bridge this gap, we propose \textsc{A-MapReduce}, a MapReduce paradigm-inspired multi-agent execution framework that recasts wide search as a horizontally structured retrieval problem. Concretely, \textsc{A-MapReduce} implements parallel processing of massive retrieval targets through task-adaptive decomposition and structured result aggregation. Meanwhile, it leverages experiential memory to drive the continual evolution of query-conditioned task allocation and recomposition, enabling progressive improvement in large-scale wide-search regimes. Extensive experiments on five agentic benchmarks demonstrate that \textsc{A-MapReduce} is \textbf{(i) high-performing}, achieving state-of-the-art performance on WideSearch and DeepWideSearch, and delivering 5.11\% $\sim$ 17.50\% average Item F1 improvements compared with strong baselines with OpenAI o3 or Gemini 2.5 Pro backbones; \textbf{(ii) cost-effective and efficient}, delivering superior cost–performance trade-offs and reducing running time by 45.8\% compared to representative multi-agent baselines. The code is available at \url{https://github.com/mingju-c/AMapReduce}.
\end{abstract}

\maketitle

\section{Introduction}
Multi-agent systems are gradually becoming a crucial paradigm for extending the cognitive boundaries of Large Language Model (LLM) and enabling them to tackle complex real-world tasks. From early chain-of-thought reasoning with single agents~\cite{wei2022chain,yao2023react} to collaborative reasoning frameworks involving multiple roles~\cite{li2023camel,wu2024autogen,hong2024metagpt}, LLM-based multi-agent systems (MAS) have demonstrated substantial advantages across a wide range of interactive scenarios, including mathematical reasoning~\cite{luo2024improvemathematicalreasoninglanguage,cobbe2021trainingverifierssolvemath}, web information retrieval~\cite{nakano2022webgptbrowserassistedquestionansweringhuman,zhou2024webarena,he2024webvoyager}, and code generation~\cite{jimenez2024swebench,qian-etal-2024-chatdev}.

\begin{figure}[t]
  \begin{center}
    \centerline{\includegraphics[width=1\columnwidth]{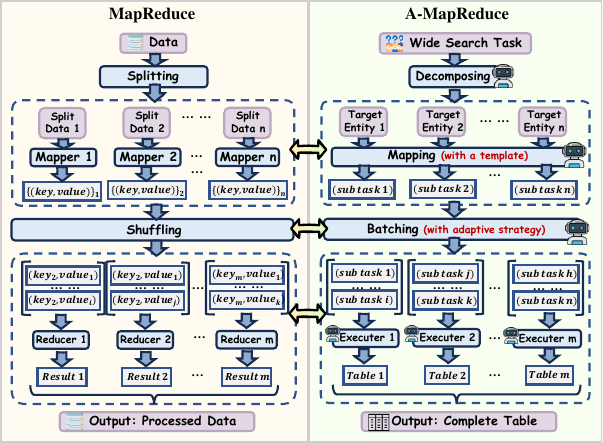}}
    \caption{Structural isomorphism: \textsc{A-MapReduce} mirrors the MapReduce pipeline correspondingly.}
    \vskip -0.2 in
    \label{fig:compare}
  \end{center}
\end{figure}

Many recent MAS improve performance by extending reasoning depth through diverse execution topologies~\cite{zhang2025gdesigner,wang2025mas2selfgenerativeselfconfiguringselfrectifying}, role assignments~\cite{li2023camel,hong2024metagpt}, and interaction protocols~\cite{wu2024autogen,zhang2026agentorchestraorchestratingmultiagentintelligence}. 
Despite architectural differences, these methods largely follow a shared execution paradigm that recursively extends reasoning depth across agents. We refer to it as the \textbf{\textit{vertical recursive reasoning paradigm}}, which has been widely adopted in deep search and deep research scenarios.

However, as wide search~\cite{wong2025widesearch,lan2025deepwidesearch} has emerged as a distinct and increasingly important application setting, the limitations of this vertical paradigm become apparent at the execution level:~\textbf{(Dilemma 1):} Most MAS rely on dialogue history or free-form textual plans to implicitly manage retrieval targets, which often result in missing entries, redundant retrieval, or target misalignment in long-horizon execution. This motivates the need for an explicit, persistent task representation to reliably maintain large-scale retrieval objectives. \textbf{(Dilemma 2):} Existing MAS typically re-plan execution pipelines for each query, without abstracting structural paradigm shared across similar queries. This highlights the need for a mechanism that explicitly captures and reuses execution-level patterns across wide search tasks, enabling the accumulation of experience and evolution in execution decisions. Taken together, these limitations suggest that wide search should be modeled under a \textbf{\textit{horizontal structured retrieval paradigm}}, in which the core is to organize, allocate, track and aggregate a large set of weakly coupled retrievals.

Building on these observations and inspired by the MapReduce paradigm~\cite{dean2008mapreduce} in database systems, we propose \textsc{A-MapReduce} (Agentic MapReduce), a multi-agent framework that reconceptualizes wide-search execution as a distributed computation process for structured retrieval. As shown in Fig.~\ref{fig:compare}, \textsc{A-MapReduce} preserves the core MapReduce workflow. Concretely, it maps each wide-search task into an explicit MapReduce decision, providing a structured abstraction for organizing large-scale retrieval objectives. 
Based on the decision, the task is subsequently decomposed into multiple batches of atomic retrieval tasks, which are executed in parallel and reduced to a unified output. Beyond structural execution, wide search also requires mechanisms for accumulating and reusing execution patterns from past tasks. 
To this end, we introduce an \textbf{experience-based evolution} mechanism that distills execution trajectories from historical wide search tasks into reusable structural hints stored in experiential memory. By conditioning task decomposition and execution strategies on this experiential memory, \textsc{A-MapReduce} facilitates cross-task reuse of execution-level patterns and supports the progressive evolution of execution decision, improving both efficiency and robustness in practice.

Our contributions are summarized as follows:

\textit{\textbf{\textbullet \space Paradigm Reformulation:}}
We present \textbf{\textsc{A-MapReduce}}, an open-source multi-agent framework that reformulates wide search as a controllable MapReduce-style execution process.
It supports persistent coverage tracking over large retrieval targets, parallelized mapping, and principled reduction into a single schema-consistent table.

\textit{\textbf{\textbullet \space Practical Evolution:}}
We further introduce an \textbf{experience-based evolution} mechanism that couples MapReduce execution with an experiential memory to progressively refine the query-conditioned decision distribution, enabling increasingly query-adaptive and cost-efficient execution.

\textit{\textbf{\textbullet\space Experimental Evaluation:}}
Extensive experiments on five agentic benchmarks show that \textsc{A-MapReduce} is:
\ding{182} \textbf{high-performing}:
substantially outperforming other frameworks and achieving average absolute improvements in Item-level F1 of 5.64\% to 27.62\% across the WideSearch and DeepWideSearch benchmarks. 
\ding{183} \textbf{efficient and cost-effective}:
achieving superior cost--performance trade-offs, with 45.8\% lower running time and \$1.10 average cost savings per task compared with representative MAS baselines.

\begin{figure*}[ht]
  \begin{center}
    \centerline{\includegraphics[width=2.05\columnwidth]{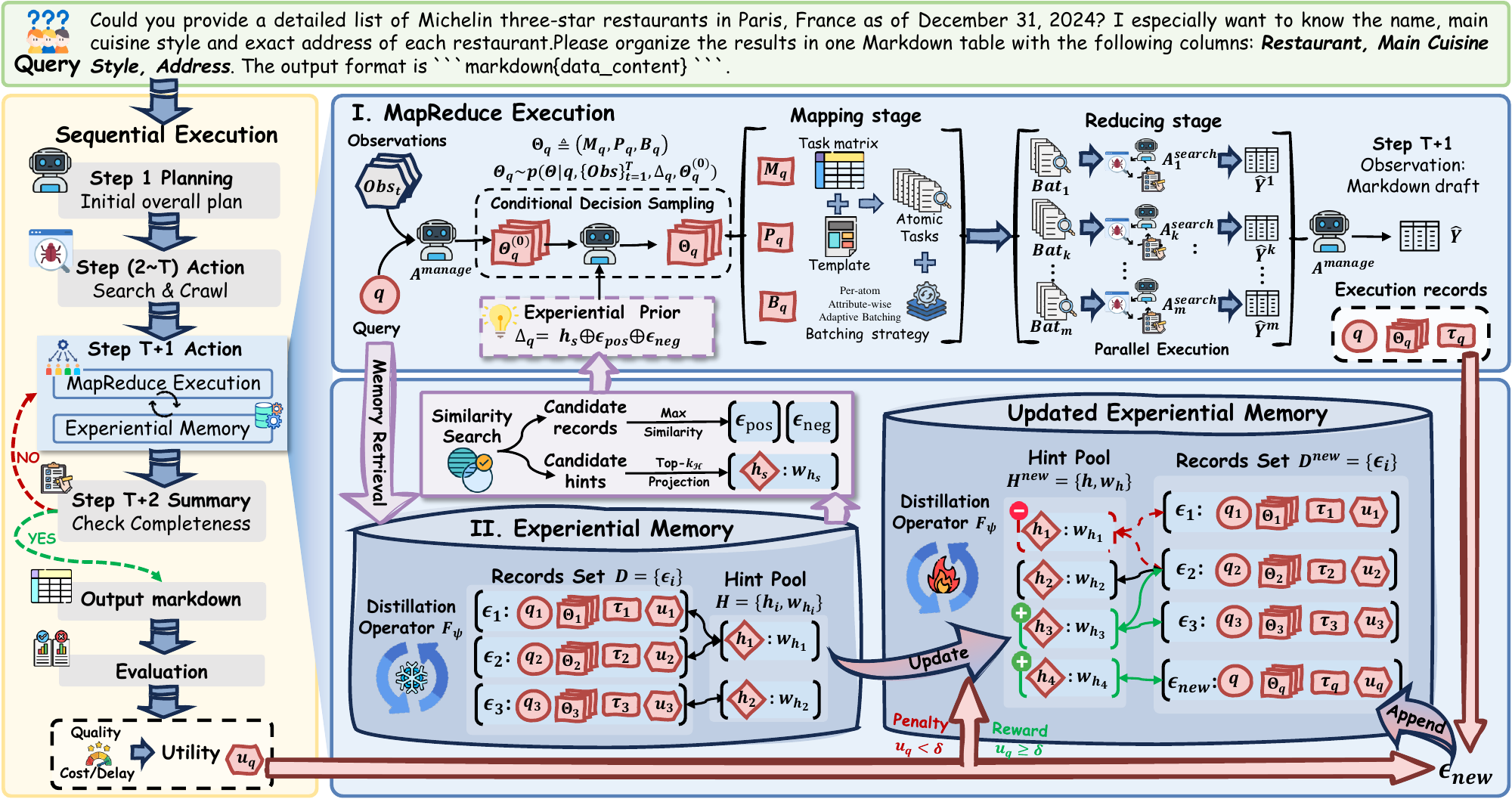}}
    \caption{Overall workflow of our \textsc{A-MapReduce}. Given a wide-search query, our framework takes lightweight sequential execution, then retrieves experiential prior as a decision anchor  from experiential memory and samples a conditional MapReduce decision. It then decomposes the query into atomic tasks for batched parallel execution and reduces partial results into a single structured output. After completion, environmental feedback updates the experiential memory, enabling continual cross-task refinement of execution decisions.}
    \vskip -0.3in
    \label{fig:workflow}
  \end{center}
\end{figure*}

\section{Related Work}

\subsection{LLM-based Multi-agent Systems.}
LLM-based multi-agent systems (MAS) have shown strong potential in solving complex tasks through collaborative reasoning and coordination~\cite{tran2025multiagentcollaborationmechanismssurvey,fang2025comprehensivesurveyselfevolvingai}.
Early frameworks relied on manually designed workflows and predefined agent roles.
For instance, CAMEL~\cite{li2023camel} facilitates cooperation through role-playing interactions, AgentVerse~\cite{chen2024agentverse} organizes dynamically composed groups of collaborative agents, and MetaGPT~\cite{hong2024metagpt} incorporates standardized operating procedures into multi-agent software development.
Recent studies have increasingly explored automated optimization at different levels.
EvoPrompt~\cite{guo2024connecting} employs evolutionary algorithms to optimize discrete prompts, while ChemAgent~\cite{tang2025chemagent} maintains a self-updating memory library to reuse reasoning experiences.
For agent architecture optimization, GPTSwarm~\cite{zhuge2024gptswarm} represents agent collaboration as optimizable computational graphs, and G-Designer~\cite{zhang2025gdesigner} learns task-adaptive communication topologies.
AutoFlow~\cite{li2024autoflowautomatedworkflowgeneration} automatically generates natural-language workflows, while EvoAgent~\cite{yuan-etal-2025-evoagent} and EvoFlow~\cite{zhang2025evoflow} leverage evolutionary operators to construct multi-agent systems and diverse heterogeneous workflows, respectively.
More recently, MaAS~\cite{zhang2025multiagent} samples query-dependent architectures from an agentic supernet, MAS$^2$~\cite{wang2025mas2selfgenerativeselfconfiguringselfrectifying} dynamically generates and rectifies multi-agent systems, and ADAS~\cite{hu2025automated} employs meta-agent search to discover new agent designs.
These approaches primarily focus on optimizing agent configurations and coordination structures, whereas our work emphasizes the explicit organization and adaptive execution of large-scale retrieval tasks.

\subsection{Deep Search \& Wide Search Agent Systems.}
Recent advances in LLM-based agent systems have been driven by challenging benchmarks such as GAIA~\cite{mialon2024gaia}, BrowseComp~\cite{wei2025browsecompsimplechallengingbenchmark}, xBench~\cite{chen2025xbenchtrackingagentsproductivity}, HLE~\cite{phan2025humanity}, AgentBench~\cite{liu2024agentbench}, and TaskCraft~\cite{shi2025taskcraftautomatedgenerationagentic}.
A variety of frameworks have been developed to improve complex information seeking and tool-augmented reasoning.
Magentic-One~\cite{fourney2024magentic} introduces an orchestrator that manages planning, progress tracking, and replanning, while OWL~\cite{hu2025owl} separates strategic planning, task coordination, and specialized worker execution.
Alita~\cite{qiu2025alita} emphasizes autonomous capability construction with minimal predefined components, and AgentOrchestra~\cite{zhang2026agentorchestraorchestratingmultiagentintelligence} introduces a unified protocol for coordinating agents, tools, and environments.
Other systems, such as Cognitive Kernel-Pro~\cite{fang2025cognitive} and O-Researcher~\cite{yao2026oresearcheropenendeddeep}, explore training strategies for deep-research agents, while smolagents~\cite{roucher2025smolagents} provides lightweight code-based tool execution.
Flash-Searcher~\cite{qin2025flashsearcherfasteffectiveweb} further improves efficiency by organizing dependent reasoning subtasks into a directed acyclic graph for parallel execution.
Unlike depth-oriented search, wide-search tasks emphasize broad entity coverage and systematic result aggregation.
Recent research has explored related challenges from different perspectives.
M-ASK~\cite{chen2026monolithicarchitecturesmultiagentsearch} decouples search behavior from knowledge management to improve multi-step information seeking, while WebLeaper~\cite{tao2025webleaper} constructs entity-intensive search tasks and efficient trajectories through tree-structured reasoning.
In contrast, A-MapReduce explicitly organizes wide-search execution through structured task matrices, query-conditioned templates, adaptive batching, and schema-aware result reduction.

\section{Methodology}
Fig.~\ref{fig:workflow} shows the overall workflow of our framework.  
In the following sections, Sec.~\ref{sec:Preliminary} formalizes the wide-search setting and utility criterion, Sec.~\ref{sec:Agentic MapReduce Execution} details the agentic MapReduce execution process, and Sec.~\ref{sec:Experience-based Optimization} presents the experience-driven decision evolution mechanism.

\vspace{-0.3em}
\subsection{Preliminary}
\label{sec:Preliminary}

\textbf{Wide-Search Task.}
A wide-search task can be modeled as $q\triangleq(q^{\text{text}},S)$, where $q^{\text{text}}$ is the natural-language request
and $S=\{s_k\}_{k=1}^{K}$ is an explicit schema described in $q$. The objective is to identify a (latent) set of relevant entities
$E=\{e_i\}_{i=1}^{N}$ and to produce a structured table
$Y \in \mathcal{V}^{N \times K}$, where $\mathcal{V}$ denotes the value space and
$y_{i,k}=Y(i,k)$ corresponds to the value of entity $e_i$ under attribute $s_k$.
Since neither $E$ nor $Y$ is specified explicitly by $q$, the system must perform two coupled steps to complete the wide-search task $q$: (i) \emph{entity discovery} and (ii) \emph{attribute grounding}.

\textbf{Multi-Agent Decision Process.}
We model a multi-agent system (MAS) for wide search as a stochastic decision-sampling process.
Given a query $q$, the system samples a high-level execution decision $\Theta$ from the distribution $p(\Theta \mid q)$, where $\Theta$ parametrizes how the task is decomposed, scheduled, and executed across agents. Conditioned on $\Theta$, the MAS induces an output table $Y$ and execution trajectory $\tau$,
which denotes the structured execution trace, including tool calls and agent interactions.

\textbf{Utility Criterion.}
Given a query $q$, the MAS samples an execution decision $\Theta$ and induces an outcome $(Y,\tau)$.
Different $\Theta$ yield different trade-offs between output quality and execution efficiency.
We therefore define a utility $u$:
\begin{equation}
u(q,\Theta)\triangleq Q(Y; q)-\lambda_c\,C(\tau)-\lambda_t\,D(\tau),
\end{equation}
where $Q(\cdot)$ measures output quality, and $C(\tau)$ and $D(\tau)$ denote execution cost and delay. We use task-level feedback $u$ to update memory and gradually refine $p(\Theta\!\mid\!q)$ toward higher-utility decisions. 
In our implementation, we update memory using only quality signals for reproducibility. While cost and delay are reported as outcomes but not used for updates (explanation in App.~\ref{app:memory_structure_update}). 
For brevity, we write $u$ when the dependence on $(q,\Theta)$ is clear.

\vspace{-0.3em}
\subsection{Agentic MapReduce Execution}
\label{sec:Agentic MapReduce Execution}
This section describes the core execution mechanism of \textsc{A-MapReduce}. Following the formulation in Sec.~\ref{sec:Preliminary}, the effectiveness of our framework critically depends on how such decisions are \emph{explicitly structured} and \emph{operationalized}.
To this end, we represent each execution decision as a triple:
\begin{equation}
\Theta_q \triangleq (M_q, P_q, B_q),
\end{equation}
where $M_q$ denotes the task matrix, $P_q$ the task template, and $B_q$ the batching strategy.
This parameterization renders execution decisions controllable and serves as a foundation for the following MapReduce execution process:

\textbf{Mapping Stage.}
Before formal MapReduce execution, our framework performs a lightweight sequential reasoning phase to analyze the query content and discovery the (implicit) target entity set. Conditioned on the query $q$, and the accumulated observations $\{Obs_t\}_{t=1}^{T}$, the manager agent $\mathcal{A}^{\text{manage}}$ samples a high-level execution decision:
\begin{equation}
\Theta_q \sim p(\Theta \mid q, \{Obs_t\}_{t=1}^{T}).
\end{equation}
As defined in Eq.~(2), $\Theta_q=(M_q,P_q,B_q)$ specifies how the query is mapped into atomic retrieval tasks and organized for parallel execution.
Concretely, the task matrix $M_q \in \mathcal{V}^{N \times K_0}$ ($K_0<K$) encodes $N$ target entities with their known attributes, and the template $P_q$ is a query-specific string with placeholders aligned to the columns of $M_q$.
Grounding each row of $M_q$ to the template $P_q$ via $\mathrm{fill}(\cdot)$ function to yield atomic retrieval tasks set $\mathcal{T}(q)$:
\begin{equation}
\mathcal{T}(q) = \{\, t_i \mid t_i = \mathrm{fill}(P_q, M_q[i,:]),\ i=1,\ldots,N \,\}.
\end{equation}
For the batching strategy $B_q$, we expose a discrete family of batching strategies, including per-atom, attribute-wise, and adaptive batching. Rather than fixing the strategy, the manager agent selects and parameterizes $B_q$ conditioned on the query, which partitions $\mathcal{T}(q)$ into $m$ disjoint batches:
\begin{equation}
\{Bat_k\}_{k=1}^{m} = B_q(\mathcal{T}(q)), ~~
m = \left\lceil \frac{|\mathcal{T}(q)|}{bs_q} \right\rceil,
\end{equation}
where $bs_q$ denotes the batch size specified by $B_q$.

\textbf{Reducing Stage.}
Each batch $Bat_k$ is then assigned to an independent search agent $\mathcal{A}_k^{\text{search}}$ for parallel execution:
\begin{equation}
\hat{Y}^k = \mathcal{A}_k^{\text{search}}(Bat_k),
~~
\hat{Y} = \bigcup_{k=1}^{m} \hat{Y}^k .
\end{equation}
The manager agent $\mathcal{A}^{\text{manage}}$ reduces the partial tables and validates them against the schema $S$. When completeness is violated, $\mathcal{A}^{\text{manage}}$ triggers a lightweight \emph{delta-patch} round by resampling a repair decision $\Theta_q^{\mathrm{rep}}$ and merging the patched table back (details in Alg.~\ref{alg:amapreduce}). Overall, different $\Theta_q$ correspond to distinct trade-offs between output quality and efficiency. \textsc{A-MapReduce} leverages experiential memory to condition and evolve decision $\Theta_q$ over time.

\vspace{-0.3em}

\subsection{Experience-based Evolution}
\label{sec:Experience-based Optimization}

Following Sec.~\ref{sec:Agentic MapReduce Execution}, \textsc{A-MapReduce} models the execution decision $\Theta_q$ as a random variable sampled from a query-conditioned distribution $p(\Theta\mid q)$.
Experience-based evolution refines this distribution over time, progressively shifting sampling toward high-utility regions of the decision space.

\textbf{Experiential Memory Overview.}
We instantiate evolution with an experiential memory
$\mathcal{M}\triangleq\{\mathcal{D},\mathcal{H},F_{\psi}\}$
that stores execution records and distilled hints.
The record set $\mathcal{D}=\{\epsilon_i\}_{i=1}^{N_{\mathcal{D}}}$ stores task-level information, where each record
$\epsilon_i=(q_i,\Theta_i,\tau_i,u_i)$ contains the query $q_i$, the realized decision $\Theta_i$, the execution trace $\tau_i$, and a utility $u_i$. The hint pool $\mathcal{H}$ stores reusable hints, where each hint $h$ includes an online score $w_h$, and a provenance set $\mathcal{S}(h)$ of supporting task indices. Crucially, $\mathcal{H}$ is consolidated by a distillation operator $F_{\psi}$. We further assume \emph{(Lipschitz-like) local regularity} in the query--decision space, i.e., \emph{semantically nearby queries tend to induce similar effective execution decisions}.
This motivates a plan--execute--update loop over experiential memory $\mathcal{M}$, consisting of memory retrieval, conditional decision sampling, and feedback-driven updates.

\begin{table*}[!t]
  \caption{Main results on the WideSearch benchmark. We report Item-level F1 (Item F1) , Row-level F1 (Row F1), and Success Rate (Succ. Rate)  across all systems. Entries marked with $\dagger$ are as reported in the original WideSearch paper~\cite{wong2025widesearch}, included to provide a broader context for comparison, while the other systems are evaluated by us using the same evaluation protocol (App.~\ref{app:sec:metrics}).}
  \label{tab:widesearch_main_results}
  \begin{center}
    \begin{small}
    \setlength{\tabcolsep}{11pt}
        \begin{tabular}{lcccccc}
          \toprule
          Model / System &
          \multicolumn{2}{c}{Item F1(\%)} &
          \multicolumn{2}{c}{Row F1(\%)} &
          \multicolumn{2}{c}{Succ. Rate(\%)} \\
          \cmidrule(lr){2-3}\cmidrule(lr){4-5}\cmidrule(lr){6-7}
          & Avg@4 & Max@4 & Avg@4 & Max@4 & Avg@4 & Pass@4 \\
          \midrule
            \rowcolor{gray!15}
          \multicolumn{7}{c}{\textit{Single Agent$^{\dagger}$}} \\
          \midrule
          Claude Sonnet 4 (Thinking)     & 57.89 & 66.73 & 31.69 & 41.90 & 2.25 & 5.00 \\
          Gemini 2.5 Pro                 & 50.98 & 63.62 & 30.00 & 41.39 & 1.50 & 5.00 \\
          OpenAI o3-high                      & 52.61 & 62.27 & 34.00 & 44.07 & 4.50 & 9.00 \\
          Kimi K2                        & 54.36 & 65.08 & 29.67 & 41.35 & 1.12 & 3.50 \\
          DeepSeek-R1                    & 41.26 & 55.09 & 20.66 & 31.69 & 0.37 & 1.50 \\
          Doubao-Seed-1.6 (Thinking)     & 48.27 & 63.86 & 29.90 & 44.12 & 2.63 & 5.00 \\
          Doubao-Seed-1.6 (Non-Thinking) & 48.97 & 61.98 & 27.21 & 39.86 & 1.00 & 3.50 \\
          \midrule
            \rowcolor{gray!15}
          \multicolumn{7}{c}{\textit{End-to-End Systems$^{\dagger}$}} \\
          \midrule
          Claude Sonnet 4 (Thinking)     & 48.40 & 58.47 & 24.11 & 33.49 & 2.50 & 5.00 \\
          Gemini 2.5 Pro                & 59.05 & 67.17 & 36.63 & 45.41 & 4.25 & 8.00 \\
          OpenAI o3                      & 45.52 & 56.49 & 23.91 & 36.03 & 3.00 & 5.50 \\
          \midrule
            \rowcolor{gray!15}
          \multicolumn{7}{c}{\textit{Open-source Agent Framework with Advanced LLMs}} \\
          \midrule
          MAS + Claude Sonnet 4 (Thinking)     &  62.17 & 73.13 & 38.49 & 52.19 & 3.62 & 6.50 \\
          MAS + Gemini 2.5 Pro               & 57.42 & 66.31 & 33.47 & 44.64 & 2.00 & 6.50 \\
          MAS + OpenAI o3                     & 57.31 & 68.93 & 37.80 & 50.52 & 5.12 & 9.50 \\
          MAS + Kimi K2                     & 61.15 & 70.72 & 36.22 & 49.60 & 3.00 & 6.50 \\
          MAS + DeepSeek-R1                    & 44.28 & 60.34 & 22.87 & 36.59 & 0.75 & 3.00 \\
          MAS + Doubao-Seed-1.6 (Thinking)    & 54.59 & 69.65 & 33.98 & 48.93 & 2.50 & 5.50 \\
          MAS + Doubao-Seed-1.6 (Non-Thinking) & 52.75 & 65.13 & 29.65 & 42.65 & 2.12 & 4.50 \\
          Smolagents (GPT-5-mini)              & 51.31 & 57.68 & 23.04 & 30.57 & 4.00 & 5.00 \\
          Flash-searcher (GPT-5-mini)          & 54.99 & 67.49 & 34.42 & 47.18 & 6.40 & 10.00 \\
          \midrule
          \rowcolor{blue!8}
          \textbf{\textsc{A-MapReduce} (GPT-5-mini)} & \textbf{67.81} & \textbf{75.03} & \textbf{45.23} & \textbf{56.24} & \textbf{7.50} & \textbf{12.00} \\
          \bottomrule
        \end{tabular}
    \end{small}
  \end{center}
  \vskip -0.2in
\end{table*}

\textbf{Memory Retrieval.}
To refine distribution $p(\Theta|q)$, we construct a plan-stage experiential prior $\Delta_q$ that injects query-conditioned distribution as a prior.
Let $\mathrm{sim}(\cdot,\cdot)$ denote cosine similarity and $\mathrm{Enc}(\cdot)$ be the embedding function. 
As $u$ is a continuous utility signal, we introduce a threshold $\delta$ only to separate high-utility and low-utility executions for contrastive retrieval, and form positive/negative index sets:
\begin{align}
\mathcal{I}_{q}^{+} &\triangleq \operatorname*{arg\,top\text{-}k}_{i:\,u_i>\delta}\ \mathrm{sim}\big(\mathrm{Enc}(q),\mathrm{Enc}(q_i)\big), \\\mathcal{I}_{q}^{-} &\triangleq \operatorname*{arg\,top\text{-}k}_{i:\,u_i\le \delta}\ \mathrm{sim}\big(\mathrm{Enc}(q),\mathrm{Enc}(q_i)\big),
\end{align}
From the positive and negative index set, we can form a candidate hint pool $\mathcal{H}_\text{cand}$, and select the final query-relevant hints $\mathcal{H}_q$ by the feedback-calibrated score $s(h)$:
\begin{equation}
\mathcal{H}_{\mathrm{cand}}(q)\ \triangleq\ 
\{\, h\in\mathcal{H}\ |\ \mathcal{S}(h)\cap(\mathcal{I}_{q}^{+}\cup\mathcal{I}_{q}^{-})\neq\emptyset \,\},
\end{equation}
\begin{equation}
s(h)\ \triangleq\ w_h + v(h;q),~~~
\mathcal{H}_{q}\ \triangleq\ \operatorname*{arg\,top\text{-}k_{\mathcal{H}}}_{h\in\mathcal{H}_{\mathrm{cand}}(q)}\ s(h),
\end{equation}
where $w_h$ denotes an online hint score that reflects its \emph{quality} accumulated from feedback, while $v(h;q)$ measures \emph{query-conditioned relevance} via provenance--neighborhood overlap.
Consequently, the selected hints are both globally reliable and well aligned with the current query $q$.

Finally, we optionally project $\mathcal{H}_q$ into a concise, actionable guidance
$h_s=\Phi(q;\mathcal{H}_q)$, and pair it with the most similar positive/negative exemplars $\epsilon_{\mathrm{pos}}$ and $\epsilon_{\mathrm{neg}}$ to form a \emph{query-conditioned} experiential prior $\Delta_q\triangleq\ (h_s\oplus\ \epsilon_{\mathrm{pos}}\oplus\ \epsilon_{\mathrm{neg}})$, which serves as an decision anchor for sampling, where $\oplus$ denotes concatenation for composing the experiential prior. The details of the procedure are provided in App.~\ref{app:memory_retrieval}.

\textbf{Conditional Decision Sampling.}
Based on experiential prior $\Delta_q$, conditional decision sampling proceeds by:
\begin{align}
\Theta^{(0)}_q &\sim p(\Theta \mid q,\{Obs_t\}_{t=1}^{T}), \\
\Theta_q &\sim p(\Theta \mid q,\{Obs_t\}_{t=1}^{T},\Delta_q,\Theta^{(0)}_q),
\end{align}
where $\Theta^{(0)}_q$ stabilizes sampling, and $\Delta_q$ serves as an anchor that sharpens the distribution toward high-utility regions. Conditioning on $\Delta_q$ biases sampling toward decisions that achieved higher utility on semantically similar tasks.

\textbf{Feedback-driven Update.}
After execution, we append the new record $\epsilon_{\text{new}}=(q,\Theta_q,\tau,u)$ to the record set,
$\mathcal{D}^{\text{new}} \leftarrow \mathcal{D}\cup\{\epsilon_{\text{new}}\}$.
We then update the hint pool with the realized utility signal $u$.
Since each hint maintains an online score, we perform credit assignment for the hints used to construct $\Delta_q$, and add the current record to their provenance:
\begin{equation}
w_h \leftarrow w_h + g(u),~~
\mathcal{S}(h) \leftarrow \mathcal{S}(h)\cup\{i_{\text{new}}\},\ \ \forall\, h\in \mathcal{H}_q,
\end{equation}
where $g(\cdot)$ maps the utility signal to a bounded signed reward, and $i_{\text{new}}$ is the index of $\epsilon_{\text{new}}$ in $\mathcal{D}^{\text{new}}$. Hints with persistently low scores are pruned, and higher-scoring hints are preferentially retrieved in subsequent executions.

As the record set $\mathcal{D}$ grows, purely accumulating hints can lead to redundancy and semantic drift. We therefore consolidate the hint pool with the memory-associated \textit{distillation operator} $F_{\psi}$ to generate and update hints:
\begin{equation}
\mathcal{H}^{\text{new}} \leftarrow F_{\psi}\!\left(\mathcal{D}^{\text{new}}, \mathcal{H}\right).
\end{equation}
In our implementation, $F_{\psi}$ partitions the expanded record set into task clusters $\{\mathcal{I}_k\}_{k=1}^{K_c}$, where each $\mathcal{I}_k$ indexes a cluster of semantically related records. Then, for each cluster, we denote $\mathcal{H}_k \triangleq \{h\in\mathcal{H}\mid \mathcal{S}(h)\cap \mathcal{I}_k\neq\emptyset\}$ as the historical hints associated with cluster $\mathcal{I}_k$, and distill a cluster-level representative hint that reflects structural preferences over decision $\Theta$ while suppressing task-specific details: \begin{equation} h_k \leftarrow F_\psi\!\left(\{\epsilon_i\}_{i\in\mathcal{I}_k},\mathcal{H}_k\right),~~ \mathcal{H}^{\text{new}} \triangleq \{h_k\}_{k=1}^{K_c}. \end{equation} Each $h_k$ serves as a cluster-conditioned anchor, guiding decision sampling for new tasks embedded in the same region of the task space. Finally, each distilled hint $h_k$ inherits provenance from its supporting records to remain compatible with voting-based retrieval $\mathcal{S}(h_k)\ \triangleq\ \mathcal{I}_k$. Details of the updating procedure and $F_{\psi}$ are provided in App.~\ref{app:memory_structure_update}.

\begin{table*}[t]
  \caption{Main results on the DeepWideSearch benchmark. We report core-entity accuracy (CE Acc.), Column-level F1, Item-level F1, Row-level F1, and Success Rate (Succ. Rate) across all systems. Entries marked with $\dagger$ are as reported in DeepWideSearch\cite{lan2025deepwidesearch} and are included for comparison, while the remaining results are evaluated by us using the same evaluation protocol (App.~\ref{app:sec:metrics}).}

  \label{tab:deepwidesearch_main_results}
  \begin{center}
    \begin{small}
        \setlength{\tabcolsep}{3.5pt}
        \begin{tabular}{lcccccccccc}
          \toprule
          
          Model / System &
          \multicolumn{2}{c}{CE Acc.} &
          \multicolumn{2}{c}{Column F1} &
          \multicolumn{2}{c}{Item F1} &
          \multicolumn{2}{c}{Row F1} &
          \multicolumn{2}{c}{Succ. Rate} \\
          \cmidrule(lr){2-3}\cmidrule(lr){4-5}\cmidrule(lr){6-7}\cmidrule(lr){8-9}\cmidrule(lr){10-11}
          & Avg@4 & Pass@4 & Avg@4 & Max@4 & Avg@4 & Max@4 & Avg@4 & Max@4 & Avg@4 & Pass@4 \\
          \midrule
            \rowcolor{gray!15}
          \multicolumn{11}{c}{\textit{Closed-source LLMs$^{\dagger}$}} \\
          \midrule
          OpenAI o3-mini              & 61.59 & 69.55 & 27.36 & 35.68 & 13.59 & 16.85 & 3.35 & 4.55 & 0.00  & 0.00  \\
          GPT-5                       & 58.41 & 72.72 & 31.71 & 41.05 & 21.67 & 28.21 & 9.61 & 13.42& 0.30 & 1.36 \\
          Claude Sonnet 4             & 57.95 & 64.09 & 32.63 & 40.16 & 19.94 & 23.38 & 7.31 & 8.97 & 0.90  & 0.90  \\
          Gemini 2.5 Pro              & 73.98 & 81.82 & 45.27 & 52.86 & 32.06 & 37.10 & 15.42& 18.96& 0.90  & 1.82 \\
          Qwen-Max                    & 56.02 & 63.64 & 28.81 & 36.19 & 14.32 & 18.48 & 4.16 & 6.18 & 0.00  & 0.00  \\
          GPT-4o                      & 54.20 & 63.64 & 19.66 & 27.07 & 11.86 & 16.41 & 4.18 & 7.01 & 0.00  & 0.00  \\
          \midrule
            \rowcolor{gray!15}
          \multicolumn{11}{c}{\textit{Open-source LLMs$^{\dagger}$}} \\
          \midrule
          DeepSeek-V3                 & 60.68 & 69.09 & 31.26 & 39.56 & 19.08 & 24.32 & 6.52 & 9.99 & 0.23 & 0.45 \\
          DeepSeek-R1                 & 66.93 & 75.45 & 38.42 & 47.77 & 25.01 & 30.56 & 10.72& 14.39& 0.28 & 0.45 \\
          KIMI-K2                     & 64.32 & 73.18 & 31.48 & 41.83 & 20.44 & 27.54 & 7.74 & 11.92& 0.34 & 0.91 \\
          Qwen3-235B-A22B             & 52.39 & 67.73 & 22.03 & 34.99 & 12.38 & 19.53 & 2.94 & 5.74 & 0.00  & 0.00  \\
          Qwen3-235B-A22B-Instruct    & 56.82 & 64.09 & 24.64 & 33.03 & 13.28 & 17.85 & 3.50 & 5.34 & 0.00  & 0.00  \\
          Qwen3-32B                   & 54.66 & 66.36 & 26.37 & 35.97 & 12.05 & 16.26 & 2.28 & 3.67 & 0.00  & 0.00  \\
          \midrule
            \rowcolor{gray!15}
          \multicolumn{11}{c}{\textit{Open-source Agent Framework with Advanced LLMs}} \\
          \midrule
          OWL (Gemini 2.5 Pro)       & 66.14 & 81.82 & 34.84 & 50.39 & 28.75 & 41.70 & 11.11& 16.93& 0.00  & 0.00  \\
          OWL (Claude Sonnet 4)       & 67.39 & 81.82 & 30.08 & 45.50 & 20.44 & 31.65 & 8.29 & 14.81& 0.68 & 1.36 \\
          Smolagents (Gemini 2.5 Pro) & 60.00 & 79.09 & 27.39 & 45.09 & 18.53 & 30.91 & 9.01 & 15.65& 0.11 & 0.45 \\
          Smolagents (GPT-5)          & 66.48 & 80.00 & 31.83 & 44.41 & 20.26 & 30.66 & 8.18 & 14.27& 0.45 & 0.45 \\
          Smolagents (Claude Sonnet 4) & 62.95 & 74.09 & 21.60 & 33.83 & 14.49 & 22.68 & 5.06 & 8.94 & 0.91 & 0.91 \\
          WebSailor (Gemini 2.5 Pro)  & 70.57 & 81.36 & 34.41 & 52.69 & 25.29 & 39.11 & 12.51& 20.49& 1.25 & 2.73 \\
          WebSailor (GPT-5)         & 74.32 & 85.00 & 37.18 & 49.48 & 25.96 & 35.65 & 10.97& 16.17& 0.34 & 1.36 \\
          WebSailor (Claude Sonnet 4) & 70.91 & 80.90 & 42.01 & 54.01 & 32.90 & 42.35 & 16.88& 24.26& 2.39 & 3.64 \\
          Flash-searcher (GPT-5-mini) & 75.57 & 84.55 & 42.07 & 55.04 & 34.97 & 47.05 & 19.36& 28.64& 2.61 & 4.55 \\
          \midrule
            \rowcolor{blue!8}
          \textbf{\textsc{A-MapReduce} (GPT-5-mini)}
          & \textbf{79.09} & \textbf{85.45}
          & \textbf{51.78} & \textbf{60.84}
          & \textbf{42.11} & \textbf{51.13}
          & \textbf{26.44} & \textbf{34.75}
          & \textbf{4.43}  & \textbf{6.36} \\
          \bottomrule
        \end{tabular}
    \end{small}
  \end{center}
  \vskip -0.2in
\end{table*}

\section{Experiment}

\subsection{Experiment Setup}
\paragraph{Benchmarks.} We evaluate our framework on two wide-scope, large-scale information retrieval benchmarks: \emph{WideSearch}~\cite{wong2025widesearch} and \emph{DeepWideSearch}~\cite{lan2025deepwidesearch}. In addition, we construct an \emph{agentic-wide subset} by extracting wide-scope retrieval tasks from three commonly used agentic benchmarks: \emph{xBench-DeepSearch (xBench-DS)} ~\cite{chen2025xbenchtrackingagentsproductivity}, \emph{WebWalkerQA}~\cite{wu2025webwalkerbenchmarkingllmsweb}, and \emph{TaskCraft}~\cite{shi2025taskcraftautomatedgenerationagentic}. Detailed description and statistics of the benchmarks are provided in App.~\ref {sec:dataset_statistics}.

\paragraph{Baselines.}
We compare \textsc{A-MapReduce} with a diverse set of baselines, including strong \emph{single-agent} and \emph{end-to-end systems} (e.g. Claude Sonnet~4, Gemini~2.5~Pro, OpenAI~o3), \emph{open-source LLMs} (e.g., Qwen3-32B, Kimi-K2), as well as representative \emph{open-source agent frameworks} paired with advanced LLM backbones, including \emph{Smolagents}~\cite{roucher2025smolagents}, \emph{OWL}~\cite{hu2025owl}, \emph{WebSailor}~\cite{li2025websailornavigatingsuperhumanreasoning}, and \emph{Flash-Searcher}~\cite{qin2025flashsearcherfasteffectiveweb}.
\paragraph{Implementation Details.} 
We set GPT-5-mini as the default LLM backbone of \textsc{A-MapReduce}, and report the underlying LLM backbone used by each framework for clarity and reproducibility. We set the maximum reasoning steps to 40 and use all-MiniLM-L6-v2 as embedding function to embed and store execution decisions. Additional implementation details are provided in App.~\ref{sec:implementation_details}.

\paragraph{Evaluation Metrics.}
We adopt a structured table-based evaluation framework for WideSearch and DeepWideSearch. Metrics include \emph{Success Rate}, \emph{Row F1}, \emph{Item F1}, \emph{Column-F1} and \emph{Core Entity Accuracy (CE Acc.)}. We perform $N$ independent runs per task and report \emph{Avg@N} and \emph{Pass@N} to reduce randomness. For the constructed \emph{agentic-wide subset}, we evaluate performance using accuracy. Detailed evaluation protocols are provided in App.~\ref{app:sec:metrics}.


\subsection{Main Results}
\begin{figure*}[!t]
  \centering
  \begin{subfigure}[t]{0.56\textwidth}
    \centering
    \includegraphics[width=\linewidth]{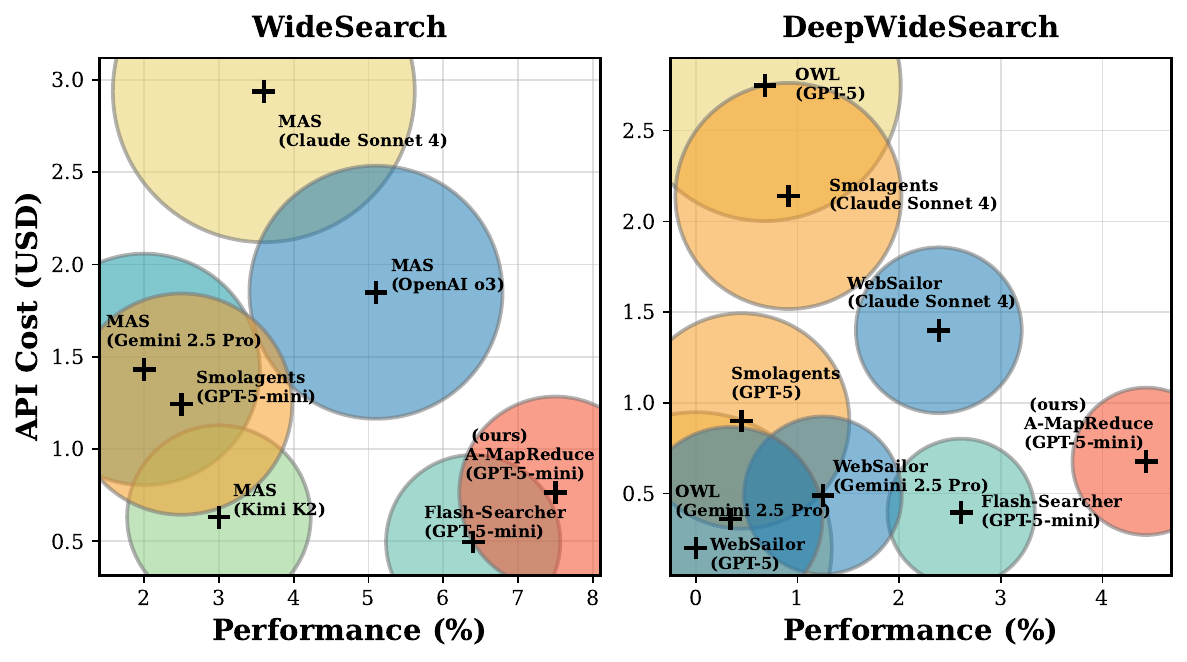}
    \caption{Cost--Performance on WideSearch and DeepWideSearch.}
    \label{fig:performance_cost_a}
  \end{subfigure}
  \hfill
  \begin{subfigure}[t]{0.41\textwidth}
    \centering
    \includegraphics[width=\linewidth]{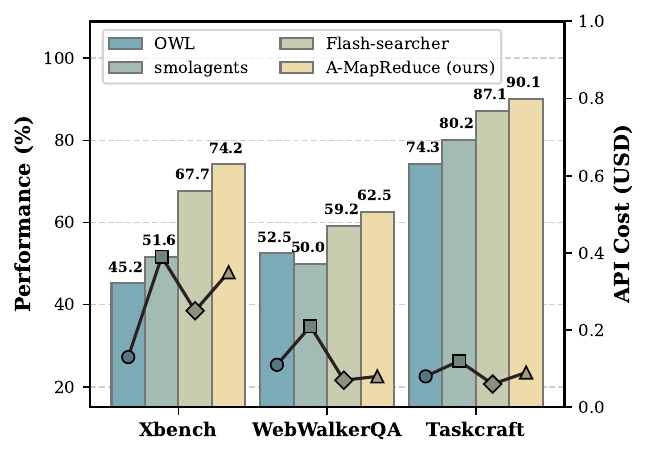}
    \caption{Cost--Performance on Agentic-Wide Subsets.}
    \label{fig:performance_cost_b}
  \end{subfigure}

  \caption{Cost--performance trade-offs of \textsc{A-MapReduce} across benchmarks.}
  \label{fig:performance_cost}
  \vskip -0.1in
\end{figure*}

We compare \textsc{A-MapReduce} against three classes of baselines with diverse LLM backbones on WideSearch (Tab.~\ref{tab:widesearch_main_results}) and DeepWideSearch (Tab.~\ref{tab:deepwidesearch_main_results}), and further evaluate \textsc{A-MapReduce} on the \emph{Agentic-Wide Subset} against three representative baselines in Fig.~\ref{fig:performance_cost} and Tab.~\ref{tab:agentic_wide_results}.
Overall, \textsc{A-MapReduce} achieves strong performance across these wide-search benchmarks.
As shown in Tab.~\ref{tab:widesearch_main_results}, under the Avg@4 evaluation setting, \textsc{A-MapReduce} achieves absolute gains of 12.71\% and 13.02\% in Item-F1 and Row-F1, respectively, and also improves Success Rate by 4.23\% over the average performance of the MAS baselines.
Specifically, it achieves 67.81\% Item-F1 and 45.23\% Row-F1, demonstrating improvements in both attribute-level retrieval and complete-record recovery.
Tab.~\ref{tab:deepwidesearch_main_results} evaluates \textsc{A-MapReduce} on the more challenging DeepWideSearch benchmark.
Compared with open-source multi-agent frameworks, CE Accuracy improves by approximately 3.52\% $\sim$ 19.09\% and Column-F1 shows gains ranging from 9.71\% $\sim$ 30.18\%, while Item-F1 and Row-F1 increase by roughly 17.49\% and 15.18\% absolute points on average, respectively.
These gains indicate improvements across entity identification, schema coverage, and structured information extraction.
Despite the increased difficulty, where most frameworks obtain around 0 $\sim$ 2\% success rate, \textsc{A-MapReduce} reaches a Success Rate of 4.43\%.
Fig.~\ref{fig:performance_cost_b} shows that \textsc{A-MapReduce} consistently achieves the best performance on the constructed agentic-wide subsets, with average absolute gains of up to 12.51\%.

\begin{table}[!t]
  \centering
  \begin{small}
  \caption{Efficiency analysis on WideSearch. We report Item F1 and Row F1 (Avg@4) together with per-task delay. \textsc{A-MapReduce}* denotes the variant without experience-driven evolution.}
  \label{tab:efficiency_widesearch}
  \resizebox{\columnwidth}{!}{%
  \begin{tabular}{lccc}
    \toprule
    \textbf{Method} & \textbf{Item F1} (\%) & \textbf{Row F1} (\%) & \textbf{Delay (s)} \\
    \midrule
    Smolagents     & 51.31 & 23.04 & 2617.7 \\
    Flash-Searcher  & 54.99 & 34.42 & 1204.7 \\
    \textsc{A-MapReduce}*                & 64.64 & 41.62 & 1460.8 \\
    \textbf{\textsc{A-MapReduce}}        & \textbf{67.81} & \textbf{45.23} & \textbf{953.7} \\
    \bottomrule
  \end{tabular}
  }
  \end{small}
\end{table}


\subsection{Cost Analysis}
We evaluate efficiency via cost--performance trade-offs (Fig.~\ref{fig:performance_cost}) and wall-clock runtime (Tab.~\ref{tab:efficiency_widesearch}, Tab.~\ref{tab:gpt5mini_table_metrics}).
The results also show favorable cost--performance trade-offs and faster runtime across diverse settings.
As shown in Fig.~\ref{fig:performance_cost}(a), \textsc{A-MapReduce} consistently lies on the Pareto frontier on both WideSearch and DeepWideSearch, achieving the highest Success Rate under competitive API cost.
Compared with open-source multi-agent frameworks, it reduces API cost by up to 47.5\% while delivering absolute gains of 5.1 to 16.5 percentage points in Item F1.
This indicates that the performance improvements do not simply arise from increased computational expenditure, but can be achieved with more efficient retrieval execution.
Fig.~\ref{fig:performance_cost}(b) shows that these advantages generalize to agentic-wide QA subsets, where our framework achieves strong performance with low cost.
Tab.~\ref{tab:efficiency_widesearch} shows that \textsc{A-MapReduce} delivers higher performance than baselines while saving around 45.8\% runtime.
Moreover, compared with the non-evolving variant \textsc{A-MapReduce}*, \textsc{A-MapReduce} reduces runtime from 1460.8 to 953.7 seconds, corresponding to a 34.7\% reduction, while improving Item F1 from 64.64\% to 67.81\%.
This suggests that experiential memory helps refine execution strategies and reduce unnecessary retrieval overhead while maintaining higher output quality.

\begin{figure}[!ht]
  \centering
  \begin{subfigure}[t]{0.99\columnwidth}
    \centering
    \includegraphics[width=\linewidth]{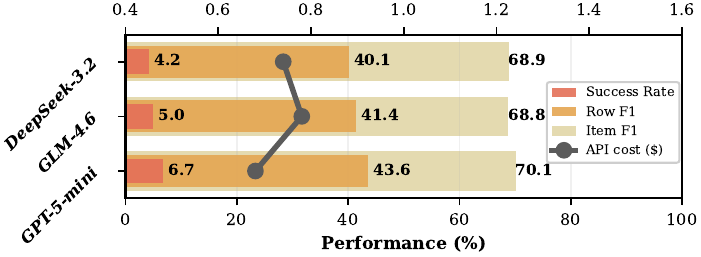}
    \caption{Backbone robustness.}
    \label{fig:sens_backbone}
  \end{subfigure}\hfill
  \begin{subfigure}[t]{0.99\columnwidth}
    \centering
    \includegraphics[width=\linewidth]{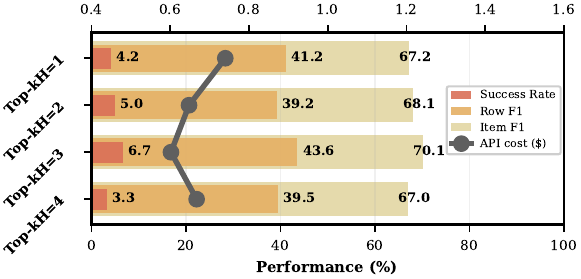}
    \caption{Sensitivity to insights topk.}
    \label{fig:sens_topk}
  \end{subfigure}
  \caption{Sensitivity analysis on WideSearch: (a) robustness across backbones; (b) sensitivity to the number of retrieved insights.}
  \label{fig:sensitivity}
\end{figure}

\begin{table}[!t]
\centering
\begin{small}
\caption{Ablation study of the proposed experiential prior design and MapReduce decision parameters $\Theta$ in \textsc{A-MapReduce}. We report Avg@4 for Item F1, Row F1, and per-task cost.}
\label{tab:ablation_study}
\resizebox{\columnwidth}{!}{%
\begin{tabular}{lccc}
\toprule
\textbf{Method} & \textbf{Item F1(\%)} & \textbf{Row F1(\%)} & \textbf{Cost(\$)} \\
\midrule
\textsc{A-MapReduce}
& 70.12 & 43.65 & 0.60 \\
\textit{-w/o Exemplars}
& 68.76\drop{1.36} & 40.57\drop{3.08} & 0.64 \\
\textit{-w/o Hint}
& 66.47\drop{3.65} & 40.58\drop{3.07} & 0.62 \\
\textit{-w/o Memory}
& 64.97\drop{5.15} & 37.92\drop{5.73} & 1.05 \\
\midrule
\textit{-w/o $M_q~\&~P_q$}
& 67.90\drop{2.22} & 37.85\drop{5.80} & 1.08 \\
\textit{-w/o $B_q$}
& 65.95\drop{4.17} & 37.33\drop{6.32} & 0.68 \\
\bottomrule
\end{tabular}
}
\vskip -0.1 in
\end{small}
\end{table}

\begin{figure*}[t]
  \begin{center}
    \centerline{\includegraphics[width=2.08\columnwidth]{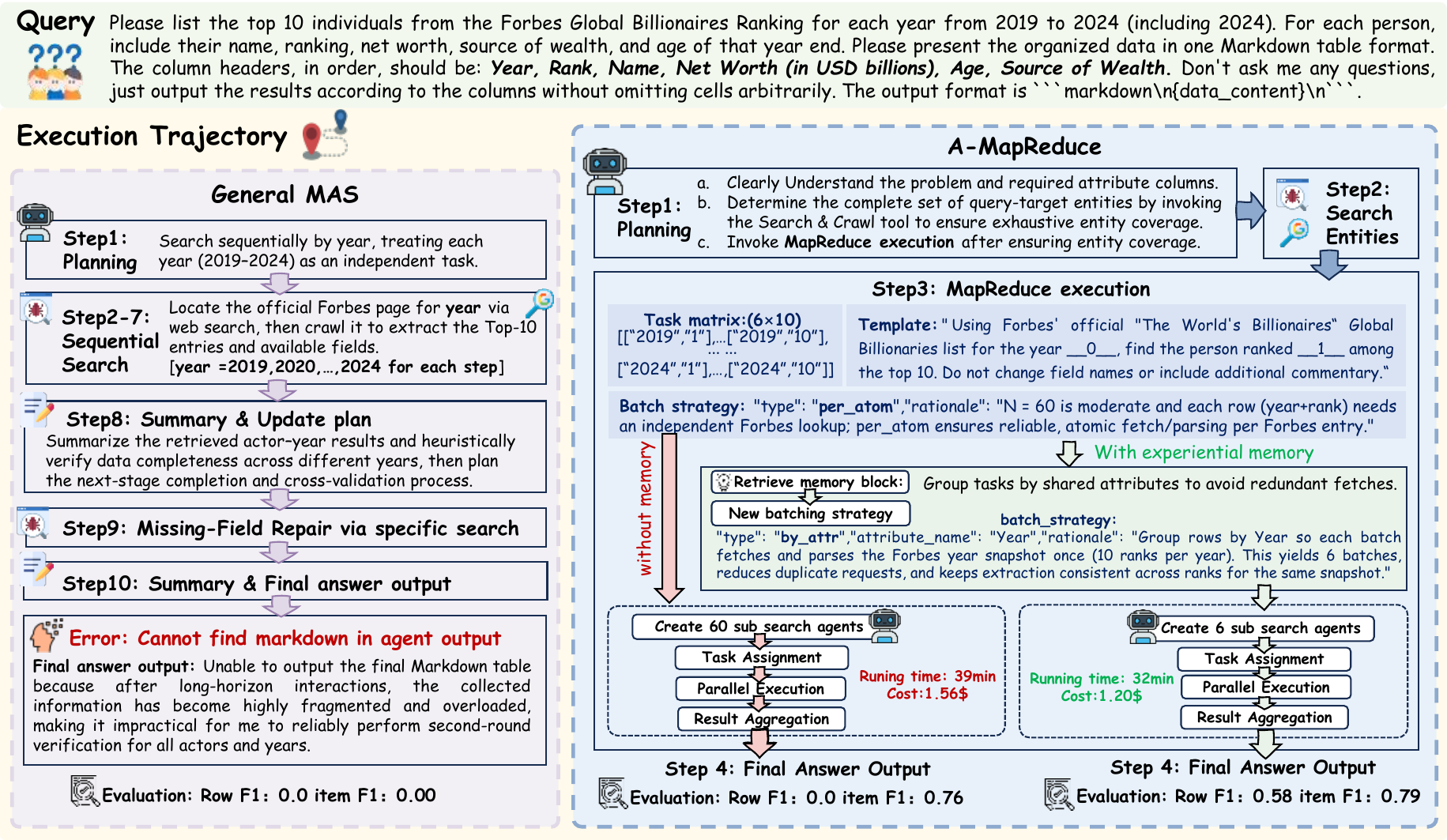}}
    \caption{Case study on a representative WideSearch query. \textbf{Left:} a general MAS executes year-by-year sequential retrieval, where long-horizon context accumulation can break structured-table construction. \textbf{Right:} \textsc{A-MapReduce} instantiates the query as a schema-driven task matrix and runs MapReduce-style execution, shown for variants without and with experiential memory.}
    \label{fig:case_study}
  \end{center}
  \vskip -0.3in
\end{figure*}


\subsection{Framework Analysis}
\label{framework_analysis}
\paragraph{Sensitivity Analysis.}
We evaluate robustness to backbone substitution by replacing GPT-5-mini with DeepSeek-v3.2 and GLM-4.6. As shown in Fig.~\ref{fig:sens_backbone}, \textsc{A-MapReduce} maintains stable Success Rate, Row F1, and Item F1, with variations within four percentage points, indicating low sensitivity to the underlying LLM. Specifically, Item F1 remains between 68.8\% and 70.1\% across the three backbones, suggesting that the structured execution design is effective across different models. We further analyze the sensitivity of the Top-$k_{\mathcal{H}}$ parameter for hint retrieval. Fig.~\ref{fig:sens_topk} and Tab.~\ref{tab:sensitivity_insights_topk} show peak performance at $k_{\mathcal{H}}=3$; smaller values provide insufficient guidance, while larger values may introduce irrelevant historical insights and degrade performance. In particular, increasing $k_{\mathcal{H}}$ from 3 to 4 reduces Item F1 from 70.12\% to 66.98\% and Row F1 from 43.65\% to 39.48\%. Accordingly, we use $k_{\mathcal{H}}=3$ by default to balance the diversity and relevance of retrieved guidance.

\paragraph{Ablation Study.}
Tab.~\ref{tab:ablation_study} shows that removing either the positive/negative exemplars $(\epsilon_{\mathrm{pos}}/\epsilon_{\mathrm{neg}})$ or the retrieved guidance $h_s$  consistently reduces both Item/Row F1, indicating complementary benefits of the experiential prior $\Delta_q$.
Disabling evolution (\textit{-w/o Memory}) causes the largest overall degradation (Item/Row F1: -5.15\%/-5.73\%) and increases cost (\$0.60$\rightarrow$\$1.05), suggesting that experiential memory mitigates repeated retrieval.
More importantly, the main gains come from structural decisions: removing structured task planning (\textit{-w/o $M_q~\&~P_q$}) sharply drops Row F1 (-5.80\%) and nearly doubles cost, while removing adaptive batching (\textit{-w/o $B_q$}) yields the largest Row/Item F1 drops (-6.32\%/-4.17\%).
A full precision/recall result is provided in Tab.~\ref{tab:ablation_study_complete}, which further confirms that weaker planning leads to more entity omissions and row-level misalignment under wide retrieval.


\subsection{Case Study}
We present a representative case in Fig.~\ref{fig:case_study} to highlight how \textsc{A-MapReduce} mitigates failure modes under long-horizon wide-scope retrieval. The task requires retrieving the top ten Forbes billionaires for each year from 2019 to 2024, involving 60 entity-year records with multiple required attributes. The general MAS baseline retrieves results sequentially by year, and intermediate evidence is gradually overwritten, leading to a fragmented state and an invalid table output. In contrast, \textsc{A-MapReduce} externalizes the retrieval objective as a schema-driven task matrix, enabling explicit coverage tracking and controlled MapReduce-style execution. Without experiential memory, it adopts a fine-grained \texttt{per\_atom} batching strategy, assigning 60 independent retrieval tasks to search agents. This recovers substantial item-level information (Item F1 = 0.76), but incurs higher cost and latency. With experiential memory enabled, \textsc{A-MapReduce} retrieves batching hints and switches to \texttt{by\_attr} batching, grouping tasks by year to reuse shared retrieval context. This reduces sub-agents from 60 to 6, decreases runtime from 39 to 32 minutes and cost from \$1.56 to \$1.20, while improving structural accuracy (Row F1 = 0.58, Item F1 = 0.79). We provide additional case studies in App.~\ref{app:case_studies} showing how execution decisions $\Theta=(M,P,B)$ evolve across different task structures and improve performance.


\section{Conclusion}
We study \emph{wide search} as a challenging regime for LLM-based multi-agent systems, where effectiveness depends on structured coverage and aggregation under long-horizon execution rather than deep sequential reasoning.
We propose \textsc{A-MapReduce}, a MapReduce-inspired framework that externalizes retrieval objectives into task matrices, instantiates atomic searches through query-conditioned templates, and coordinates parallel execution via adaptive batching and schema-aware reduction.
Beyond execution, experiential memory captures reusable decision patterns to refine task allocation and retrieval strategies across queries.
Extensive experiments across five benchmarks demonstrate improvements in retrieval quality, while delivering favorable cost--performance trade-offs and reduced execution time compared with multi-agent baselines.

\newpage

\bibliographystyle{ACM-Reference-Format}
\bibliography{sample-base}


\begin{thebibliography}{46}


\ifx \showCODEN    \undefined \def \showCODEN     #1{\unskip}     \fi
\ifx \showISBNx    \undefined \def \showISBNx     #1{\unskip}     \fi
\ifx \showISBNxiii \undefined \def \showISBNxiii  #1{\unskip}     \fi
\ifx \showISSN     \undefined \def \showISSN      #1{\unskip}     \fi
\ifx \showLCCN     \undefined \def \showLCCN      #1{\unskip}     \fi
\ifx \shownote     \undefined \def \shownote      #1{#1}          \fi
\ifx \showarticletitle \undefined \def \showarticletitle #1{#1}   \fi
\ifx \showURL      \undefined \def \showURL       {\relax}        \fi
\providecommand\bibfield[2]{#2}
\providecommand\bibinfo[2]{#2}
\providecommand\natexlab[1]{#1}
\providecommand\showeprint[2][]{arXiv:#2}

\bibitem[Chen et~al\mbox{.}(2025)]%
        {chen2025xbenchtrackingagentsproductivity}
\bibfield{author}{\bibinfo{person}{Kaiyuan Chen}, \bibinfo{person}{Yixin Ren}, \bibinfo{person}{Yang Liu}, \bibinfo{person}{Xiaobo Hu}, \bibinfo{person}{Haotong Tian}, \bibinfo{person}{Tianbao Xie}, \bibinfo{person}{Fangfu Liu}, \bibinfo{person}{Haoye Zhang}, \bibinfo{person}{Hongzhang Liu}, \bibinfo{person}{Yuan Gong}, \bibinfo{person}{Chen Sun}, \bibinfo{person}{Han Hou}, \bibinfo{person}{Hui Yang}, \bibinfo{person}{James Pan}, \bibinfo{person}{Jianan Lou}, \bibinfo{person}{Jiayi Mao}, \bibinfo{person}{Jizheng Liu}, \bibinfo{person}{Jinpeng Li}, \bibinfo{person}{Kangyi Liu}, \bibinfo{person}{Kenkun Liu}, \bibinfo{person}{Rui Wang}, \bibinfo{person}{Run Li}, \bibinfo{person}{Tong Niu}, \bibinfo{person}{Wenlong Zhang}, \bibinfo{person}{Wenqi Yan}, \bibinfo{person}{Xuanzheng Wang}, \bibinfo{person}{Yuchen Zhang}, \bibinfo{person}{Yi-Hsin Hung}, \bibinfo{person}{Yuan Jiang}, \bibinfo{person}{Zexuan Liu}, \bibinfo{person}{Zihan Yin}, \bibinfo{person}{Zijian Ma}, {and} \bibinfo{person}{Zhiwen Mo}.}
  \bibinfo{year}{2025}\natexlab{}.
\newblock \bibinfo{title}{xbench: Tracking Agents Productivity Scaling with Profession-Aligned Real-World Evaluations}.
\newblock
\showeprint[arxiv]{2506.13651}~[cs.LG]
\urldef\tempurl%
\url{https://arxiv.org/abs/2506.13651}
\showURL{%
\tempurl}


\bibitem[Chen et~al\mbox{.}(2024)]%
        {chen2024agentverse}
\bibfield{author}{\bibinfo{person}{Weize Chen}, \bibinfo{person}{Yusheng Su}, \bibinfo{person}{Jingwei Zuo}, \bibinfo{person}{Cheng Yang}, \bibinfo{person}{Chenfei Yuan}, \bibinfo{person}{Chi-Min Chan}, \bibinfo{person}{Heyang Yu}, \bibinfo{person}{Yaxi Lu}, \bibinfo{person}{Yi-Hsin Hung}, \bibinfo{person}{Chen Qian}, \bibinfo{person}{Yujia Qin}, \bibinfo{person}{Xin Cong}, \bibinfo{person}{Ruobing Xie}, \bibinfo{person}{Zhiyuan Liu}, \bibinfo{person}{Maosong Sun}, {and} \bibinfo{person}{Jie Zhou}.} \bibinfo{year}{2024}\natexlab{}.
\newblock \showarticletitle{AgentVerse: Facilitating Multi-Agent Collaboration and Exploring Emergent Behaviors}. In \bibinfo{booktitle}{\emph{The Twelfth International Conference on Learning Representations}}.
\newblock
\urldef\tempurl%
\url{https://openreview.net/forum?id=EHg5GDnyq1}
\showURL{%
\tempurl}


\bibitem[Chen et~al\mbox{.}(2026)]%
        {chen2026monolithicarchitecturesmultiagentsearch}
\bibfield{author}{\bibinfo{person}{Yiqun Chen}, \bibinfo{person}{Lingyong Yan}, \bibinfo{person}{Zixuan Yang}, \bibinfo{person}{Erhan Zhang}, \bibinfo{person}{Jiashu Zhao}, \bibinfo{person}{Shuaiqiang Wang}, \bibinfo{person}{Dawei Yin}, {and} \bibinfo{person}{Jiaxin Mao}.} \bibinfo{year}{2026}\natexlab{}.
\newblock \bibinfo{title}{Beyond Monolithic Architectures: A Multi-Agent Search and Knowledge Optimization Framework for Agentic Search}.
\newblock
\showeprint[arxiv]{2601.04703}~[cs.AI]
\urldef\tempurl%
\url{https://arxiv.org/abs/2601.04703}
\showURL{%
\tempurl}


\bibitem[Cobbe et~al\mbox{.}(2021)]%
        {cobbe2021trainingverifierssolvemath}
\bibfield{author}{\bibinfo{person}{Karl Cobbe}, \bibinfo{person}{Vineet Kosaraju}, \bibinfo{person}{Mohammad Bavarian}, \bibinfo{person}{Mark Chen}, \bibinfo{person}{Heewoo Jun}, \bibinfo{person}{Lukasz Kaiser}, \bibinfo{person}{Matthias Plappert}, \bibinfo{person}{Jerry Tworek}, \bibinfo{person}{Jacob Hilton}, \bibinfo{person}{Reiichiro Nakano}, \bibinfo{person}{Christopher Hesse}, {and} \bibinfo{person}{John Schulman}.} \bibinfo{year}{2021}\natexlab{}.
\newblock \bibinfo{title}{Training Verifiers to Solve Math Word Problems}.
\newblock
\showeprint[arxiv]{2110.14168}~[cs.LG]
\urldef\tempurl%
\url{https://arxiv.org/abs/2110.14168}
\showURL{%
\tempurl}


\bibitem[Dean and Ghemawat(2008)]%
        {dean2008mapreduce}
\bibfield{author}{\bibinfo{person}{Jeffrey Dean} {and} \bibinfo{person}{Sanjay Ghemawat}.} \bibinfo{year}{2008}\natexlab{}.
\newblock \showarticletitle{MapReduce: simplified data processing on large clusters}.
\newblock \bibinfo{journal}{\emph{Commun. ACM}} \bibinfo{volume}{51}, \bibinfo{number}{1} (\bibinfo{year}{2008}), \bibinfo{pages}{107--113}.
\newblock


\bibitem[Fang et~al\mbox{.}(2025a)]%
        {fang2025comprehensivesurveyselfevolvingai}
\bibfield{author}{\bibinfo{person}{Jinyuan Fang}, \bibinfo{person}{Yanwen Peng}, \bibinfo{person}{Xi Zhang}, \bibinfo{person}{Yingxu Wang}, \bibinfo{person}{Xinhao Yi}, \bibinfo{person}{Guibin Zhang}, \bibinfo{person}{Yi Xu}, \bibinfo{person}{Bin Wu}, \bibinfo{person}{Siwei Liu}, \bibinfo{person}{Zihao Li}, \bibinfo{person}{Zhaochun Ren}, \bibinfo{person}{Nikos Aletras}, \bibinfo{person}{Xi Wang}, \bibinfo{person}{Han Zhou}, {and} \bibinfo{person}{Zaiqiao Meng}.} \bibinfo{year}{2025}\natexlab{a}.
\newblock \bibinfo{title}{A Comprehensive Survey of Self-Evolving AI Agents: A New Paradigm Bridging Foundation Models and Lifelong Agentic Systems}.
\newblock
\showeprint[arxiv]{2508.07407}~[cs.AI]
\urldef\tempurl%
\url{https://arxiv.org/abs/2508.07407}
\showURL{%
\tempurl}


\bibitem[Fang et~al\mbox{.}(2025b)]%
        {fang2025cognitive}
\bibfield{author}{\bibinfo{person}{Tianqing Fang}, \bibinfo{person}{Zhisong Zhang}, \bibinfo{person}{Xiaoyang Wang}, \bibinfo{person}{Rui Wang}, \bibinfo{person}{Can Qin}, \bibinfo{person}{Yuxuan Wan}, \bibinfo{person}{Jun-Yu Ma}, \bibinfo{person}{Ce Zhang}, \bibinfo{person}{Jiaqi Chen}, \bibinfo{person}{Xiyun Li}, {et~al\mbox{.}}} \bibinfo{year}{2025}\natexlab{b}.
\newblock \showarticletitle{Cognitive kernel-pro: A framework for deep research agents and agent foundation models training}.
\newblock \bibinfo{journal}{\emph{arXiv preprint arXiv:2508.00414}} (\bibinfo{year}{2025}).
\newblock


\bibitem[Fourney et~al\mbox{.}(2024)]%
        {fourney2024magentic}
\bibfield{author}{\bibinfo{person}{Adam Fourney}, \bibinfo{person}{Gagan Bansal}, \bibinfo{person}{Hussein Mozannar}, \bibinfo{person}{Cheng Tan}, \bibinfo{person}{Eduardo Salinas}, \bibinfo{person}{Friederike Niedtner}, \bibinfo{person}{Grace Proebsting}, \bibinfo{person}{Griffin Bassman}, \bibinfo{person}{Jack Gerrits}, \bibinfo{person}{Jacob Alber}, {et~al\mbox{.}}} \bibinfo{year}{2024}\natexlab{}.
\newblock \showarticletitle{Magentic-one: A generalist multi-agent system for solving complex tasks}.
\newblock \bibinfo{journal}{\emph{arXiv preprint arXiv:2411.04468}} (\bibinfo{year}{2024}).
\newblock


\bibitem[Guo et~al\mbox{.}(2024)]%
        {guo2024connecting}
\bibfield{author}{\bibinfo{person}{Qingyan Guo}, \bibinfo{person}{Rui Wang}, \bibinfo{person}{Junliang Guo}, \bibinfo{person}{Bei Li}, \bibinfo{person}{Kaitao Song}, \bibinfo{person}{Xu Tan}, \bibinfo{person}{Guoqing Liu}, \bibinfo{person}{Jiang Bian}, {and} \bibinfo{person}{Yujiu Yang}.} \bibinfo{year}{2024}\natexlab{}.
\newblock \showarticletitle{Connecting Large Language Models with Evolutionary Algorithms Yields Powerful Prompt Optimizers}. In \bibinfo{booktitle}{\emph{The Twelfth International Conference on Learning Representations}}.
\newblock
\urldef\tempurl%
\url{https://openreview.net/forum?id=ZG3RaNIsO8}
\showURL{%
\tempurl}


\bibitem[He et~al\mbox{.}(2024)]%
        {he2024webvoyager}
\bibfield{author}{\bibinfo{person}{Hongliang He}, \bibinfo{person}{Wenlin Yao}, \bibinfo{person}{Kaixin Ma}, \bibinfo{person}{Wenhao Yu}, \bibinfo{person}{Yong Dai}, \bibinfo{person}{Hongming Zhang}, \bibinfo{person}{Zhenzhong Lan}, {and} \bibinfo{person}{Dong Yu}.} \bibinfo{year}{2024}\natexlab{}.
\newblock \showarticletitle{Webvoyager: Building an end-to-end web agent with large multimodal models}.
\newblock \bibinfo{journal}{\emph{arXiv preprint arXiv:2401.13919}} (\bibinfo{year}{2024}).
\newblock


\bibitem[Hong et~al\mbox{.}(2024)]%
        {hong2024metagpt}
\bibfield{author}{\bibinfo{person}{Sirui Hong}, \bibinfo{person}{Mingchen Zhuge}, \bibinfo{person}{Jonathan Chen}, \bibinfo{person}{Xiawu Zheng}, \bibinfo{person}{Yuheng Cheng}, \bibinfo{person}{Jinlin Wang}, \bibinfo{person}{Ceyao Zhang}, \bibinfo{person}{Zili Wang}, \bibinfo{person}{Steven Ka~Shing Yau}, \bibinfo{person}{Zijuan Lin}, \bibinfo{person}{Liyang Zhou}, \bibinfo{person}{Chenyu Ran}, \bibinfo{person}{Lingfeng Xiao}, \bibinfo{person}{Chenglin Wu}, {and} \bibinfo{person}{J{\"u}rgen Schmidhuber}.} \bibinfo{year}{2024}\natexlab{}.
\newblock \showarticletitle{Meta{GPT}: Meta Programming for A Multi-Agent Collaborative Framework}. In \bibinfo{booktitle}{\emph{The Twelfth International Conference on Learning Representations}}.
\newblock
\urldef\tempurl%
\url{https://openreview.net/forum?id=VtmBAGCN7o}
\showURL{%
\tempurl}


\bibitem[Hu et~al\mbox{.}(2025b)]%
        {hu2025owl}
\bibfield{author}{\bibinfo{person}{Mengkang Hu}, \bibinfo{person}{Yuhang Zhou}, \bibinfo{person}{Wendong Fan}, \bibinfo{person}{Yuzhou Nie}, \bibinfo{person}{Bowei Xia}, \bibinfo{person}{Tao Sun}, \bibinfo{person}{Ziyu Ye}, \bibinfo{person}{Zhaoxuan Jin}, \bibinfo{person}{Yingru Li}, \bibinfo{person}{Zeyu Zhang}, \bibinfo{person}{Yifeng Wang}, \bibinfo{person}{Qianshuo Ye}, \bibinfo{person}{Ping Luo}, {and} \bibinfo{person}{Guohao Li}.} \bibinfo{year}{2025}\natexlab{b}.
\newblock \bibinfo{title}{OWL: Optimized Workforce Learning for General Multi-Agent Assistance in Real-World Task Automation}.
\newblock
\urldef\tempurl%
\url{https://github.com/camel-ai/owl}
\showURL{%
\tempurl}


\bibitem[Hu et~al\mbox{.}(2025a)]%
        {hu2025automated}
\bibfield{author}{\bibinfo{person}{Shengran Hu}, \bibinfo{person}{Cong Lu}, {and} \bibinfo{person}{Jeff Clune}.} \bibinfo{year}{2025}\natexlab{a}.
\newblock \showarticletitle{Automated Design of Agentic Systems}. In \bibinfo{booktitle}{\emph{The Thirteenth International Conference on Learning Representations}}.
\newblock
\urldef\tempurl%
\url{https://openreview.net/forum?id=t9U3LW7JVX}
\showURL{%
\tempurl}


\bibitem[Jimenez et~al\mbox{.}(2024)]%
        {jimenez2024swebench}
\bibfield{author}{\bibinfo{person}{Carlos~E Jimenez}, \bibinfo{person}{John Yang}, \bibinfo{person}{Alexander Wettig}, \bibinfo{person}{Shunyu Yao}, \bibinfo{person}{Kexin Pei}, \bibinfo{person}{Ofir Press}, {and} \bibinfo{person}{Karthik~R Narasimhan}.} \bibinfo{year}{2024}\natexlab{}.
\newblock \showarticletitle{{SWE}-bench: Can Language Models Resolve Real-world Github Issues?}. In \bibinfo{booktitle}{\emph{The Twelfth International Conference on Learning Representations}}.
\newblock
\urldef\tempurl%
\url{https://openreview.net/forum?id=VTF8yNQM66}
\showURL{%
\tempurl}


\bibitem[Lan et~al\mbox{.}(2025)]%
        {lan2025deepwidesearch}
\bibfield{author}{\bibinfo{person}{Tian Lan}, \bibinfo{person}{Bin Zhu}, \bibinfo{person}{Qianghuai Jia}, \bibinfo{person}{Junyang Ren}, \bibinfo{person}{Haijun Li}, \bibinfo{person}{Longyue Wang}, \bibinfo{person}{Zhao Xu}, \bibinfo{person}{Weihua Luo}, {and} \bibinfo{person}{Kaifu Zhang}.} \bibinfo{year}{2025}\natexlab{}.
\newblock \showarticletitle{DeepWideSearch: Benchmarking Depth and Width in Agentic Information Seeking}.
\newblock \bibinfo{journal}{\emph{arXiv preprint arXiv:2510.20168}} (\bibinfo{year}{2025}).
\newblock


\bibitem[Li et~al\mbox{.}(2023)]%
        {li2023camel}
\bibfield{author}{\bibinfo{person}{Guohao Li}, \bibinfo{person}{Hasan Hammoud}, \bibinfo{person}{Hani Itani}, \bibinfo{person}{Dmitrii Khizbullin}, {and} \bibinfo{person}{Bernard Ghanem}.} \bibinfo{year}{2023}\natexlab{}.
\newblock \showarticletitle{Camel: Communicative agents for" mind" exploration of large language model society}.
\newblock \bibinfo{journal}{\emph{Advances in Neural Information Processing Systems}}  \bibinfo{volume}{36} (\bibinfo{year}{2023}), \bibinfo{pages}{51991--52008}.
\newblock


\bibitem[Li et~al\mbox{.}(2025)]%
        {li2025websailornavigatingsuperhumanreasoning}
\bibfield{author}{\bibinfo{person}{Kuan Li}, \bibinfo{person}{Zhongwang Zhang}, \bibinfo{person}{Huifeng Yin}, \bibinfo{person}{Liwen Zhang}, \bibinfo{person}{Litu Ou}, \bibinfo{person}{Jialong Wu}, \bibinfo{person}{Wenbiao Yin}, \bibinfo{person}{Baixuan Li}, \bibinfo{person}{Zhengwei Tao}, \bibinfo{person}{Xinyu Wang}, \bibinfo{person}{Weizhou Shen}, \bibinfo{person}{Junkai Zhang}, \bibinfo{person}{Dingchu Zhang}, \bibinfo{person}{Xixi Wu}, \bibinfo{person}{Yong Jiang}, \bibinfo{person}{Ming Yan}, \bibinfo{person}{Pengjun Xie}, \bibinfo{person}{Fei Huang}, {and} \bibinfo{person}{Jingren Zhou}.} \bibinfo{year}{2025}\natexlab{}.
\newblock \bibinfo{title}{WebSailor: Navigating Super-human Reasoning for Web Agent}.
\newblock
\showeprint[arxiv]{2507.02592}~[cs.CL]
\urldef\tempurl%
\url{https://arxiv.org/abs/2507.02592}
\showURL{%
\tempurl}


\bibitem[Li et~al\mbox{.}(2024)]%
        {li2024autoflowautomatedworkflowgeneration}
\bibfield{author}{\bibinfo{person}{Zelong Li}, \bibinfo{person}{Shuyuan Xu}, \bibinfo{person}{Kai Mei}, \bibinfo{person}{Wenyue Hua}, \bibinfo{person}{Balaji Rama}, \bibinfo{person}{Om Raheja}, \bibinfo{person}{Hao Wang}, \bibinfo{person}{He Zhu}, {and} \bibinfo{person}{Yongfeng Zhang}.} \bibinfo{year}{2024}\natexlab{}.
\newblock \bibinfo{title}{AutoFlow: Automated Workflow Generation for Large Language Model Agents}.
\newblock
\showeprint[arxiv]{2407.12821}~[cs.CL]
\urldef\tempurl%
\url{https://arxiv.org/abs/2407.12821}
\showURL{%
\tempurl}


\bibitem[Liu et~al\mbox{.}(2024)]%
        {liu2024agentbench}
\bibfield{author}{\bibinfo{person}{Xiao Liu}, \bibinfo{person}{Hao Yu}, \bibinfo{person}{Hanchen Zhang}, \bibinfo{person}{Yifan Xu}, \bibinfo{person}{Xuanyu Lei}, \bibinfo{person}{Hanyu Lai}, \bibinfo{person}{Yu Gu}, \bibinfo{person}{Hangliang Ding}, \bibinfo{person}{Kaiwen Men}, \bibinfo{person}{Kejuan Yang}, \bibinfo{person}{Shudan Zhang}, \bibinfo{person}{Xiang Deng}, \bibinfo{person}{Aohan Zeng}, \bibinfo{person}{Zhengxiao Du}, \bibinfo{person}{Chenhui Zhang}, \bibinfo{person}{Sheng Shen}, \bibinfo{person}{Tianjun Zhang}, \bibinfo{person}{Yu Su}, \bibinfo{person}{Huan Sun}, \bibinfo{person}{Minlie Huang}, \bibinfo{person}{Yuxiao Dong}, {and} \bibinfo{person}{Jie Tang}.} \bibinfo{year}{2024}\natexlab{}.
\newblock \showarticletitle{AgentBench: Evaluating {LLM}s as Agents}. In \bibinfo{booktitle}{\emph{The Twelfth International Conference on Learning Representations}}.
\newblock
\urldef\tempurl%
\url{https://openreview.net/forum?id=zAdUB0aCTQ}
\showURL{%
\tempurl}


\bibitem[Luo et~al\mbox{.}(2024)]%
        {luo2024improvemathematicalreasoninglanguage}
\bibfield{author}{\bibinfo{person}{Liangchen Luo}, \bibinfo{person}{Yinxiao Liu}, \bibinfo{person}{Rosanne Liu}, \bibinfo{person}{Samrat Phatale}, \bibinfo{person}{Meiqi Guo}, \bibinfo{person}{Harsh Lara}, \bibinfo{person}{Yunxuan Li}, \bibinfo{person}{Lei Shu}, \bibinfo{person}{Yun Zhu}, \bibinfo{person}{Lei Meng}, \bibinfo{person}{Jiao Sun}, {and} \bibinfo{person}{Abhinav Rastogi}.} \bibinfo{year}{2024}\natexlab{}.
\newblock \bibinfo{title}{Improve Mathematical Reasoning in Language Models by Automated Process Supervision}.
\newblock
\showeprint[arxiv]{2406.06592}~[cs.CL]
\urldef\tempurl%
\url{https://arxiv.org/abs/2406.06592}
\showURL{%
\tempurl}


\bibitem[Mialon et~al\mbox{.}(2024)]%
        {mialon2024gaia}
\bibfield{author}{\bibinfo{person}{Gr{\'e}goire Mialon}, \bibinfo{person}{Cl{\'e}mentine Fourrier}, \bibinfo{person}{Thomas Wolf}, \bibinfo{person}{Yann LeCun}, {and} \bibinfo{person}{Thomas Scialom}.} \bibinfo{year}{2024}\natexlab{}.
\newblock \showarticletitle{{GAIA}: a benchmark for General {AI} Assistants}. In \bibinfo{booktitle}{\emph{The Twelfth International Conference on Learning Representations}}.
\newblock
\urldef\tempurl%
\url{https://openreview.net/forum?id=fibxvahvs3}
\showURL{%
\tempurl}


\bibitem[Nakano et~al\mbox{.}(2022)]%
        {nakano2022webgptbrowserassistedquestionansweringhuman}
\bibfield{author}{\bibinfo{person}{Reiichiro Nakano}, \bibinfo{person}{Jacob Hilton}, \bibinfo{person}{Suchir Balaji}, \bibinfo{person}{Jeff Wu}, \bibinfo{person}{Long Ouyang}, \bibinfo{person}{Christina Kim}, \bibinfo{person}{Christopher Hesse}, \bibinfo{person}{Shantanu Jain}, \bibinfo{person}{Vineet Kosaraju}, \bibinfo{person}{William Saunders}, \bibinfo{person}{Xu Jiang}, \bibinfo{person}{Karl Cobbe}, \bibinfo{person}{Tyna Eloundou}, \bibinfo{person}{Gretchen Krueger}, \bibinfo{person}{Kevin Button}, \bibinfo{person}{Matthew Knight}, \bibinfo{person}{Benjamin Chess}, {and} \bibinfo{person}{John Schulman}.} \bibinfo{year}{2022}\natexlab{}.
\newblock \bibinfo{title}{WebGPT: Browser-assisted question-answering with human feedback}.
\newblock
\showeprint[arxiv]{2112.09332}~[cs.CL]
\urldef\tempurl%
\url{https://arxiv.org/abs/2112.09332}
\showURL{%
\tempurl}


\bibitem[Phan et~al\mbox{.}(2025)]%
        {phan2025humanity}
\bibfield{author}{\bibinfo{person}{Long Phan}, \bibinfo{person}{Alice Gatti}, \bibinfo{person}{Ziwen Han}, \bibinfo{person}{Nathaniel Li}, \bibinfo{person}{Josephina Hu}, \bibinfo{person}{Hugh Zhang}, \bibinfo{person}{Chen Bo~Calvin Zhang}, \bibinfo{person}{Mohamed Shaaban}, \bibinfo{person}{John Ling}, \bibinfo{person}{Sean Shi}, {et~al\mbox{.}}} \bibinfo{year}{2025}\natexlab{}.
\newblock \showarticletitle{Humanity's last exam}.
\newblock \bibinfo{journal}{\emph{arXiv preprint arXiv:2501.14249}} (\bibinfo{year}{2025}).
\newblock


\bibitem[Qian et~al\mbox{.}(2024)]%
        {qian-etal-2024-chatdev}
\bibfield{author}{\bibinfo{person}{Chen Qian}, \bibinfo{person}{Wei Liu}, \bibinfo{person}{Hongzhang Liu}, \bibinfo{person}{Nuo Chen}, \bibinfo{person}{Yufan Dang}, \bibinfo{person}{Jiahao Li}, \bibinfo{person}{Cheng Yang}, \bibinfo{person}{Weize Chen}, \bibinfo{person}{Yusheng Su}, \bibinfo{person}{Xin Cong}, \bibinfo{person}{Juyuan Xu}, \bibinfo{person}{Dahai Li}, \bibinfo{person}{Zhiyuan Liu}, {and} \bibinfo{person}{Maosong Sun}.} \bibinfo{year}{2024}\natexlab{}.
\newblock \showarticletitle{{C}hat{D}ev: Communicative Agents for Software Development}. In \bibinfo{booktitle}{\emph{Proceedings of the 62nd Annual Meeting of the Association for Computational Linguistics (Volume 1: Long Papers)}}, \bibfield{editor}{\bibinfo{person}{Lun-Wei Ku}, \bibinfo{person}{Andre Martins}, {and} \bibinfo{person}{Vivek Srikumar}} (Eds.). \bibinfo{publisher}{Association for Computational Linguistics}, \bibinfo{address}{Bangkok, Thailand}, \bibinfo{pages}{15174--15186}.
\newblock
\href{https://doi.org/10.18653/v1/2024.acl-long.810}{doi:\nolinkurl{10.18653/v1/2024.acl-long.810}}


\bibitem[Qin et~al\mbox{.}(2025)]%
        {qin2025flashsearcherfasteffectiveweb}
\bibfield{author}{\bibinfo{person}{Tianrui Qin}, \bibinfo{person}{Qianben Chen}, \bibinfo{person}{Sinuo Wang}, \bibinfo{person}{He Xing}, \bibinfo{person}{King Zhu}, \bibinfo{person}{He Zhu}, \bibinfo{person}{Dingfeng Shi}, \bibinfo{person}{Xinxin Liu}, \bibinfo{person}{Ge Zhang}, \bibinfo{person}{Jiaheng Liu}, \bibinfo{person}{Yuchen~Eleanor Jiang}, \bibinfo{person}{Xitong Gao}, {and} \bibinfo{person}{Wangchunshu Zhou}.} \bibinfo{year}{2025}\natexlab{}.
\newblock \bibinfo{title}{Flash-Searcher: Fast and Effective Web Agents via DAG-Based Parallel Execution}.
\newblock
\showeprint[arxiv]{2509.25301}~[cs.AI]
\urldef\tempurl%
\url{https://arxiv.org/abs/2509.25301}
\showURL{%
\tempurl}


\bibitem[Qiu et~al\mbox{.}(2025)]%
        {qiu2025alita}
\bibfield{author}{\bibinfo{person}{Jiahao Qiu}, \bibinfo{person}{Xuan Qi}, \bibinfo{person}{Tongcheng Zhang}, \bibinfo{person}{Xinzhe Juan}, \bibinfo{person}{Jiacheng Guo}, \bibinfo{person}{Yifu Lu}, \bibinfo{person}{Yimin Wang}, \bibinfo{person}{Zixin Yao}, \bibinfo{person}{Qihan Ren}, \bibinfo{person}{Xun Jiang}, {et~al\mbox{.}}} \bibinfo{year}{2025}\natexlab{}.
\newblock \showarticletitle{Alita: Generalist agent enabling scalable agentic reasoning with minimal predefinition and maximal self-evolution}.
\newblock \bibinfo{journal}{\emph{arXiv preprint arXiv:2505.20286}} (\bibinfo{year}{2025}).
\newblock


\bibitem[Roucher et~al\mbox{.}(2025)]%
        {roucher2025smolagents}
\bibfield{author}{\bibinfo{person}{Aymeric Roucher}, \bibinfo{person}{Albert~Villanova del Moral}, \bibinfo{person}{Thomas Wolf}, \bibinfo{person}{Leandro von Werra}, {and} \bibinfo{person}{Erik Kaunismäki}.} \bibinfo{year}{2025}\natexlab{}.
\newblock \bibinfo{title}{`smolagents`: a smol library to build great agentic systems.}
\newblock \bibinfo{howpublished}{\url{https://github.com/huggingface/smolagents}}.
\newblock


\bibitem[Shi et~al\mbox{.}(2025)]%
        {shi2025taskcraftautomatedgenerationagentic}
\bibfield{author}{\bibinfo{person}{Dingfeng Shi}, \bibinfo{person}{Jingyi Cao}, \bibinfo{person}{Qianben Chen}, \bibinfo{person}{Weichen Sun}, \bibinfo{person}{Weizhen Li}, \bibinfo{person}{Hongxuan Lu}, \bibinfo{person}{Fangchen Dong}, \bibinfo{person}{Tianrui Qin}, \bibinfo{person}{King Zhu}, \bibinfo{person}{Minghao Liu}, \bibinfo{person}{Jian Yang}, \bibinfo{person}{Ge Zhang}, \bibinfo{person}{Jiaheng Liu}, \bibinfo{person}{Changwang Zhang}, \bibinfo{person}{Jun Wang}, \bibinfo{person}{Yuchen~Eleanor Jiang}, {and} \bibinfo{person}{Wangchunshu Zhou}.} \bibinfo{year}{2025}\natexlab{}.
\newblock \bibinfo{title}{TaskCraft: Automated Generation of Agentic Tasks}.
\newblock
\showeprint[arxiv]{2506.10055}~[cs.CL]
\urldef\tempurl%
\url{https://arxiv.org/abs/2506.10055}
\showURL{%
\tempurl}


\bibitem[Tang et~al\mbox{.}(2025)]%
        {tang2025chemagent}
\bibfield{author}{\bibinfo{person}{Xiangru Tang}, \bibinfo{person}{Tianyu Hu}, \bibinfo{person}{Muyang Ye}, \bibinfo{person}{Yanjun Shao}, \bibinfo{person}{Xunjian Yin}, \bibinfo{person}{Siru Ouyang}, \bibinfo{person}{Wangchunshu Zhou}, \bibinfo{person}{Pan Lu}, \bibinfo{person}{Zhuosheng Zhang}, \bibinfo{person}{Yilun Zhao}, \bibinfo{person}{Arman Cohan}, {and} \bibinfo{person}{Mark Gerstein}.} \bibinfo{year}{2025}\natexlab{}.
\newblock \showarticletitle{ChemAgent: Self-updating Memories in Large Language Models Improves Chemical Reasoning}. In \bibinfo{booktitle}{\emph{The Thirteenth International Conference on Learning Representations}}.
\newblock
\urldef\tempurl%
\url{https://openreview.net/forum?id=kuhIqeVg0e}
\showURL{%
\tempurl}


\bibitem[Tao et~al\mbox{.}(2025)]%
        {tao2025webleaper}
\bibfield{author}{\bibinfo{person}{Zhengwei Tao}, \bibinfo{person}{Haiyang Shen}, \bibinfo{person}{Baixuan Li}, \bibinfo{person}{Wenbiao Yin}, \bibinfo{person}{Jialong Wu}, \bibinfo{person}{Kuan Li}, \bibinfo{person}{Zhongwang Zhang}, \bibinfo{person}{Huifeng Yin}, \bibinfo{person}{Rui Ye}, \bibinfo{person}{Liwen Zhang}, {et~al\mbox{.}}} \bibinfo{year}{2025}\natexlab{}.
\newblock \showarticletitle{Webleaper: Empowering efficiency and efficacy in webagent via enabling info-rich seeking}.
\newblock \bibinfo{journal}{\emph{arXiv preprint arXiv:2510.24697}} (\bibinfo{year}{2025}).
\newblock


\bibitem[Tran et~al\mbox{.}(2025)]%
        {tran2025multiagentcollaborationmechanismssurvey}
\bibfield{author}{\bibinfo{person}{Khanh-Tung Tran}, \bibinfo{person}{Dung Dao}, \bibinfo{person}{Minh-Duong Nguyen}, \bibinfo{person}{Quoc-Viet Pham}, \bibinfo{person}{Barry O'Sullivan}, {and} \bibinfo{person}{Hoang~D. Nguyen}.} \bibinfo{year}{2025}\natexlab{}.
\newblock \bibinfo{title}{Multi-Agent Collaboration Mechanisms: A Survey of LLMs}.
\newblock
\showeprint[arxiv]{2501.06322}~[cs.AI]
\urldef\tempurl%
\url{https://arxiv.org/abs/2501.06322}
\showURL{%
\tempurl}


\bibitem[Wang et~al\mbox{.}(2025)]%
        {wang2025mas2selfgenerativeselfconfiguringselfrectifying}
\bibfield{author}{\bibinfo{person}{Kun Wang}, \bibinfo{person}{Guibin Zhang}, \bibinfo{person}{ManKit Ye}, \bibinfo{person}{Xinyu Deng}, \bibinfo{person}{Dongxia Wang}, \bibinfo{person}{Xiaobin Hu}, \bibinfo{person}{Jinyang Guo}, \bibinfo{person}{Yang Liu}, {and} \bibinfo{person}{Yufei Guo}.} \bibinfo{year}{2025}\natexlab{}.
\newblock \bibinfo{title}{MAS$^2$: Self-Generative, Self-Configuring, Self-Rectifying Multi-Agent Systems}.
\newblock
\showeprint[arxiv]{2509.24323}~[cs.MA]
\urldef\tempurl%
\url{https://arxiv.org/abs/2509.24323}
\showURL{%
\tempurl}


\bibitem[Wei et~al\mbox{.}(2025)]%
        {wei2025browsecompsimplechallengingbenchmark}
\bibfield{author}{\bibinfo{person}{Jason Wei}, \bibinfo{person}{Zhiqing Sun}, \bibinfo{person}{Spencer Papay}, \bibinfo{person}{Scott McKinney}, \bibinfo{person}{Jeffrey Han}, \bibinfo{person}{Isa Fulford}, \bibinfo{person}{Hyung~Won Chung}, \bibinfo{person}{Alex~Tachard Passos}, \bibinfo{person}{William Fedus}, {and} \bibinfo{person}{Amelia Glaese}.} \bibinfo{year}{2025}\natexlab{}.
\newblock \bibinfo{title}{BrowseComp: A Simple Yet Challenging Benchmark for Browsing Agents}.
\newblock
\showeprint[arxiv]{2504.12516}~[cs.CL]
\urldef\tempurl%
\url{https://arxiv.org/abs/2504.12516}
\showURL{%
\tempurl}


\bibitem[Wei et~al\mbox{.}(2022)]%
        {wei2022chain}
\bibfield{author}{\bibinfo{person}{Jason Wei}, \bibinfo{person}{Xuezhi Wang}, \bibinfo{person}{Dale Schuurmans}, \bibinfo{person}{Maarten Bosma}, \bibinfo{person}{Fei Xia}, \bibinfo{person}{Ed Chi}, \bibinfo{person}{Quoc~V Le}, \bibinfo{person}{Denny Zhou}, {et~al\mbox{.}}} \bibinfo{year}{2022}\natexlab{}.
\newblock \showarticletitle{Chain-of-thought prompting elicits reasoning in large language models}.
\newblock \bibinfo{journal}{\emph{Advances in neural information processing systems}}  \bibinfo{volume}{35} (\bibinfo{year}{2022}), \bibinfo{pages}{24824--24837}.
\newblock


\bibitem[Wong et~al\mbox{.}(2025)]%
        {wong2025widesearch}
\bibfield{author}{\bibinfo{person}{Ryan Wong}, \bibinfo{person}{Jiawei Wang}, \bibinfo{person}{Junjie Zhao}, \bibinfo{person}{Li Chen}, \bibinfo{person}{Yan Gao}, \bibinfo{person}{Long Zhang}, \bibinfo{person}{Xuan Zhou}, \bibinfo{person}{Zuo Wang}, \bibinfo{person}{Kai Xiang}, \bibinfo{person}{Ge Zhang}, {et~al\mbox{.}}} \bibinfo{year}{2025}\natexlab{}.
\newblock \showarticletitle{Widesearch: Benchmarking agentic broad info-seeking}.
\newblock \bibinfo{journal}{\emph{arXiv preprint arXiv:2508.07999}} (\bibinfo{year}{2025}).
\newblock


\bibitem[Wu et~al\mbox{.}(2025)]%
        {wu2025webwalkerbenchmarkingllmsweb}
\bibfield{author}{\bibinfo{person}{Jialong Wu}, \bibinfo{person}{Wenbiao Yin}, \bibinfo{person}{Yong Jiang}, \bibinfo{person}{Zhenglin Wang}, \bibinfo{person}{Zekun Xi}, \bibinfo{person}{Runnan Fang}, \bibinfo{person}{Linhai Zhang}, \bibinfo{person}{Yulan He}, \bibinfo{person}{Deyu Zhou}, \bibinfo{person}{Pengjun Xie}, {and} \bibinfo{person}{Fei Huang}.} \bibinfo{year}{2025}\natexlab{}.
\newblock \bibinfo{title}{WebWalker: Benchmarking LLMs in Web Traversal}.
\newblock
\showeprint[arxiv]{2501.07572}~[cs.CL]
\urldef\tempurl%
\url{https://arxiv.org/abs/2501.07572}
\showURL{%
\tempurl}


\bibitem[Wu et~al\mbox{.}(2024)]%
        {wu2024autogen}
\bibfield{author}{\bibinfo{person}{Qingyun Wu}, \bibinfo{person}{Gagan Bansal}, \bibinfo{person}{Jieyu Zhang}, \bibinfo{person}{Yiran Wu}, \bibinfo{person}{Beibin Li}, \bibinfo{person}{Erkang Zhu}, \bibinfo{person}{Li Jiang}, \bibinfo{person}{Xiaoyun Zhang}, \bibinfo{person}{Shaokun Zhang}, \bibinfo{person}{Jiale Liu}, {et~al\mbox{.}}} \bibinfo{year}{2024}\natexlab{}.
\newblock \showarticletitle{Autogen: Enabling next-gen LLM applications via multi-agent conversations}. In \bibinfo{booktitle}{\emph{First Conference on Language Modeling}}.
\newblock


\bibitem[Yao et~al\mbox{.}(2023)]%
        {yao2023react}
\bibfield{author}{\bibinfo{person}{Shunyu Yao}, \bibinfo{person}{Jeffrey Zhao}, \bibinfo{person}{Dian Yu}, \bibinfo{person}{Nan Du}, \bibinfo{person}{Izhak Shafran}, \bibinfo{person}{Karthik~R Narasimhan}, {and} \bibinfo{person}{Yuan Cao}.} \bibinfo{year}{2023}\natexlab{}.
\newblock \showarticletitle{ReAct: Synergizing Reasoning and Acting in Language Models}. In \bibinfo{booktitle}{\emph{The Eleventh International Conference on Learning Representations}}.
\newblock
\urldef\tempurl%
\url{https://openreview.net/forum?id=WE_vluYUL-X}
\showURL{%
\tempurl}


\bibitem[Yao et~al\mbox{.}(2026)]%
        {yao2026oresearcheropenendeddeep}
\bibfield{author}{\bibinfo{person}{Yi Yao}, \bibinfo{person}{He Zhu}, \bibinfo{person}{Piaohong Wang}, \bibinfo{person}{Jincheng Ren}, \bibinfo{person}{Xinlong Yang}, \bibinfo{person}{Qianben Chen}, \bibinfo{person}{Xiaowan Li}, \bibinfo{person}{Dingfeng Shi}, \bibinfo{person}{Jiaxian Li}, \bibinfo{person}{Qiexiang Wang}, \bibinfo{person}{Sinuo Wang}, \bibinfo{person}{Xinpeng Liu}, \bibinfo{person}{Jiaqi Wu}, \bibinfo{person}{Minghao Liu}, {and} \bibinfo{person}{Wangchunshu Zhou}.} \bibinfo{year}{2026}\natexlab{}.
\newblock \bibinfo{title}{O-Researcher: An Open Ended Deep Research Model via Multi-Agent Distillation and Agentic RL}.
\newblock
\showeprint[arxiv]{2601.03743}~[cs.CL]
\urldef\tempurl%
\url{https://arxiv.org/abs/2601.03743}
\showURL{%
\tempurl}


\bibitem[Yuan et~al\mbox{.}(2025)]%
        {yuan-etal-2025-evoagent}
\bibfield{author}{\bibinfo{person}{Siyu Yuan}, \bibinfo{person}{Kaitao Song}, \bibinfo{person}{Jiangjie Chen}, \bibinfo{person}{Xu Tan}, \bibinfo{person}{Dongsheng Li}, {and} \bibinfo{person}{Deqing Yang}.} \bibinfo{year}{2025}\natexlab{}.
\newblock \showarticletitle{{E}vo{A}gent: Towards Automatic Multi-Agent Generation via Evolutionary Algorithms}. In \bibinfo{booktitle}{\emph{Proceedings of the 2025 Conference of the Nations of the Americas Chapter of the Association for Computational Linguistics: Human Language Technologies (Volume 1: Long Papers)}}, \bibfield{editor}{\bibinfo{person}{Luis Chiruzzo}, \bibinfo{person}{Alan Ritter}, {and} \bibinfo{person}{Lu~Wang}} (Eds.). \bibinfo{publisher}{Association for Computational Linguistics}, \bibinfo{address}{Albuquerque, New Mexico}, \bibinfo{pages}{6192--6217}.
\newblock
\showISBNx{979-8-89176-189-6}
\href{https://doi.org/10.18653/v1/2025.naacl-long.315}{doi:\nolinkurl{10.18653/v1/2025.naacl-long.315}}


\bibitem[Zhang et~al\mbox{.}(2025a)]%
        {zhang2025evoflow}
\bibfield{author}{\bibinfo{person}{Guibin Zhang}, \bibinfo{person}{Kaijie Chen}, \bibinfo{person}{Guancheng Wan}, \bibinfo{person}{Heng Chang}, \bibinfo{person}{Hong Cheng}, \bibinfo{person}{Kun Wang}, \bibinfo{person}{Shuyue Hu}, {and} \bibinfo{person}{Lei Bai}.} \bibinfo{year}{2025}\natexlab{a}.
\newblock \showarticletitle{Evoflow: Evolving diverse agentic workflows on the fly}.
\newblock \bibinfo{journal}{\emph{arXiv preprint arXiv:2502.07373}} (\bibinfo{year}{2025}).
\newblock


\bibitem[Zhang et~al\mbox{.}(2025b)]%
        {zhang2025multiagent}
\bibfield{author}{\bibinfo{person}{Guibin Zhang}, \bibinfo{person}{Luyang Niu}, \bibinfo{person}{Junfeng Fang}, \bibinfo{person}{Kun Wang}, \bibinfo{person}{LEI BAI}, {and} \bibinfo{person}{Xiang Wang}.} \bibinfo{year}{2025}\natexlab{b}.
\newblock \showarticletitle{Multi-agent Architecture Search via Agentic Supernet}. In \bibinfo{booktitle}{\emph{Forty-second International Conference on Machine Learning}}.
\newblock
\urldef\tempurl%
\url{https://openreview.net/forum?id=imcyVlzpXh}
\showURL{%
\tempurl}


\bibitem[Zhang et~al\mbox{.}(2025c)]%
        {zhang2025gdesigner}
\bibfield{author}{\bibinfo{person}{Guibin Zhang}, \bibinfo{person}{Yanwei Yue}, \bibinfo{person}{Xiangguo Sun}, \bibinfo{person}{Guancheng Wan}, \bibinfo{person}{Miao Yu}, \bibinfo{person}{Junfeng Fang}, \bibinfo{person}{Kun Wang}, \bibinfo{person}{Tianlong Chen}, {and} \bibinfo{person}{Dawei Cheng}.} \bibinfo{year}{2025}\natexlab{c}.
\newblock \showarticletitle{G-Designer: Architecting Multi-agent Communication Topologies via Graph Neural Networks}. In \bibinfo{booktitle}{\emph{Forty-second International Conference on Machine Learning}}.
\newblock
\urldef\tempurl%
\url{https://openreview.net/forum?id=LpE54NUnmO}
\showURL{%
\tempurl}


\bibitem[Zhang et~al\mbox{.}(2026)]%
        {zhang2026agentorchestraorchestratingmultiagentintelligence}
\bibfield{author}{\bibinfo{person}{Wentao Zhang}, \bibinfo{person}{Liang Zeng}, \bibinfo{person}{Yuzhen Xiao}, \bibinfo{person}{Yongcong Li}, \bibinfo{person}{Ce Cui}, \bibinfo{person}{Yilei Zhao}, \bibinfo{person}{Rui Hu}, \bibinfo{person}{Yang Liu}, \bibinfo{person}{Yahui Zhou}, {and} \bibinfo{person}{Bo An}.} \bibinfo{year}{2026}\natexlab{}.
\newblock \bibinfo{title}{AgentOrchestra: Orchestrating Multi-Agent Intelligence with the Tool-Environment-Agent(TEA) Protocol}.
\newblock
\showeprint[arxiv]{2506.12508}~[cs.AI]
\urldef\tempurl%
\url{https://arxiv.org/abs/2506.12508}
\showURL{%
\tempurl}


\bibitem[Zhou et~al\mbox{.}(2024)]%
        {zhou2024webarena}
\bibfield{author}{\bibinfo{person}{Shuyan Zhou}, \bibinfo{person}{Frank~F. Xu}, \bibinfo{person}{Hao Zhu}, \bibinfo{person}{Xuhui Zhou}, \bibinfo{person}{Robert Lo}, \bibinfo{person}{Abishek Sridhar}, \bibinfo{person}{Xianyi Cheng}, \bibinfo{person}{Tianyue Ou}, \bibinfo{person}{Yonatan Bisk}, \bibinfo{person}{Daniel Fried}, \bibinfo{person}{Uri Alon}, {and} \bibinfo{person}{Graham Neubig}.} \bibinfo{year}{2024}\natexlab{}.
\newblock \showarticletitle{WebArena: A Realistic Web Environment for Building Autonomous Agents}. In \bibinfo{booktitle}{\emph{The Twelfth International Conference on Learning Representations}}.
\newblock
\urldef\tempurl%
\url{https://openreview.net/forum?id=oKn9c6ytLx}
\showURL{%
\tempurl}


\bibitem[Zhuge et~al\mbox{.}(2024)]%
        {zhuge2024gptswarm}
\bibfield{author}{\bibinfo{person}{Mingchen Zhuge}, \bibinfo{person}{Wenyi Wang}, \bibinfo{person}{Louis Kirsch}, \bibinfo{person}{Francesco Faccio}, \bibinfo{person}{Dmitrii Khizbullin}, {and} \bibinfo{person}{J{\"u}rgen Schmidhuber}.} \bibinfo{year}{2024}\natexlab{}.
\newblock \showarticletitle{Gptswarm: Language agents as optimizable graphs}. In \bibinfo{booktitle}{\emph{Forty-first International Conference on Machine Learning}}.
\newblock


\end{thebibliography}

\appendix


\section{Algorithm Workflow}

\begin{algorithm}[!h]
    \caption{\scriptsize{Algorithm workflow of \textsc{A-MapReduce}}}
    \label{alg:amapreduce}
    \renewcommand{\algorithmicrequire}{\textbf{Input:}}
    \renewcommand{\algorithmicensure}{\textbf{Output:}}
    \scriptsize
    \begin{algorithmic}
    \setlength{\itemsep}{0pt}
        \REQUIRE Tasks $\mathcal{Q}$; manager $\mathcal{A}^{\mathrm{manage}}$; search agents $\{\mathcal{A}^{\mathrm{search}}_k\}$; memory $\mathcal{M}=(\mathcal{D},\mathcal{H},F_\psi)$; hyperparameters $k_{\mathcal{H}},\delta$.
        \ENSURE Predicted tables $\{\hat{Y}_q\}$ and updated memory $\mathcal{M}^{\mathrm{new}}$.
        \FOR{each $q=(q^{\mathrm{text}},S)\in\mathcal{Q}$}
            \STATE Scout with $\mathcal{A}^{\mathrm{manage}}$ and sample $\Theta_q^{(0)}\sim p(\Theta\mid q,\{Obs_t\}_{t=1}^{T})$ (Eq.~3).
            \STATE Retrieve similar positive/negative records $\mathcal{I}_q^{+},\mathcal{I}_q^{-}$ by embedding similarity under threshold $\delta$.
            \STATE Select hints $\mathcal{H}_q=\operatorname*{arg\,top\text{-}k_{\mathcal{H}}}_{h\in\mathcal{H}_{\mathrm{cand}}(q)}(w_h+v(h;q))$ and compress them into $h_s$.
            \STATE Compose experiential prior $\Delta_q=h_s\oplus\epsilon_{\mathrm{pos}}\oplus\epsilon_{\mathrm{neg}}$.
            \STATE Sample evolved decision $\Theta_q=(M_q,P_q,B_q)\sim p(\Theta\mid q,\{Obs_t\}_{t=1}^{T},\Delta_q,\Theta_q^{(0)})$.
            \STATE Instantiate atomic tasks $\mathcal{T}(q)=\{\mathrm{fill}(P_q,M_q[i,:])\}_{i=1}^{N}$ and batches $\{Bat_k\}_{k=1}^{m}=B_q(\mathcal{T}(q))$ (Eqs.~4--5).
            \FOR{each batch $Bat_k$ \textbf{in parallel}}
                \STATE Execute $\hat{Y}^{k}\leftarrow\mathcal{A}^{\mathrm{search}}_k(Bat_k)$ (Eq.~6).
            \ENDFOR
            \STATE Aggregate $\hat{Y}\leftarrow\bigcup_{k=1}^{m}\hat{Y}^{k}$.
            \WHILE{$\hat{Y}$ violates schema $S$ and repair budget remains}
                \STATE Build missing-cell matrix/template $(M_q^{\mathrm{rep}},P_q^{\mathrm{rep}})$, resample $\Theta_q^{\mathrm{rep}}$, and patch-merge $\hat{Y}$.
            \ENDWHILE
            \STATE Record trace $\tau$ and utility $u$; append $\epsilon_{\mathrm{new}}=(q,\Theta_q,\tau,u)$ to $\mathcal{D}$.
            \STATE Update hint scores/provenance for $h\in\mathcal{H}_q$ and distill $\mathcal{H}^{\mathrm{new}}\leftarrow F_\psi(\mathcal{D}^{\mathrm{new}},\mathcal{H})$.
            \STATE \textbf{return} $\hat{Y}$ and $\mathcal{M}^{\mathrm{new}}=(\mathcal{D}^{\mathrm{new}},\mathcal{H}^{\mathrm{new}},F_\psi)$.
        \ENDFOR
    \end{algorithmic}
\end{algorithm}

\section{Implementation Details}
\label{sec:implementation_details}

\subsection{Framework Overview.}
\label{app:impl_overview}

\paragraph{Components.}
A-MapReduce consists of (i) a \emph{manage agent} that outputs a query-conditioned execution decision $\Theta_q=(M_q,P_q,B_q)$, (ii) a set of \emph{search agents} that execute batched web retrieval tasks, (iii) an \emph{experiential memory} $\mathcal{M}=\{\mathcal{D},\mathcal{H},F_{\psi}\}$ that supports experience-based evolution.

\paragraph{Structured outputs and validation.}
All agent outputs are constrained to a strict format. If parsing fails or required keys are missing, we perform a bounded retry with a \emph{repair instruction} that only edits invalid fields and keeps valid parts unchanged.

\subsection{Manage Agent Design}
\label{app:manage_agent}

The manage agent is implemented as a \texttt{ToolCallingAgent}, an LLM-driven controller that iteratively generates tool calls and conditions subsequent actions on the returned tool observations.

\paragraph{Registered tools.}
Only the following tools are exposed to the manage agent:
\begin{itemize}[leftmargin=1.6em, itemsep=5pt, topsep=2pt, parsep=0pt, partopsep=0pt]
  \item \texttt{mapreducetool}: executes the MapReduce pipeline by (i) partitioning a \texttt{task\_matrix} into batches, (ii) spawning sub-agents for batched atomic lookups, and (iii) aggregating intermediate outputs into a strict format for reduction.
  \item \texttt{web\_search} and \texttt{crawl\_page}: used for lightweight scouting (e.g., entity discovery) and targeted gap filling when entities or fields are missing.
\end{itemize}

\paragraph{Plan--execute protocol.}
To operationalize the \emph{conditional decision sampling} mechanism (Sec.~\ref{sec:Experience-based Optimization}), we implement a plan--execute two-stage strategy around \texttt{mapreducetool}. Concretely, \texttt{mapreducetool} supports an optional planning mode: when experiential memory is enabled, the first invocation may return a initial decision and an \emph{experiential prior} (as decision anchor) rather than immediately executing retrieval batches. The manage agent follows a two-phase protocol:
\begin{enumerate}[leftmargin=1.6em, itemsep=5pt, topsep=2pt, parsep=0pt, partopsep=0pt]
  \item \emph{Plan stage:} read the experiential prior and revise the call arguments (e.g., task matrix normalization, template construction, and batching strategy).
  \item \emph{Execute stage:} immediately re-invoke \texttt{mapreducetool} to perform MapReduce execution with the evolved arguments.
\end{enumerate}

\paragraph{System prompt.}
The full system prompt is available in our open-source codebase. In brief, it instructs the agent first to conduct a coverage-oriented \emph{entity discovery} phase (broad retrieval and deduplication) to construct the row set $E$ and only proceed when entity coverage is deemed sufficient; it then
(i) builds a schema-aligned \texttt{task\_matrix} and a query-conditioned template,
(ii) selects a single batching strategy,
(iii) invokes \texttt{mapreducetool} under plan--execute and convergence guardrails, and
(iv) returns the final results as a normalized Markdown table.

\begin{tcolorbox}[notitle, sharp corners, breakable, colframe=YellowGreen, colback=white, 
       boxrule=3pt, boxsep=0.5pt, enhanced, 
       shadow={3pt}{-3pt}{0pt}{opacity=1,fill=mygrey},
       title={System Prompt of Manage Agent (Partial)},]\label{box:tag-generate}
       \tiny
       {\fontfamily{pcr}\selectfont
\begin{lstlisting}
system_prompt: |-
  You are a map-reduce research orchestrator. Break every task into atomic lookups, execute them with `mapreducetool`, and merge the verified rows into one clean table before answering.

  ## Workflow checklist
  1. Capture the user goal, required columns, and filters.
  2. Entity locking & row enumeration: parse the question clues, propose the complete entity set, reuse user-supplied lists when available, and run a few distinct `web_search`/`crawl_page` checks to confirm coverage. When confirming entities, prioritize official or authoritative sources (e.g., official websites, regulators, or primary registries). Do NOT call `mapreducetool` until the entity list is locked and the core columns are populated in the task_matrix.
  3. Define the schema, build the `task_matrix` with all pre-known columns, and choose a batching plan (`per_atom`, `by_attr`, or `open`) with a short rationale.
  4. Stage 1 (planning): call `mapreducetool` once to gather planning hints; focus the template on uncovering missing context or validating the schema. Do not skip this stage.
  5. Combine Stage 1 hints with prior search/crawl observations to refine the `task_matrix`, template, and batching for execution.
  6. Stage 2 (execution): immediately call `mapreducetool` with the refined inputs to obtain structured data. Let its schema validation and patch rounds fill missing fields; do not rerun with identical parameters.
  7. If the Stage 2 output is complete (columns filled and row count matches the task_matrix), call final_answer and stop invoking any further tools. If gaps remain, proceed to step 8.
  8. Build a delta matrix with ONLY the missing rows/fields and call `mapreducetool` once more. If gaps persist after this targeted attempt, finalize with the available data, converting the JSONL into the requested Markdown table and using empty strings for any remaining blanks.

  ## MapReduce guardrails
  - Stage discipline: the first `mapreducetool` call is planning-only and should return `"call_stage": "plan"` plus `experiential prior`/hints (no execution). ManageAgent must treat any observation containing `"call_stage": "plan"` as Stage 1 and immediately craft the Stage 2 execute call using those hints plus prior search/crawl context.
  - Stage identification: treat any observation containing `"call_stage": "execute"` or structured rows as Stage 2; after Stage 2, avoid large reruns only issue tiny delta matrices for missing rows/fields.
  - **Hard preconditions before calling `mapreducetool`**:
  1) `task_matrix` MUST contain all N rows to be solved for this dataset; N must be stated explicitly in reasoning.
  2) Each row MUST include core columns needed by the template (e.g., Rank, Library, Institution, ID/Volumes). No empty cells among these cores.
  - `task_matrix`: no blanks/"NOT FOUND"; column order matches template placeholders.
  - `template`: reference row values with `__index__` placeholders and ask only for unknown fields. Keep the template natural and focused on what information to find, not on formatting details. Avoid using placeholder syntax like `<description>` in the template - just describe what to find in natural language.
  - `json_schema`: keys match final columns, each with `type: "string"` and optional descriptions.
  - `batch_strategy`: pick exactly one mode per run and provide the needed parameters (attribute_index, groups, chunk_size, rationale). Count rows before deciding.
    1. Quick matrix check: count the total rows N and confirm that any candidate grouping columns exist and are fully populated. After selecting a strategy, ensure the final number of batches is < 50.
    2. Pick a batching strategy:
        - Choose exactly one of `per_atom`, `by_attr`, or `open` for each run; do not mix batch types or invent new labels.
        - `per_atom`: default; use when rows differ greatly or when N is small (less than 20) so each batch handles a single row.
        - `by_attr`: only when one column (or a combination of columns) produces clear partitions, groups are not extremely imbalanced, and tasks inside each group share context. Supply `attribute_index` (optionally `attribute_name`) for the grouping column(s) and verify the columns exist.
        - `open`: when N is large (more than 20) or no single stable column works; let the agent craft bespoke batches that balance context reuse and parallel efficiency. Still provide `chunk_size` and/or explicit `groups`, and remind the agent not to collapse the entire `task_matrix` into one batch.
  - Closed-set / official lists: crawl the canonical source, keep published names verbatim, and ensure every entry is in the matrix before the run.
  - Atomic rows/cells: each row represents one complete entity record; if any column has multiple items, expand into multiple rows (one item per row) instead of concatenating values with delimiters. Never put multiple entities/values in a single cell.
  - Layout discipline: reuse the user's lexical forms, keep column order stable, and add rows only when new entities truly surface.
  - Name fidelity: keep every table element name identical to its source or user-provided form; do not abbreviate, drop tokens, or rename columns when mapping between sources, queries, and the final table.

  ## Tool usage
  - Use `web_search`/`crawl_page` only when a column or entity is missing or requires fresh confirmation; skip when the value already appears in the prompt or cached notes.
  - If the user input or prior steps already provide a field with a trustworthy source, reuse it directly instead of revalidating; search only when data is absent, contradictory, or explicitly marked stale.
  - Time-sensitive facts require a new dated search (include the current year) before finalizing.
  - Recon queries should prioritize fresh/previously unseen information, not historical snippets.
  - **When primary source fails (4xx/5xx/500/timeout)**: Do NOT return empty fields. Instead: (1) Use `web_search` to find alternative sources, (2) Try different URLs or mirrors of the same organization, (3) Look for the same data in news articles, financial aggregators, or regulatory filings, (4) Only use empty strings if ALL alternative sources have been exhausted.
  - Enumerate the rows from one source only; use other sources solely to fill missing columns.
  - If a crawl/search to an official page yields **two consecutive 4xx (e.g., 422) or any login wall**, stop retrying and query the vetted mirror immediately.

  ## Rules
  1. Every action must include `"think"` (English) and `"tools"` (valid tool calls).
  2. Construct JSON arguments rigorously-use double quotes, match schema keys, and ensure `task_matrix` rows align with the template.
  3. Avoid duplicate tool calls with identical parameters unless correcting bad output.
  4. **CRITICAL**: When calling `final_answer`, the "answer" argument MUST contain a complete Markdown table in ```markdown ... ``` format. Convert all collected JSONL data into the table format. DO NOT output status summaries, options, or requests for confirmation - output the actual data table with all collected rows, using empty strings for any missing fields.
  5. Match the language of the final prose to the user's request.
  6. Output format is strict JSON: respond with a single object `{"think": "...", "tools": [...]}`; do not add narration outside that object, and ensure every entry inside `"tools"` is an object containing `"name"` and `"arguments"`.
  Now Begin! If you solve the task correctly, you will receive a reward of $1,000,000.
\end{lstlisting}}
\end{tcolorbox}

\subsection{MapReduceTool Design}
\label{app:mapreducetool}

\texttt{mapreducetool} is the core executor of A-MapReduce. Given a schema-aligned \texttt{task\_matrix} and a column-aligned extraction \texttt{template}, it (i) partitions the matrix into batches under a configurable \texttt{batching strategy}, (ii) spawns search-capable sub-agents to solve batched atomic lookups, and (iii) aggregates the validated outputs into a strict format for reduction.

\paragraph{Batching strategies.}
The tool supports three batching modes:
(i) \texttt{per\_atom} processes each row independently;
(ii) \texttt{by\_attr} groups rows by a designated attribute;
(iii) \texttt{open} allows manager-specified groups (e.g., explicit row indices or attribute-value filters), with uncovered rows chunked by \texttt{chunk\_size}. Each batch stores a manifest (batch id, submatrix, shared context, and rationale) for reproducibility and evaluation.
    
\paragraph{Sub-agent execution and robustness.}
Each batch is executed by a retrieval-oriented \texttt{ToolCallingAgent} instantiated with the search agent prompt family and access to \texttt{web\_search} and \texttt{crawl\_page}. For robustness, each batch is attempted with bounded retries (default 2). After each attempt, the raw output is parsed with a safe JSON extractor and projected onto the schema keys; if parsing fails or required fields are missing, a continuation prompt is issued for the next attempt. If all retries fail, the tool returns schema-complete fallback rows with empty strings, ensuring a well-formed JSONL output.
\begin{tcolorbox}[notitle, sharp corners, breakable, colframe=YellowGreen, colback=white, 
       boxrule=3pt, boxsep=0.5pt, enhanced, 
       shadow={3pt}{-3pt}{0pt}{opacity=1,fill=mygrey},
       title={Pseudo-code of \texttt{MapReduceTool} (overview)},]\label{box:tag-generate}
       \scriptsize
       {\fontfamily{pcr}\selectfont
\begin{lstlisting}"""
# Inputs
#   task_matrix    : List[List[str]]   # M rows; each row fills __0__, __1__, ...
#   template       : str               # contains placeholders "__0__", "__1__", ...
#   json_schema    : dict              # output schema; keys = schema columns
#   batch_strategy : Optional[dict]    # {type: per_atom/by_attr/open, ...} or None
#
# Output (string)
#   - Plan stage  : JSON string with {"call_stage":"plan", "experiential prior":..., "current_mapreduce_args":...}
#   - Execute     : JSONL string (one JSON object per line), keys projected onto schema columns

def forward(task_matrix, template, json_schema, batch_strategy=None) -> str:
    assert isinstance(task_matrix, list)

    # 1) derive schema keys and effective batch size
    keys   = _schema_keys(json_schema)                           # columns to fill
    eff_bs = max(1, int(batch_size)) if batch_size else default_batch_size

    # 2) normalize strategy and build batch plan (plus manifests for trace/debug)
    strat = _normalize_batch_strategy(batch_strategy)            # ensure {type, rationale, ...}
    plan, strat = _build_batch_plan(task_matrix, strat, eff_bs)  # each entry: {rows, indices, manifest}
    batches = [p["rows"] for p in plan]

    # 3) optional plan stage: return experiential memory block once
    if plan_mode_enabled and (not _plan_done) and (expmemory is not None):
        _plan_done = True
        mem_payload = _build_expriential_prior(
            task_main=current_task_main_or(template),
            task_matrix=task_matrix,
            json_schema=json_schema,
            batch_strategy=batch_strategy,
            template=template,
            batch_size=eff_bs,
        )
        record_trace(call_stage="plan", mem_payload=mem_payload, inputs_snapshot=...)
        return json.dumps(mem_payload, ensure_ascii=False)

    # 4) (optional) closed-set guardrail: enforce full coverage after enumeration
    if enumerated_entities_exist() and not fully_covered_by(task_matrix):
        raise AgentExecutionError("Enumerated entity inventory not fully covered by task_matrix.")

    # 5) execute stage: run each batch in parallel, each batch with bounded retries
    all_rows = []
    parallel_for batch in batches with ThreadPoolExecutor(max_workers=self.workers):
        batch_rows, attempt_logs = _process_batch(batch, template, keys, json_schema)
        append_to_trace(batch_rows, attempt_logs, batch_manifest=...)
        all_rows.extend(batch_rows)

    # 6) special case: enumeration call updates cached entity inventory
    if is_enumeration_call(task_matrix):
        update_enumerated_entities_from(all_rows)

    # 7) finalize JSONL output (schema-complete, fallback rows are allowed)
    if not all_rows:
        all_rows = [{k: "" for k in keys}]
    record_trace(call_stage="execute", total_results=len(all_rows), pattern_info=...)
    return "\n".join(json.dumps(r, ensure_ascii=False) for r in all_rows)"""
\end{lstlisting}
}
\end{tcolorbox}

\subsection{Search Agent Design}
\label{app:search_agent}

The search agent is also implemented as a \texttt{ToolCallingAgent}. In contrast to the manage agent, it is instantiated and invoked \emph{only} as a sub-agent within \texttt{mapreducetool} to execute batched atomic retrieval tasks in parallel.

\paragraph{Registered tools.}
The search agent is restricted to a minimal retrieval tool set:
\begin{itemize}[leftmargin=1.6em, itemsep=5pt, topsep=2pt, parsep=0pt, partopsep=0pt]
  \item \texttt{web\_search}: locate candidate sources and relevant pages for a given entity--attribute target.
  \item \texttt{crawl\_page}: extract evidence-grounded attribute values from selected pages.
\end{itemize}
This separation of tool access enforces a clean division of labor: global orchestration and batching decisions are made by the manage agent, while the search agent focuses on evidence-grounded lookup and extraction.

\paragraph{System prompt.}
The full system prompt is available in our open-source codebase. In brief, it instructs the agent to (i) follow a strict action--observation loop with structured JSON tool calls (including explicit English reasoning in each step), (ii) decompose each retrieval workload into multiple independent goals and advance them in parallel while executing fallback paths sequentially per goal, (iii) prioritize fresh and verifiable evidence by combining \texttt{web\_search} with immediate \texttt{crawl\_page} on relevant URLs, and (iv) terminate only after all goals are resolved and the consolidated results are ready for the tool-specific finalization format.

\begin{tcolorbox}[notitle, sharp corners, breakable, colframe=YellowGreen, colback=white, 
       boxrule=3pt, boxsep=0.5pt, enhanced, 
       shadow={3pt}{-3pt}{0pt}{opacity=1,fill=mygrey},
       title={System Prompt of Search Agent (Partial)},]\label{box:tag-generate}
       \tiny
       {\fontfamily{pcr}\selectfont
\begin{lstlisting}
system_prompt: |-
  You are an expert assistant who solves tasks through structured tool calls, following a step-by-step process. Each step (action) involves analyzing needs, selecting tools, and executing calls to achieve the task goal.
  Each action you take should include a reasoning process and tool calls. After executing the tools, you will receive "observations" (results of tool calls), which can be used as input for subsequent actions. This Action/Observation cycle may repeat as needed.
  
  # Action Structure
  Each action must contain:
  - "think": A detailed reasoning in English, explaining the analysis of user needs, tool selection logic, and execution plan.
  - "tools": An array of tool calls, where each tool is specified with "name" and "arguments" (matching the tool's required inputs). Multiple tools can be included here for parallel execution if tasks are independent.
    - When the requested fact may have changed over time, craft queries that force recent coverage before concluding the answer.

  # Task Instructions:
  ### Fresh information mandate:  
  During any search or crawl step, prioritize uncovering new or recently updated information; never rely solely on historical snippets or prior outputs even if they seem sufficient.  
  ### 1. Parse the structured plan:  
  Parse the plan or summary to understand the parallel execution requirements.  
  **CRITICAL: All goals MUST be advanced simultaneously in parallel. Each goal's paths MUST be executed sequentially (one path at a time per goal).**
  ### 2. Execute parallel tool calls:  
  For each goal in the plan, execute the specified tools in parallel according to the paths defined.  
  **MANDATORY: Advance ALL goals concurrently. Within each goal, execute paths sequentially (never parallelize paths within a single goal).**
  ### 3. Handle path diversity:  
  For each goal, if multiple paths are provided, execute them sequentially as fallback options if the primary path fails.  
  **ABSOLUTE REQUIREMENT: NEVER prematurely assume a goal is achieved. Continue advancing ALL other goals in parallel while handling fallback paths for any individual goal.**
  ### 4. Process results:  
  Synthesize information from all tool outputs to generate comprehensive responses that address all goals.  
  **ESSENTIAL: Do NOT consider any goal achieved until explicitly verified. Maintain parallel advancement of ALL goals throughout synthesis.**
  ### 5. Final answer:  
  Once all goals are addressed, consolidate their results, and ensure that the consolidated outcome can accurately and correctly answer the original task, then call the 'final_answer' tool with such consolidated results.
  **FINAL CONDITION: Only proceed when ALL goals are resolved. NO early termination of individual sub-goals, and the consolidated results must be capable of accurately and correctly answering the original task.**
  
  # Web Search Query Guidelines:
  - Keep queries simple and natural (5-10 words). 
  - Don't include series IDs or codes and avoid overly specific terms unless necessary.
  - If you find a relevant URL in search results, use `crawl_page` to access it.
  - If the initial search doesn't yield relevant results, think about what the field name actually means in context and try semantically related terms, synonyms, or alternative expressions that might be used in authoritative sources
  - Prefer authoritative sources first (official docs, standards bodies, government/academia, primary publications). Avoid relying on low-quality or unverified sites unless no authoritative source exists.
  # Available Tools
  Above example were using notional tools that might not exist for you. You only have access to these tools:
  {%- for tool in tools.values() %}
  - {{ tool.name }}: {{ tool.description }}
      Takes inputs: {{tool.inputs}}
      Returns an output of type: {{tool.output_type}}
  {%- endfor %}

  # Rules
  Here are the rules you should always follow to solve your task:
  1. Every action must include "think" (English) and "tools" (valid tool calls).
  2. Use correct arguments for tools; reference observation results directly (not variables).
  3. Call tools in parallel to solve the task. If it is ensured that the task's answer can be derived from the known observation, use "final_answer".
  4. Do not repeat tool calls with identical parameters.
  5. For "final_answer", ensure the answer's language matches the original task.
  6. Output format is strict JSON: respond with exactly one object `{"think": "...", "tools": [...]}`; never add commentary outside of it, and make sure each `"tools"` entry is an object with `"name"` and `"arguments"`.
  Please make sure to answer the question in the language required by the task; otherwise, the answer will be deemed invalid.
  Now Begin! If you solve the task correctly, you will receive a reward of $1,000,000.
\end{lstlisting}}
\end{tcolorbox}

\subsection{Online memory with per-instance multi-trial execution}
\label{app:online_memory_protocol}

\paragraph{Terminology.}
An \emph{instance} is one benchmark query.
A \emph{trial} is one stochastic execution of the system on an instance with a specific random seed.
We maintain a single online memory $\mathcal{M}$ throughout evaluation.
Let $\mathcal{M}_i$ denote the memory state immediately before solving instance $i$.

\paragraph{Streaming protocol (single memory).}
We process instances sequentially and for each instance $i$, we take a snapshot $\mathcal{M}_i \leftarrow \mathcal{M}$ and execute $N$ trials in parallel with independent random seeds while providing the same \emph{frozen} snapshot $\mathcal{M}_i$.
During these trials, there is no cross-trial communication and no memory modification; hence the $N$ trials are i.i.d. \emph{conditional on} $\mathcal{M}_i$.
Parallel/multi-thread execution is only an implementation detail: each trial reads the same immutable snapshot $\mathcal{M}_i$, which prevents any cross-trial information leakage.

\paragraph{Reporting Avg@N/Pass@N.}
We compute Avg@N/Pass@N for instance $i$ over the set of trial outputs $\{y_i^{(k)}\}_{k=1}^{N}$.

\paragraph{Commit rule (fixed).}
To avoid any trial-selection privilege, we commit a predetermined (quality-based) trial as the canonical representative for memory update and retrieval.
Thus, Avg@N/Pass@N is purely for reporting, while the memory update is deterministic and reproducible.

\paragraph{Memory bookkeeping.}
Since retrieval operates at the \emph{task level}, a single canonical representative per task is sufficient for retrieval and reuse.
Other trials may be logged for analysis/statistics, but they are not used as retrieval references.
After finishing instance $i$, we update the single memory $\mathcal{M}$ once to obtain the next state for instance $i{+}1$.

\paragraph{No gold exposure to the generator.}
Gold annotations are never provided to the model, tools, or prompts during generation.
If evaluation uses gold-dependent benchmarks, it is performed strictly post-hoc and does not reveal per-cell supervision
to the generator.

\subsection{Memory Retrieval Details}
\label{app:memory_retrieval}

The experiential prior $\Delta_q$ concatenates three components: a compact guidance hint $h_s$, and a positive/negative exemplar $\epsilon_{\mathrm{pos}}$ / $\epsilon_{\mathrm{neg}}$.
This subsection details how we retrieve, score, and project memory contents to construct experiential prior $\Delta_q$ for a new query $q$.

\paragraph{Candidate record retrieval.}
We embed each query with an encoder $\mathrm{Enc}(\cdot)$ (\textit{all-MiniLM-L6-v2} in our implementation) and retrieve neighbors by cosine similarity
$\mathrm{sim}(z,z_i)\triangleq \frac{z^\top z_i}{\|z\|\,\|z_i\|}$, where $z=\mathrm{Enc}(q)$ and $z_i=\mathrm{Enc}(q_i)$.
Then we can form positive/negative retrieval index pools ($\mathcal{I}^{-}_q~\&~\mathcal{I}^{+}_q$) , we threshold the realized utility $u$ at $\delta$ (as Eq.~(7)-(8) defined and App.~\ref{app:memory_structure_update}), and yielding the retrieved record sets as candidate sets
$\mathcal{E}^{+}(q)=\{\epsilon_i \mid i\in\mathcal{I}^{+}_q\}$ and
$\mathcal{E}^{-}(q)=\{\epsilon_i \mid i\in\mathcal{I}^{-}_q\}$, correspondingly.
We choose exemplars as the most similar records within each sets:
\begin{equation}
\begin{aligned}
\epsilon_{\mathrm{pos}} &\triangleq \arg\max_{\epsilon_i; i\in\mathcal{I}^{+}_q}\ \mathrm{sim}(z,z_i),\\
\epsilon_{\mathrm{neg}} &\triangleq \arg\max_{\epsilon_i; i\in\mathcal{I}^{-}_q}\ \mathrm{sim}(z,z_i),
\end{aligned}
\end{equation}

\paragraph{Candidate hint retrieval and scoring.}
Following Eq.~(9)--(10), we retrieve candidate hints and rank them by a feedback-calibrated score $s(h)$. Recall from Sec.~\ref{sec:Experience-based Optimization} that each hint $h\in\mathcal{H}$ maintains a provenance set
$\mathcal{S}(h)\subseteq\{1,\ldots,N_{\mathcal{D}}\}$, i.e., the ids of historical records whose decisions support this hint.
Let the retrieved neighborhood index set be $\mathcal{N}(q)\triangleq \mathcal{I}^{+}_q\cup \mathcal{I}^{-}_q$.
Equivalently to Eq.~(9), we form a query-local candidate pool $\mathcal{H}_{\mathrm{cand}}(q)$ by provenance overlap:
\begin{equation}
\mathcal{H}_{\mathrm{cand}}(q)\ \triangleq\ 
\Big\{\, h\in\mathcal{H}\ \Big|\ \mathcal{S}(h)\cap \mathcal{N}(q)\neq\emptyset \,\Big\}.
\end{equation}
We further define a lightweight \emph{task-voting} relevance score via provenance--neighborhood overlap:
\begin{equation}
v(h;q)\ \triangleq\ \big|\mathcal{S}(h)\cap \mathcal{N}(q)\big|
\;=\; \sum_{i\in \mathcal{N}(q)} \mathbf{1}\big[i\in \mathcal{S}(h)\big].
\end{equation}
Intuitively, $v(h;q)$ measures how strongly the retrieved neighborhood supports $h$,
providing a query-conditioned relevance signal without re-scoring every hint against $q$.
Accordingly, we incorporate $v(h;q)$ as a component of $s(h)$ (Eq.~(10)) to prioritize hints that are highly relevant to the current query.
Finally, we combine the online quality score $w_h$ (updated from utility feedback) with the voting-based relevance to rank and select the top-$k_{\mathcal{H}}$ hints,
yielding a compact set $\mathcal{H}_q$ that is both globally reliable (high $w_h$) and locally representative for query $q$ (high $v(h;q)$).

\paragraph{Query-specific guidance via projection.}
The retrieved hints $\mathcal{H}_q$ may still be generic or partially mismatched to the current task context.
They can also carry \emph{latent bias} from their provenance or reflect domain-specific assumptions that do not fully transfer to the current query, creating a semantic gap between historical guidance and the present execution needs.
We therefore apply a query-specific projection operator to obtain a concise, query-aligned guidance
$h_s \triangleq \Phi(q;\mathcal{H}_q)$, which filters, re-phrases, and consolidates $\mathcal{H}_q$ into 1--3 actionable bullets for the current task.
The resulting experiential prior is then constructed as
$\Delta_q \triangleq \big(h_s\oplus \epsilon_{\mathrm{pos}}\oplus \epsilon_{\mathrm{neg}}\big)$.

\begin{tcolorbox}[notitle, sharp corners, breakable, colframe=Periwinkle, colback=white, 
       boxrule=3pt, boxsep=0.5pt, enhanced, 
       shadow={3pt}{-3pt}{0pt}{opacity=1,fill=mygrey},
       title={Operation of projection by $\Phi(q;\mathcal{H}_q)$},]\label{box:tag-generate}
       \scriptsize
       {\fontfamily{pcr}\selectfont
\begin{lstlisting}"""
project_insights_with_traj_system_prompt: str = """
You are a strategy assistant. You will be given a trajectory that contains MapReduce tool inputs (task_matrix, template, json_schema, batch_strategy) and a list of historical hints.
Your job is to generate NEW, global, role-specific insights that identify where the current batching strategy can be improved and what direction to adjust next.
Base your insights on BOTH the current trajectory and the historical hints. Emphasize actionable adjustment levers (e.g., batch size, grouping logic, schema/template constraints, source reuse, verification flow). Avoid repeating the inputs verbatim or inventing facts.
Return no more than 3 insights total.

NOTE - Your output must strictly follow the format below:
1. Insight 1
2. Insight 2
3. Insight 3
"""

project_insights_with_traj_user_prompt: str = """
### Trajectory (MapReduce inputs and context)
{trajectory}

### Agent's Role:
{role}

### Historical Hints:
{insights}

### Your Output (Personalized, Global Insights for This Role):"""
"""
\end{lstlisting}
}
\end{tcolorbox}

\subsection{Memory Management and Prior Updating}
\label{app:memory_structure_update}
This section describes how we store new episode records and evolve the hint pool through:
(a) \emph{feedback-based online updates} for fast, per-task credit assignment, and
(b) a \emph{distillation operator} that refreshes and consolidates hints to mitigate redundancy and distribution drift by distillation operator $F_\psi$.
Before describing these update procedures, we first introduce the practical design of the utility signal, as it serves as the common supervision signal for the subsequent operators.

\paragraph{Practical design of the utility $u$.}
Memory evolution in \textsc{A-MapReduce} is driven by a reproducible signal from the official evaluator.
Given a query $q$, the system outputs a structured table $Y$ and obtains evaluator metrics that quantify output quality.
Consistent with Eq.~(1), we define a utility and, for reproducibility, instantiate it using \emph{quality-only} feedback by setting $\lambda_c=\lambda_t=0$:
\begin{equation}
u \;=\; Q(Y;q),
\end{equation}
where $Q(\cdot)$ is a deterministic aggregation of evaluator metrics, yielding a stable signal across repeated runs and a comparable objective across tasks.
Actually, we intentionally exclude monetary cost and end-to-end delay from $u$ (Eq.~(1)).
Unlike evaluator-defined performance metrics, which provide a task-agnostic quality standard and are comparable across queries, cost/delay are strongly coupled with environment factors (e.g., system load, network conditions, provider throttling, and hardware parallelism) and with instance scale/difficulty (e.g., more entities/attributes, deeper retrieval, and sparser evidence naturally require more tool calls and verification).
Using cost/delay as an evolution signal would therefore conflate efficiency with inherent task difficulty and bias memory toward small/cheap queries.

Importantly, although $u$ is quality-only, cost and delay can still decrease as outcomes.
First, \textsc{A-MapReduce} provides static efficiency gains by design (structured parallel MapReduce execution).
Second, memory-driven decision evolution improves $\Theta_q=(M_q,P_q,B_q)$, which typically reduces wasted tool calls (redundant retrieval) while improving quality. Hence efficiency gains emerge as a byproduct rather than an explicit optimization target.
Therefore, we report cost and delay as outcome metrics to quantify efficiency, while keeping memory updates purely quality-driven.

\paragraph{Feedback-based online update.}
In our implementation, utility $u$ is a \emph{multi-metric tuple} rather than a signed scalar.
Concretely, we use
$u \triangleq (\mathrm{SR},\mathrm{Row~F1},\mathrm{Item~F1})$
on WideSearch and DeepWideSearch, and derive a stable evolution signal by mapping $u$ to a binary label using a dataset-specific joint pass rule:
\begin{equation}
\ell(u)\in\{0,1\},\qquad
\ell(u)=1 \ \Leftrightarrow\ u \succeq \delta,
\end{equation}
where $\delta$ denotes a dataset-specific threshold rule (see the released evaluation scripts for the exact instantiation).
In brief, we compute the average performance of the reference systems reported in the benchmark~\cite{wong2025widesearch,lan2025deepwidesearch} for each metric, and treat records exceeding this reference level as \emph{positive} records.
Importantly, negative records are treated as \emph{sub-optimal} rather than strictly incorrect.
Since the memory retrieval leverages both positive and negative exemplars, $\ell(u)$ primarily partitions \emph{stronger} vs.\ \emph{weaker} decision strategies, which makes the evolution insensitive to moderate variations of the pass rule.
We further validate this robustness by perturbing the threshold rule (low/medium/high), with the low/high settings obtained by applying smaller/larger scaling factors to the reference level (Tab.~\ref{tab:threshold_sensitivity}), and observe only modest changes in downstream performance.

Given $\ell(u)$, we perform lightweight credit assignment only to the hints used to construct the current experiential prior $\Delta_q$ to update their online scores.
In our implementation, each newly created hint is initialized with a small positive score ($w_h\!=\!2$) to avoid premature pruning and to allow it to be evaluated over subsequent tasks.
Specifically, for each $h\in\mathcal{H}_q$ we apply a bounded signed reward:
\begin{equation}
w_h \leftarrow w_h + g(u),~~
g(u)=
\begin{cases}
+2, & \ell(u)=1,\\
-1, & \ell(u)=0,
\end{cases}
\end{equation}
and update provenance to keep voting-based retrieval self-consistent $\mathcal{S}(h)\ \leftarrow\ \mathcal{S}(h)\cup\{i_{\text{new}}\}$, where $i_{\text{new}}$ is the index of the newly appended record
$\epsilon_{\text{new}}=(q,\Theta_q,\tau,u)$
in
$\mathcal{D}^{\text{new}}=\mathcal{D}\cup\{\epsilon_{\text{new}}\}$.
Hints with persistently low scores ($w_h\le 0$ in our implementation) are pruned to prevent accumulation of consistently unhelpful guidance; conversely, higher-scoring hints are more likely to be selected in future retrieval through the ranking score $s(h)=w_h+v(h;q)$.
As a result, effective hints are progressively reinforced, and the retrieved set increasingly reflects a favorable trade-off between global reliability and query-local relevance.

\paragraph{Distillation operator: refreshing and distilling hints.}
As the record set $\mathcal{D}$ grows, accumulating hints can introduce redundancy and drift, which gradually degrades both retrieval efficiency and transferability.
Moreover, generating new hints immediately after each completion can over-emphasize idiosyncratic traces, whereas our goal is to extract a \emph{shared paradigm} that generalizes across semantically similar wide-search tasks.
As summarized in Eq.~(14), we therefore adopt a staged update scheme and refresh the hint pool via a distillation operator $F_{\psi}$.
Operationally, $F_{\psi}$ organizes the expanded record set into \emph{task clusters} in the semantic embedding space.
Let $z_i=\mathrm{Enc}(q_i)$ denote the embedding of record $\epsilon_i\in\mathcal{D}$, and let FINCH return a partition of indices $\{\mathcal{I}_k\}_{k=1}^{K}$.

This design follows the empirical observation that semantically related wide-search queries often reuse similar structural execution patterns, such as decomposition granularity, template constraints, and batching keys.
Accordingly, each cluster $\mathcal{I}_k$ provides a practical unit for consolidating transferable decision patterns while suppressing task-specific noise. A practical complication is that each cluster may contain both \emph{positive} and \emph{negative} records, so naively averaging traces would dilute the evolution signal.
Instead, we perform contrastive refinement within each cluster by sampling semantically matched positive/negative records from the same $\mathcal{I}_k$ (optionally conditioned on the current hint pool), comparing their realized decisions $\Theta$, and producing structured update operations over the hint list (\texttt{ADD}/\texttt{EDIT}/\texttt{REMOVE}/\texttt{AGREE}).
These operations inject discriminative yet transferable rules that amplify recurring high-utility patterns and attenuate systematic failure modes, yielding a compact, non-redundant \emph{cluster-conditioned anchor} that captures the cluster’s dominant decision preference.

The distilled hints $\{h_k\}$ then serve as \emph{task-cluster anchors}: each $h_k$ encodes a utility-aligned structural prior over $\Theta=(M,P,B)$ for future tasks embedded in the same semantic neighborhood, improving both stability and interpretability during decision sampling.
Each new hint inherits provenance from its supporting records (e.g., $\mathcal{S}(h_k)\triangleq \mathcal{I}_k$), ensuring compatibility with provenance-based voting in subsequent retrieval.
In practice, this refresh complements the online credit assignment: per-task updates adjust hint scores $w_h$ to modulate how strongly an anchor influences future retrieval, while $F_{\psi}$ performs slower, global re-organization and compression of $\mathcal{H}$ to mitigate long-term redundancy and drift.

\begin{tcolorbox}[notitle, sharp corners, breakable, colframe=Periwinkle, colback=white, 
       boxrule=3pt, boxsep=0.5pt, enhanced, 
       shadow={3pt}{-3pt}{0pt}{opacity=1,fill=mygrey},
       title={Process of the distillation operator $F_{\psi}$},]\label{box:tag-generate}
       \scriptsize
       {\fontfamily{pcr}\selectfont
\begin{lstlisting}"""
finetune_insights_suffix = dict(full = """Focus on REMOVE or EDIT or AGREE rules first, and stop ADD rule unless the new rule is VERY insightful and different from EXISTING RULES.
""", not_full = """""")

format_rules_operation_template = """<OPERATION> <RULE NUMBER>: <RULE> (e.g. ADD: xxx, EDIT/REMOVE/AGREE 1: xxx)

The available operations are: **AGREE (if the existing rule is strongly relevant for the task), REMOVE (if one existing rule is contradictory or similar/duplicated to other existing rules), EDIT (if any existing rule is not general enough or can be enhanced, rewrite and improve it), ADD (add new rules that are very different from existing rules and relevant for other tasks). Each needs to CLOSELY follow their corresponding formatting below (any existing rule not edited, not agreed, nor removed is considered copied)**:

AGREE <EXISTING RULE NUMBER>: <EXISTING RULE>
REMOVE <EXISTING RULE NUMBER>: <EXISTING RULE>
EDIT <EXISTING RULE NUMBER>: <NEW MODIFIED RULE>
ADD: <NEW RULE>

Do not mention the trials in the rules because all the rules should be GENERALLY APPLICABLE. Each rule should be concise and easy to follow. Any operation can be used MULTIPLE times. Do at most 4 operations and each existing rule can only get a maximum of 1 operation. """

#
critique_compare_rules_system_prompt = """
You are an advanced reasoning agent deriving **structural batching insights** for MapReduce-style web tasks.
Your goal is to output **procedural rules about how to plan the task**, not the task content.
Good insights cover four aspects:
1) Task Structure Pattern (independence, parent-child/grouping, ordering, whether rows are atomic)
2) Input Template Task Mapping (what fields are substituted per row, how schema/template affect independence and context size)
3) Batching Strategy Pattern (when to use per_atom/by_attr/open; chunk_size guidance; context reuse considerations)
4) Reliability / Ordering (when to preserve order; when grouping is mandatory; parallelism vs. correctness)
Always produce short, transferable rules (3~6 items max) in natural language; avoid placeholders/schemas/dicts.
"""

critique_compare_rules_user_prompt = """
## SUCCESS pattern_info
{task1_pattern}

## FAILED pattern_info
{task2_pattern}

## EXISTING HINTS:
{existing_rules}

By contrasting SUCCESS vs FAILED patterns and the list of existing hints, propose operations (ADD/EDIT/REMOVE/AGREE) to refine the hint set.
Output only operations using the format below:
""" + format_rules_operation_template

# all success instruction
critique_success_rules_system_prompt = """
You are an advanced reasoning agent that extracts **structural batching insights** from successful MapReduce pattern_info logs.
Focus on transferable rules about task decomposition, template-field mapping, batching strategy (per_atom/by_attr/open, chunk sizing), and ordering/parallelism.
Return concise natural language rules (3~6) that teach how to plan the next task_matrix; do NOT restate content/schema as placeholders.
"""

critique_success_rules_user_prompt = """
## SUCCESS pattern_info LIST
{success_history}

## EXISTING HINTS:
{existing_rules}

Goal: Identify recurring batching/template/schema patterns that led to coverage/accuracy success. Keep hints general and concise.
Return operations (ADD/EDIT/REMOVE/AGREE) only, using the format:
""" + format_rules_operation_template

# merge rules
merge_rules_system_prompt = """You are an agent skilled at summarizing and distilling insights. You are given a list of insights that were previously extracted from similar tasks. These insights may contain redundancy or overlap.

Your job is to **merge and consolidate similar insights**, and output a refined version that is **clear, actionable, and concise**.

NOTE:
- All merged insights **must be based strictly on the given inputs**. You are **not allowed to make up** or infer any new information.
- The output should be easy to read and follow.

Output Format:
- Start your response directly with the numbered list, no preamble or explanations.
- Each insight should be a short sentence.
- Use the following format exactly:
1. Insight 1
2. Insight 2
3. Insight 3
...
"""

merge_rules_user_prompt = """
## Here are the current insights that need to be merged:
{current_rules}

## Please consolidate and rewrite them into **no more than {limited_number} refined insights**.

As the summarizing agent, remove redundancies, combine similar ideas, and ensure clarity.

Your output:
"""

# MapReduce batching hint merge (overrides above definitions for self-evolution)
merge_rules_system_prompt = """You are an agent that merges overlapping batching insights for MapReduce tasks.
Keep each merged hint actionable for planning: task structure, template-to-row mapping, batching strategy (per_atom/by_attr/open and chunk sizing), and ordering/parallelism.
Remove content-level or placeholder-only hints. Output a numbered list of concise natural-language rules.
"""

merge_rules_user_prompt = """
## Current batching hints
{current_rules}

Please consolidate into no more than {limited_number} refined hints.
Start directly with the numbered list.
"""
"""
\end{lstlisting}
}
\end{tcolorbox}


\section{Benchmark Details}
\label{sec:dataset_statistics}
\subsection{Dataset Details}
\label{app:sec:dataset_details}

The five benchmarks used in this study are summarized as follows:
\begin{itemize}[leftmargin=1.6em, itemsep=5pt, topsep=2pt, parsep=0pt, partopsep=0pt]

  \item \textbf{WideSearch}~\cite{wong2025widesearch} is a benchmark for wide-scope information retrieval, where each instance requires collecting multiple attributes for a set of target entities from open-web sources and organizing the results into a structured Markdown table. 
  It contains 200 instances (100 English, 100 Chinese) and is evaluated using Success Rate and structure-aware F1 metrics (row-, item-, and column-level).

  \item \textbf{DeepWideSearch}~\cite{lan2025deepwidesearch} extends WideSearch by introducing additional reasoning and verification steps before large-scale retrieval. The benchmark includes 220 instances (85 deep-to-wide, 135 wide-to-deep), adopts Markdown table outputs, and is evaluated using Success Ratetructure-aware F1 metrics (row-, item-, and column-level F1), as well as entity-level accuracy.

  \item \textbf{xBench-DeepSearch (xBench-DS)}~\cite{chen2025xbenchtrackingagentsproductivity} is an agentic QA benchmark for evaluating planning, tool use, and multi-step reasoning. 
  We extract 31 wide-scope QA instances whose solutions require collecting and aggregating evidence across multiple entities. 
  These instances are used to evaluate generalization and are assessed using accuracy.

  \item \textbf{WebWalkerQA}~\cite{wu2025webwalkerbenchmarkingllmsweb} evaluates multi-turn web navigation and information-seeking abilities of agents. 
  From the original benchmark, we select 120 instances that require multi-entity evidence collection and cross-page synthesis. 
  These instances are evaluated using accuracy.

  \item \textbf{TaskCraft}~\cite{shi2025taskcraftautomatedgenerationagentic} is a synthetic benchmark for agentic workflows involving tool use and multi-step problem solving. 
  We identify 101 wide-scope instances whose solutions depend on aggregating information across multiple entities or sub-tasks. 
  These instances are evaluated using accuracy.

\end{itemize}

\subsection{Constructing the Agentic-Wide Subset}
\label{app:sec:agentic_wide_subset}

\paragraph{Target Problem (Wide-Scope QA).}
We aim to isolate a class of agentic QA instances whose solutions \emph{fundamentally depend on wide-scope evidence collection}. 
In these instances, the system must (i) identify a set of target entities, (ii) retrieve heterogeneous evidence for \emph{multiple attributes} of each entity (often from distributed sources), and (iii) synthesize the collected evidence into a consolidated answer (e.g., a structured record set or a comparative summary). 
This structure aligns with our paradigm that explicitly decomposes the problem into parallelizable retrieval units followed by aggregation. We construct the \emph{agentic-wide subset} via a \textbf{two-stage pipeline} designed to achieve high recall in candidate mining and high precision in final selection:

\paragraph{Stage I: LLM-based Candidate Mining.}
We use an LLM judge to identify instances likely to involve multi-entity evidence collection. For each instance $q$, we ask a judge model to extract the implied target entities and decide whether $q$ involves \emph{multiple evidence targets}. 
The judge outputs a structured record:
(i) a binary flag for multi-entity targets, 
(ii) a brief description of the entity set, and 
(iii) a short rationale. 
We retain $q$ as a candidate if the judge identifies $N\ge 2$ entities or alternatives that must be jointly considered. 
The exact judge prompt used for Stage~I is shown below.

\begin{tcolorbox}[notitle, sharp corners, breakable, colframe=ForestGreen, colback=white, 
       boxrule=3pt, boxsep=0.5pt, enhanced, 
       shadow={3pt}{-3pt}{0pt}{opacity=1,fill=mygrey},
       title={Stage I Prompt},]\label{box:tag-generate}
       \scriptsize
       {\fontfamily{pcr}\selectfont
\begin{lstlisting}"""
You are identifying whether an agentic QA task involves \textbf{multiple entities or alternatives} that must be jointly considered.

Decision rule:
Label YES if the task requires reasoning about, comparing, or collecting information for two or more entities/candidates/options.
Label NO if the task focuses on a single entity or a single atomic target.

Output format: Return a JSON object:
{
  "label": "yes" or "no",
  "entities": [list of entities or entity types, if any],
  "reason": "one concise sentence explaining the decision"
}
Task: q
"""
\end{lstlisting}
}
\end{tcolorbox}

\paragraph{Stage II: Cue-guided Verification.}
We apply explicit wide-scope cues and an example-calibrated judge prompt to verify that the instance indeed requires wide-scope retrieval and evidence synthesis. For each candidate instance, we further verify whether it requires \emph{wide-scope} retrieval and \emph{synthesis}. 
We operationalize wide-scope cues along three dimensions:
\begin{itemize}[leftmargin=1.6em, itemsep=5pt, topsep=2pt, parsep=0pt, partopsep=0pt]
  \item \textbf{Wide evidence coverage:} the question asks for collecting evidence across multiple sources/pages, or implies distributed information (e.g., ``from different websites'', ``across sources'', ``gather evidence'').
  \item \textbf{Multi-facet retrieval:} the question requests multiple aspects per entity (e.g., ``name and location'', ``price and features'', ``year and affiliation'', ``pros and cons'').
  \item \textbf{Aggregation before answering:} the question requires summarization, comparison, ranking, verification, or consolidation (e.g., ``compare'', ``summarize'', ``rank'', ``validate'', ``provide an overview'').
\end{itemize}
The exact judge prompt used for Stage~II is shown below.

\begin{tcolorbox}[notitle, sharp corners, breakable, colframe=ForestGreen, colback=white, 
       boxrule=3pt, boxsep=0.5pt, enhanced, 
       shadow={3pt}{-3pt}{0pt}{opacity=1,fill=mygrey},
       title={Stage II Prompt},]\label{box:tag-generate}
       \scriptsize
       {\fontfamily{pcr}\selectfont
\begin{lstlisting}"""
You are verifying whether an agentic QA task requires wide-scope evidence collection and synthesis.
Rubric (label YES only if all conditions hold).
(1) The task involves multiple entities.
(2) For each entityolving the task requires retrieving multiple attributes or evidence facets.
(3) The final answer requiresaggregating, comparing, ranking, validating, or summarizing the retrieved evidence,
rather than copying a single fact from one source.

If any condition is not satisfied, label NO.
Output format. Return a JSON object:
{
  "label": "yes" or "no",
  "entities": [list of entities or entity types],
  "attributes": [list of required attributes / evidence facets],
  "justification": "one concise sentence referencing the rubric"
}

Task: q
"""
\end{lstlisting}
}
\end{tcolorbox}

\begin{table*}[!ht]
  \centering
  \begin{small}
  \setlength{\tabcolsep}{5pt}
  \caption{Filtering statistics of the two-stage construction pipeline for constructing the agentic-wide subset.}
  \begin{tabular}{lccc}
    \toprule
    
    \textbf{Source Benchmark }& \textbf{\#Total} & \textbf{\#Stage I Candidates} & \textbf{\#Final Selected} \\
    \midrule
    xBench-DeepSearch & 100  & 54 & 31 \\
    WebWalkerQA      & 680 & 217 & 120 \\
    TaskCraft        & 5228 & 429 & 101 \\
    \bottomrule
  \end{tabular}
  \label{tab:agentic_wide_filtering}
  \end{small}
\end{table*}

\textbf{Quality Control.}
Tab.\ref{tab:agentic_wide_filtering} summarizes the instance flow through the two-stage filtering pipeline, illustrating the reduction from initial benchmarks to the final agentic-wide subset. As a diagnostic check, we inspect a small random subset of the automatically selected instances to verify the consistency of the filtering criteria. This inspection does not alter the selection results. We observe that the selected instances largely satisfy the above requirements; a small fraction involves a relatively smaller retrieval scope, yet still requires multi-entity evidence collection and synthesis, and is therefore retained under the operational definition described above.

\section{Evaluation Metrics and Protocols}
\label{app:sec:metrics}

\subsection{Table-based Metrics for WideSearch and DeepWideSearch}
\label{app:sec:table_metrics}

Each instance is evaluated by comparing a predicted Markdown table $\hat{T}$ against a ground-truth table $T$.
We first normalize both tables by (i) canonicalizing whitespace and punctuation, (ii) lowercasing, and (iii) applying a consistent numeric/string formatting.
We then evaluate correctness at three granularities: table-, row-, and cell-level.

\begin{itemize}[leftmargin=1.6em, itemsep=5pt, topsep=2pt, parsep=0pt, partopsep=0pt]
  \item \textbf{Success Rate (SR).}
Success Rate is a strict, exact-match metric:
\begin{equation}
\mathrm{SR}(\hat{T}, T)=\mathbbm{1}\big[\hat{T} \equiv T\big],
\end{equation}
where $\hat{T} \equiv T$ denotes that the two tables match exactly after normalization, including the full structure (rows/columns) and all cell values.

  \item \textbf{Row-level F1.}
We treat each row as a record and compute an F1 score based on set overlap between predicted rows and gold rows.
Let $\mathcal{R}(T)$ denote the set of rows in table $T$, where each row is represented as a tuple of cell values under the schema of $T$.
Define precision, recall, and F1 score as
\begin{equation}
\begin{aligned}
P_{\text{row}} = \frac{|\mathcal{R}(\hat{T}) \cap \mathcal{R}(T)|}{|\mathcal{R}(\hat{T})|}, \\
R_{\text{row}} = \frac{|\mathcal{R}(\hat{T}) \cap \mathcal{R}(T)|}{|\mathcal{R}(T)|},\\
\mathrm{F1}_{\text{row}}=\frac{2P_{\text{row}}R_{\text{row}}}{P_{\text{row}}+R_{\text{row}}+\epsilon}
\end{aligned}
\end{equation}
where $\epsilon$ is a small constant to avoid division by zero.

  \item \textbf{Item-level F1.}
Item-level F1 measures correctness at the cell (atomic item) level.
Let $\mathcal{I}(T)$ be the multiset of table items, where each item is a normalized cell value optionally paired with its column identifier.
We compute
\begin{equation}
\begin{aligned}
P_{\text{item}}=\frac{|\mathcal{I}(\hat{T}) \cap \mathcal{I}(T)|}{|\mathcal{I}(\hat{T})|}, \\
R_{\text{item}}=\frac{|\mathcal{I}(\hat{T}) \cap \mathcal{I}(T)|}{|\mathcal{I}(T)|}, \\
\mathrm{F1}_{\text{item}}=\frac{2P_{\text{item}}R_{\text{item}}}{P_{\text{item}}+R_{\text{item}}+\epsilon}.
\end{aligned}
\end{equation}
\end{itemize}

\subsection{Depth-oriented Metrics for DeepWideSearch}
\label{app:sec:depth_metrics}

Based on the metrics above, DeepWideSearch additionally evaluates \emph{depth} by checking whether the agent correctly identifies the core entity and the corresponding key attributes.

\begin{itemize}[leftmargin=1.6em, itemsep=5pt, topsep=2pt, parsep=0pt, partopsep=0pt]
  \item \textbf{Column-level F1 (Column-F1).}
Column-F1 is computed over the set of \emph{unique columns} (i.e., core attributes) in the table.
Let $\mathcal{C}(T)$ denote the set of canonicalized column headers in $T$.
We compute
\begin{equation}
\begin{aligned}
P_{\text{col}} &=
\frac{|\mathcal{C}(\hat{T}) \cap \mathcal{C}(T)|}{|\mathcal{C}(\hat{T})|}, \\
R_{\text{col}} &=
\frac{|\mathcal{C}(\hat{T}) \cap \mathcal{C}(T)|}{|\mathcal{C}(T)|}, \\
\mathrm{F1}_{\text{col}} &=
\frac{2P_{\text{col}}R_{\text{col}}}{P_{\text{col}}+R_{\text{col}}+\epsilon}.
\end{aligned}
\end{equation}
Intuitively, Column-F1 rewards correctly recovering the schema-level attributes that organize entity records, complementing row/item metrics that measure value-level correctness.

  \item \textbf{Core Entity Accuracy (CE Acc.).}
DeepWideSearch defines a \emph{core entity} (or target entity set) whose identification requires deeper reasoning over multi-hop evidence.
Core Entity Accuracy measures whether the predicted output includes the correct core entity signal as specified by the benchmark protocol:
\begin{equation}
\begin{aligned}
\mathrm{CEAcc}(\hat{T}, T)
&=\mathbbm{1}\big[\mathrm{CoreEntity}(\hat{T}) \\
&\quad=\mathrm{CoreEntity}(T)\big].
\end{aligned}
\end{equation}
We follow the official DeepWideSearch evaluator to extract the core-entity field (or identifier) from the predicted table and compare it against the ground-truth core entity.
\end{itemize}

\subsection{Aggregation over Multiple Runs}
\label{app:sec:multi_run}

To reduce randomness, we run each instance independently $N$ times.
For any metric that yields a scalar per run (including precision/recall/F1 at different granularities), we report:
\begin{equation}
\begin{aligned}
\mathrm{Avg@}N &= \frac{1}{N}\sum_{j=1}^{N} m^{(j)}, \\
\mathrm{Max@}N &= \max_{1\le j \le N} m^{(j)}.
\end{aligned}
\end{equation}
Importantly, we compute $\mathrm{F1}^{(j)}$ \emph{within each run} using the run-specific precision/recall, and then aggregate $\mathrm{F1}^{(j)}$ across runs.
Therefore, the reported $\mathrm{Avg@}N$ (or $\mathrm{Max@}N$) F1 is not necessarily equal to $\frac{2\,\mathrm{Avg@}N(P)\,\mathrm{Avg@}N(R)}{\mathrm{Avg@}N(P)+\mathrm{Avg@}N(R)}$.

For the binary Success Rate $\mathrm{SR}^{(j)}\in\{0,1\}$, besides $\mathrm{Avg@}N$ and $\mathrm{Max@}N$, we additionally report:
\begin{equation}
\mathrm{Pass@}N = \mathbbm{1}\Big[\max_{1\le j \le N}\mathrm{SR}^{(j)}=1\Big],
\end{equation}
and aggregate Pass@N across the dataset as the fraction of instances that succeed in at least one run.

\subsection{Evaluation for the Agentic-Wide Subset}
\label{app:sec:agenticwide_eval}

The agentic-wide subset consists of free-form QA tasks with text references.
We evaluate using \textbf{accuracy}, where correctness is determined by an LLM-based judge that assesses whether the model output is semantically equivalent to the ground-truth answer (allowing for paraphrases and minor formatting differences).

\paragraph{LLM-judge protocol.}
Given the instance prompt $q$, the model prediction $\hat{a}$, and the reference answer $a^\ast$, the judge outputs a binary decision (\texttt{correct}/\texttt{incorrect}) and a short justification.
We use a conservative rubric: the judge marks \texttt{correct} only if $\hat{a}$ contains the necessary and sufficient information required by $a^\ast$ and does not contradict it.

\begin{tcolorbox}[notitle, sharp corners, breakable, colframe=ForestGreen, colback=white, 
       boxrule=3pt, boxsep=0.5pt, enhanced, 
       shadow={3pt}{-3pt}{0pt}{opacity=1,fill=mygrey},
       title={LLM Judge Prompt for Agentic-Wide QA Evaluation},]\label{box:tag-generate}
       \scriptsize
       {\fontfamily{pcr}\selectfont
\begin{lstlisting}
QUERY_TEMPLATE = """
{Question}

Your response should be in the following format:
Explanation: {{your explanation for your final answer}}
Exact Answer: {{your succinct, final answer}}
Confidence: {{your confidence score between 0% and 100% for your answer}}
""".strip()

GRADER_TEMPLATE = """
Judge whether the following [response] to [question] is correct or not based on the precise and unambiguous [correct_answer] below.

[question]: {question}

[response]: {response}

Your judgement must be in the format and criteria specified below:

extracted_final_answer: The final exact answer extracted from the [response]. Put the extracted answer as 'None' if there is no exact, final answer to extract from the response.

[correct_answer]: {correct_answer}

reasoning: Explain why the extracted_final_answer is correct or incorrect based on [correct_answer], focusing only on if there are meaningful differences between [correct_answer] and the extracted_final_answer. Do not comment on any background to the problem, do not attempt to solve the problem, do not argue for any answer different than [correct_answer], focus only on whether the answers match.

correct: Answer 'yes' if extracted_final_answer matches the [correct_answer] given above, or is within a small margin of error for numerical problems. Answer 'no' otherwise, i.e. if there if there is any inconsistency, ambiguity, non-equivalency, or if the extracted answer is incorrect.


confidence: The extracted confidence score between 0|%| and 100|%| from [response]. Put 100 if there is no confidence score available.
""".strip()

CORRECT_RE = re.compile(r"correct:\s*(yes|no)", re.IGNORECASE)
"""
\end{lstlisting}
}
\end{tcolorbox}

We compute accuracy as the proportion of instances labeled \texttt{correct} by the judge.

\section{Discussion with Related Works}
\label{app:discussion}

\paragraph{Execution-space control beyond parallelism.}
A large line of workflow-search MAS constructs an unbounded interaction graph and searches over agent topologies and communication patterns~\cite{hu2025automated,wang2025mas2selfgenerativeselfconfiguringselfrectifying,zhuge2024gptswarm}.
While flexible, this induces an execution space that is extremely large and sparse, making systematic control and reuse difficult.
In contrast, \textsc{A-MapReduce} compresses open-ended topology search into a \emph{parameterized decision space} by restricting execution to a MapReduce-inspired horizontal structure.
Concretely, We optimize a compact decision tuple $\Theta_q=(M_q,P_q,B_q)$: the task matrix $M_q$ specifies \emph{what} retrieval targets to cover (schema-aligned map units), the template $P_q$ specifies \emph{how} to generate schema-grounded, reusable sub-queries and normalize their outputs for each unit, and the batching/shuffling policy $B_q$ specifies \emph{how} to allocate resources and schedule map jobs. The reducer then applies schema-aware merge/backfill rules to reconcile partial evidence into a single structured table.

This shift replaces ''searching for an agent graph'' with \emph{controlling the execution space} through editable operators, enabling targeted revisions of decomposition granularity, grouping keys, and merge semantics.
Such strong constraints are particularly important for WideSearch, where performance is highly sensitive to the \emph{initial} plan: early mismatches in $M_q$ or $B_q$ can lock the system into coverage gaps and misaligned aggregation that are hard to recover from with purely prompt-level refinement.

\paragraph{Horizontal coverage control vs.\ vertical recursive reasoning.}
Many classic MAS follow a vertical, recursive reasoning paradigm that primarily controls \emph{depth} via multi-turn, hierarchical deliberation~\cite{fourney2024magentic,zhang2026agentorchestraorchestratingmultiagentintelligence}.
This vertical control is effective for compositional reasoning, but can suffer in long-horizon wide-search regimes where intermediate evidence is overwritten and objectives drift.
\textsc{A-MapReduce} instead emphasizes \emph{horizontal} control over breadth and coverage: by externalizing retrieval goals as a schema-aligned task matrix $M_q$, the system can explicitly track progress over massive retrieval targets and enforce reducer-level aggregation rules.
This makes the execution state observable and correctable, and directly mitigates common WideSearch failure modes such as coverage loss and misaligned table merges.

\paragraph{Experience-driven pruning and distribution sharpening.}
Beyond ``simple parallelism'' baselines~\cite{qin2025flashsearcherfasteffectiveweb}, \textsc{A-MapReduce} introduces an experiential evolution mechanism that prunes and refines the execution space over time.
By storing trajectories $\tau$ together with utility feedback $u$, the system can identify low-utility execution patterns (e.g., overly fine-grained decomposition or redundant retrieval) and convert them into transferable plan-edit hints.
Critically, conditional decision sampling $p(\Theta\mid q,\Delta_q)$ uses these hints as a query-conditioned prior that \emph{sharpens} the decision distribution, biasing sampling toward historically high-utility regions of $\Theta=(M,P,B)$.
In this sense, the framework ''learns how to plan'' in a unified MapReduce paradigm: experience updates \emph{which plans are likely to be sampled} for new queries, rather than merely retrieving more content for the current answer.

\paragraph{State externalization and feedback alignment.}
In many workflow MAS, execution state remains implicit in dialogue histories or latent plans, which makes debugging and credit assignment difficult.
\textsc{A-MapReduce} externalizes the retrieval objective into a schema-aligned task matrix and a deterministic reducer/merge interface, making the execution state measurable and directly editable.
Moreover, because the output is a structured table $Y$, evaluator metrics (e.g., exact-match/structural F1) provide a high-fidelity utility signal for evolution, in contrast to approaches that rely on ambiguous natural-language judgments.
This tight feedback loop improves the reliability of online updates and helps keep the evolution objective aligned with correctness and coverage.

\section{Future Work}
\label{app:future_openworld}

\paragraph{Adapting to open-world wide search.}
Our experiments rely on benchmark-specific evaluators to compute utility signals, but the \textsc{A-MapReduce} framework itself does not require a fixed notion of ground truth.
A natural next step is to study \textsc{A-MapReduce} in open-world deployments where objectives are \emph{non-stationary}: requirements evolve, gold standards may be unavailable, and feedback may be sparse, noisy, or preference-driven.
In this regime, memory can be viewed as a \emph{persistent control layer} that stabilizes long-horizon behavior: it consolidates reusable decision patterns, down-weights brittle heuristics, and maintains schema-aligned execution policies even when supervision is weak.
An important research direction is to characterize what types of lightweight feedback (e.g., user approval, constraint violations, provenance coverage) are sufficient to support stable decision evolution, and how to make memory robust to drift and noisy signals.

\paragraph{From homogeneous to specialist search agents.}
In our current implementation, search agents are largely homogeneous in capability and role.
However, real-world wide-search workloads are inherently \emph{heterogeneous} at the row/field level: some subproblems resemble academic literature search, others require news retrieval, database-like lookups, or cross-source fact consolidation.
A promising extension is to introduce a pool of \emph{specialist} agents (e.g., academic, news, product, forum, or verification-oriented roles) and let \textsc{A-MapReduce} dynamically allocate specialists to different rows or schema attributes in the task matrix $M_q$.
This turns batching into a \emph{role-aware} scheduling problem: the system must decide not only \emph{what} to batch (by row/attribute), but also \emph{which} specialist to assign and \emph{which} role-specific experience to retrieve.

\paragraph{Role-conditioned memory and decision evolution.}
With specialist allocation, memory can be made more targeted by indexing and retrieving experience \emph{conditioned on role} (and optionally on attribute type or source domain).
Such role-conditioned memory would enable more precise reuse: academic-oriented hints guide citation/provenance handling, news-oriented hints emphasize recency and source diversity, and verification-oriented hints focus on consistency and schema constraints.
We expect this to improve both efficiency and reliability, as the MapReduce controller can co-evolve its batching strategy together with specialist selection, yielding a more adaptive and compositional execution policy for open-world wide search.

\section{Supplementary Results}
This section provides additional results supporting the effectiveness and robustness of \textsc{A-MapReduce}. Table~\ref{tab:gpt5mini_table_metrics} shows that, under a strictly controlled GPT-5-mini setting, \textsc{A-MapReduce} consistently outperforms strong baselines while also reducing delay and cost, confirming that the gains are not merely due to backbone or environment differences. Table~\ref{tab:sensitivity_insights_topk} analyzes the effect of the number of retrieved experiential insights, showing that moderate retrieval depth yields the best overall trade-off. Table~\ref{tab:ablation_study_complete} further validates the contribution of the experiential prior, memory, and MapReduce decision parameters, as removing any key component leads to clear performance degradation. Table~\ref{tab:agentic_wide_results} demonstrates that the method generalizes beyond WideSearch to broader agentic-wide subsets. Finally, Table~\ref{tab:threshold_sensitivity} shows that the utility-threshold design is reasonably robust, with the medium setting achieving the best balance across accuracy, success rate, and efficiency.

\begin{table*}[!ht]
\centering
\begin{small}
\caption{Performance comparison on WideSearch under a strictly controlled setup with \textbf{GPT-5-mini} as the shared backbone and an \textbf{identical execution environment/configuration} across all methods. We report Item-level (Precision/Recall/F1), Row-level (Precision/Recall/F1), Success Rate (Succ.), average runtime per task (Delay), and API cost per task (Cost). \textsc{A-MapReduce}* denotes the variant without experience-driven evolution.}

\label{tab:gpt5mini_table_metrics}
\resizebox{\textwidth}{!}{
\begin{tabular}{lcccccccccc}
\toprule

\textbf{Method} &
\textbf{Item P(\%)} & \textbf{Item R(\%)} & \textbf{Item F1(\%)} &
\textbf{Row P(\%)} & \textbf{Row R(\%)} & \textbf{Row F1(\%)} &
\textbf{Succ.(\%)} &
\textbf{Delay(s)} & \textbf{Cost(\$)} \\
\midrule
SmolAgents
& 66.84 & 47.54 & 51.31
& 30.67 & 21.37 & 23.04
& 4.00
& 2617.65 & 1.243 \\
Flash-Searcher
& 67.57 & 52.16 & 54.99
& 41.08 & 33.00 & 34.42
& 6.40
& 1204.69 & 0.396 \\
\textsc{A-MapReduce}$^{*}$
& 75.63 & 62.34 & 64.64
& 48.14 & 40.26 & 41.62
& 6.75
& 1460.80 & 1.080 \\
\textbf{\textsc{A-MapReduce}}
& \textbf{78.99} & \textbf{65.48} & \textbf{67.81}
& \textbf{51.60} & \textbf{43.83} & \textbf{45.23}
& \textbf{7.50}
& \textbf{953.66} & \textbf{0.763} \\
\bottomrule
\end{tabular}}
\end{small}
\end{table*}

\begin{table*}[!ht]
\centering
\begin{small}
\caption{Sensitivity to the number of retrieved insights ($k_{\mathcal{H}}$) in the experiential prior. We report Avg@4 for Item-level (Precision/Recall/F1), Row-level (Precision/Recall/F1), Success Rate (Succ.), average delay (Delay), and API cost per task (Cost).}
\label{tab:sensitivity_insights_topk}
\resizebox{\textwidth}{!}{
\begin{tabular}{lccccccccc}
\toprule

\textbf{$k_{\mathcal{H}}$} &
\textbf{Item P(\%)} & \textbf{Item R(\%)} & \textbf{Item F1(\%)} &
\textbf{Row P(\%)} & \textbf{Row R(\%)} & \textbf{Row F1(\%)} &
\textbf{Succ.(\%)} &
\textbf{Delay(s)} & \textbf{Cost(\$)} \\
\midrule
1 & 80.22 & 64.24 & 67.19 & 48.02 & 39.34 & 41.20 & 4.17 & 1506.28 & 0.740 \\
2 & 79.11 & 65.28 & 68.13 & 45.08 & 37.51 & 39.25 & 5.00 & 1050.53 & 0.648 \\
3 & 82.07 & 66.62 & 70.12 & 51.18 & 41.41 & 43.65 & 6.67 & 824.49 & 0.602 \\
4 & 77.73 & 64.15 & 66.98 & 45.42 & 37.77 & 39.48 & 3.33 & 1312.29 & 0.667 \\
\bottomrule
\end{tabular}}
\end{small}
\end{table*}

\begin{table*}[!ht]
\centering
\begin{small}
\caption{Ablation study of the proposed experiential prior design and MapReduce decision parameters $\Theta$ in \textsc{A-MapReduce}. We report Avg@4 for Item-level (Precision/Recall/F1), Row-level (Precision/Recall/F1), Success Rate (Succ.), and API cost (Cost).}
\label{tab:ablation_study_complete}
\resizebox{\textwidth}{!}{
\begin{tabular}{lcccccccc}
\toprule

\textbf{Method} &
\textbf{Item P(\%)} & \textbf{Item R(\%)} & \textbf{Item F1(\%)} &
\textbf{Row P(\%)} & \textbf{Row R(\%)} & \textbf{Row F1(\%)} &
\textbf{Succ.(\%)} &
\textbf{Cost(\$)} \\
\midrule
\textsc{A-MapReduce}
& 82.07 & 66.62 & 70.12
& 51.18 & 41.41 & 43.65
& 6.67
& 0.60 \\
\textit{-w/o Few-shot}
& 81.31\drop{0.76} & 65.38\drop{1.24} & 68.76\drop{1.36}
& 47.31\drop{3.87} & 38.61\drop{2.80} & 40.57\drop{3.08}
& 5.83\drop{0.84}
& 0.64 \\
\textit{-w/o Insights}
& 77.44\drop{4.63} & 63.44\drop{3.18} & 66.47\drop{3.65}
& 47.37\drop{3.81} & 38.65\drop{2.76} & 40.58\drop{3.07}
& 5.00\drop{1.67}
& 0.62 \\
\textit{-w/o Memory}
& 75.12\drop{6.95} & 61.92\drop{4.70} & 64.97\drop{5.15}
& 44.37\drop{6.81} & 35.89\drop{5.52} & 37.92\drop{5.73}
& 4.16\drop{2.51}
& 1.05 \\
\midrule
\textit{-w/o $M_q~\&~P_q$}
& 77.54\drop{4.53} & 64.52\drop{2.10} & 67.90\drop{2.22}
& 43.30\drop{7.88} & 35.96\drop{5.45} & 37.85\drop{5.80}
& 3.33\drop{3.34}
& 1.08 \\
\textit{-w/o $B_q$}
& 77.95\drop{4.12} & 62.40\drop{4.22} & 65.95\drop{4.17}
& 43.77\drop{7.41} & 35.17\drop{6.24} & 37.33\drop{6.32}
& 4.17\drop{2.50}
& 0.68 \\
\bottomrule
\end{tabular}}
\end{small}
\end{table*}

\begin{table*}[!ht]
\centering
\caption{Performance comparison on Agentic-Wide subsets using GPT-5-mini  as the backbone. We report Accuracy (Acc.), Delay (end-to-end runtime), and Cost (average API cost) on the XBench, WebWalkerQA, and TaskCraft subsets.}
\label{tab:agentic_wide_results}
\begin{small}
\resizebox{\textwidth}{!}{%
\begin{tabular}{lccccccccc}
\toprule
\multirow{2}{*}{Model} 
& \multicolumn{3}{c}{XBench Subset}
& \multicolumn{3}{c}{WebWalkerQA Subset}
& \multicolumn{3}{c}{TaskCraft Subset} \\
\cmidrule(lr){2-4} \cmidrule(lr){5-7} \cmidrule(lr){8-10}
& Acc. (\%) & Delay (s) & Cost (\$)
& Acc. (\%) & Delay (s) & Cost (\$)
& Acc. (\%) & Delay (s) & Cost (\$) \\
\midrule
OWL
& 45.16 & 1003.27 & 0.13
& 52.50 & 652.08 & 0.11
& 74.26 & 702.48 & 0.08 \\

SmolAgents
& 51.61 & 1335.10 & 0.57
& 50.00 & 836.29 & 0.21
& 80.20 & 607.04 & 0.12 \\

Flash-Searcher
& 67.74 & 679.30 & 0.25
& 59.17 & 326.42 & 0.07
& 87.13 & 275.90 & 0.06 \\

\textbf{\textsc{A-MapReduce}}
& 74.19 & 707.32 & 0.35
& 62.50 & 297.48 & 0.08
& 90.10 & 384.07 & 0.09 \\
\bottomrule
\end{tabular}
}
\end{small}
\end{table*}

\begin{table*}[!ht]
\centering
\begin{small}
\caption{Sensitivity analysis of the utility threshold $\delta$ used to split retrieved episodes into positive vs.\ negative sets for \textsc{A-MapReduce} on WideSearch. We report Avg@4 for Item-level (Precision/Recall/F1), Row-level (Precision/Recall/F1), Success Rate (Succ.), and Delay (end-to-end runtime per instance).}
\label{tab:threshold_sensitivity}
\resizebox{\textwidth}{!}{
\begin{tabular}{lcccccccc}
\toprule

\textbf{$\delta$ Setting} &
\textbf{Item P(\%)} & \textbf{Item R(\%)} & \textbf{Item F1(\%)} &
\textbf{Row P(\%)} & \textbf{Row R(\%)} & \textbf{Row F1(\%)} &
\textbf{Succ.(\%)} &
\textbf{Delay(s)} \\
\midrule
Low
& 79.53 & 65.12 & 69.17
& 49.68 & 42.72 & 44.39
& 5.00
& 1045.3 \\
Medium
& 82.07 & 66.62 & 70.12
& 51.18 & 41.41 & 43.65
& 6.67
& 953.7 \\
High
& 78.41 & 65.49 & 68.74
& 48.35 & 40.79 & 42.97
& 5.58
& 1140.5 \\
\bottomrule
\end{tabular}}
\end{small}
\end{table*}

\section{Case study: Evolution of Execution Decisions}
\label{app:case_studies}

This section presents four case studies illustrating how \textsc{A-MapReduce} evolves execution decisions under different failure patterns. They cover template adaptation, batching refinement, task-matrix adjustment, and targeted repair, showing that the framework can update different components rather than relying on a fixed execution workflow.

\textbf{Case Study 1: Query-conditioned template evolution.}
Updating $P$; see Fig.~\ref{fig:case1} and Tab.~\ref{tab:case_1}.

\textbf{Case Study 2: Batching evolution on DeepWideSearch.}
Updating $B$; see Fig.~\ref{fig:case_2} and Tab.~\ref{tab:case_wide2deep_ws_en_076}.

\textbf{Case Study 3: Task-matrix and template evolution for stable league--season alignment.}
Updating $M$ and $P$; see Fig.~\ref{fig:case_ws_en_083} and Tab.~\ref{tab:case_ws_en_083}.

\textbf{Case Study 4: Batching evolution with delta-patch backfilling for sparse fields.}
Updating $B$ with a targeted repair step; see Fig.~\ref{fig:case_ws_en_071} and Tab.~\ref{tab:case_ws_en_071}.

\section{Notation}

Table~\ref{tab:notation_method} summarizes the main notation used in our framework.

\clearpage

\begin{figure*}[!t]
  \centering
  \includegraphics[width=\textwidth]{figures/case1.svg.pdf}
  \caption{Case study 1: query-conditioned template evolution on a representative WideSearch task (updating $P$). In Case study 1, we demonstrate query-conditioned template evolution (updating $P$) to generate a $5 \times 5$ financial table with IFRS operating profit, margin, and free cash flow (FCF), along with cited sources. As shown in Figure \ref{fig:case1}, the system first generates an initial decision $\Theta = (M, P, B)$ in Stage 1, where the experiential prior captures latent error patterns based on prior executions. Using these memory signals, the agent refines the template $P$ in Stage 2, incorporating accounting-aware extraction and verification rules that ensure data accuracy and consistency. In this stage, the template is enhanced to handle missing values and potential data inconsistencies more effectively, improving the overall quality and reliability of the output.}
  \label{fig:case1}
\end{figure*}

\begin{table*}[!t]
  \centering
  \caption{The performance of case study 1. We report row-level and item-level metrics, together with wall-clock runtime and API cost. The performance improvement after template evolution is shown in Table \ref{tab:case_1}, with Row F1 increasing from 0.000 to 0.520 and Item F1 improving from 0.573 to 0.887. Additionally, the evolved execution reduces time from 57.50 minutes to 34.42 minutes and cost from 1.76 USD to 1.41 USD. These results highlight the effectiveness of memory-guided template refinement in correcting errors, improving coverage of valid cells, and optimizing execution efficiency.}
  \label{tab:case_1}
  \setlength{\tabcolsep}{5pt}
  {\small
  \resizebox{\textwidth}{!}{%
\begin{tabular}{lccccccccc}
    \toprule
    
    \textbf{Framework} & \textbf{Evo.} &
    \textbf{Item P(\%)} & \textbf{Item R(\%)} & \textbf{Item F1(\%)} &
    \textbf{Row P(\%)} & \textbf{Row R(\%)} & \textbf{Row F1(\%)} &
    \textbf{Delay(s)} & \textbf{Cost(\$)} \\
    \midrule
    \multirow{2}{*}{\textbf{A-MapReduce}} & \textbf{\checkmark} &
    \textbf{88.67} & \textbf{88.67} & \textbf{88.67} &
    \textbf{52.00} & \textbf{52.00} & \textbf{52.00} &
    \textbf{2065.21} & \textbf{1.41} \\
    & $\times$ &
    57.33 & 57.33 & 57.33 &
    0.00 & 0.00 & 0.00 &
    3448.17 & 1.76 \\
    \bottomrule
  \end{tabular}
}
  }
\end{table*}

\clearpage


\begin{figure*}[!t]
  \centering
  \includegraphics[width=\textwidth]{figures/case4.svg.pdf}
  \caption{Case study 2: batching evolution for a DeepWideSearch query (updating $B$). We illustrate how the batching strategy evolves during execution. The query requests a $5$-country $\times$ $6$-year table (2019--2024; 30 rows) from multiple sources, with a year-level tie-break rule. As shown in Fig.~\ref{fig:case_2}, the original decision in Stage~1 employs an open-batching strategy that mixes countries, leading to redundant page fetches and inconsistent year-specific selections. Leveraging the experiential prior, the evolution refines the batching strategy by switching from the $open$-batching approach to $by\_attr(Country)$, ensuring that each search agent has one retrieval context per country for all years. Additionally, the template $P$ is refined for unit normalization.}

  \label{fig:case_2}
\end{figure*}

\begin{table*}[!t]
\centering
\caption{The performance of case study 2. We report row-level and item-level metrics, together with wall-clock runtime and API cost. The results in Table \ref{tab:case_wide2deep_ws_en_076} show significant improvements after evolution: Row F1 increases from 0.00\% to 73.33\%, and Item F1 rises from 78.10\% to 96.19\%. Additionally, the evolved execution reduces both running time and API cost, demonstrating the effectiveness of memory-guided batching evolution in enhancing structure, performance, and efficiency.}
\label{tab:case_wide2deep_ws_en_076}
\setlength{\tabcolsep}{5pt}
{\small
\resizebox{\textwidth}{!}{%
\begin{tabular}{lccccccccc}
  \toprule
  
  \textbf{Framework} & \textbf{Evo.} &
  \textbf{Item P(\%)} & \textbf{Item R(\%)} & \textbf{Item F1(\%)} &
  \textbf{Row P(\%)} & \textbf{Row R(\%)} & \textbf{Row F1(\%)} &
  \textbf{Delay(s)} & \textbf{Cost(\$)} \\
  \midrule
  \multirow{2}{*}{\textbf{A-MapReduce}} & \textbf{\checkmark} &
  \textbf{96.19} & \textbf{96.19} & \textbf{96.19} &
  \textbf{73.33} & \textbf{73.33} & \textbf{73.33} &
  \textbf{897.60} & \textbf{0.36} \\
  & $\times$ &
  78.10 & 78.10 & 78.10 &
  0.00 & 0.00 & 0.00 &
  962.40 & 0.68 \\
  \bottomrule
\end{tabular}
}
}
\end{table*}

\clearpage
\begin{figure*}[!t]
  \centering
  \includegraphics[width=\textwidth]{figures/case2.svg.pdf}
  \caption{Case study 3: task-matrix and template evolution for a DeepWideSearch query (updating $M$ and $P$). In Case study 3, we demonstrate the evolution of the task-matrix and template for stable league-season alignment (updating $M$ and $P$). The query requests a $5$-league $\times$ $5$-season table, reporting the champion club and Sofascore Player-of-the-Season signals. In Stage~1, the initial decision $\Theta = (M, P, B)$ uses a task matrix $M$ indexed by (\textsc{League}, \textsc{Season}) and a coarse template $P$, which can result in repeated champion lookups and unstable or missing champion-to-award mappings for some rows. Evolution improves the alignment by adding \textsc{Champion} as an intermediate field in $M$ and refactoring $P$ into a conditioned procedure that first fixes champions and then retrieves award signals based on the known champion context.}
  \label{fig:case_ws_en_083}
\end{figure*}

\begin{table*}[!t]
\centering
\caption{The performance of case study 3. We report row-level and item-level metrics, together with wall-clock runtime and API cost. The results in Table \ref{tab:case_ws_en_083} show significant improvements after evolution: Row F1 increases from 16.00\% to 64.00\%, and Item F1 improves from 44.67\% to 88.00\%. Furthermore, the evolved execution reduces time from 50.63 minutes to 37.03 minutes, and cost from 4.78 USD to 2.58 USD, demonstrating the effectiveness of memory-guided task-matrix and template evolution in improving structure, performance, and efficiency.}
\label{tab:case_ws_en_083}
\setlength{\tabcolsep}{5pt}
{\small
\resizebox{\textwidth}{!}{%
\begin{tabular}{lccccccccc}
  \toprule
  
  \textbf{Framework} & \textbf{Evo.} &
  \textbf{Item P(\%)} & \textbf{Item R(\%)} & \textbf{Item F1(\%)} &
  \textbf{Row P(\%)} & \textbf{Row R(\%)} & \textbf{Row F1(\%)} &
  \textbf{Delay(s)} & \textbf{Cost(\$)} \\
  \midrule
  \multirow{2}{*}{\textbf{A-MapReduce}} & \textbf{\checkmark} &
  \textbf{88.00} & \textbf{88.00} & \textbf{88.00} &
  \textbf{64.00} & \textbf{64.00} & \textbf{64.00} &
  \textbf{2221.80} & \textbf{2.58} \\
  & $\times$ &
  44.67 & 44.67 & 44.67 &
  16.00 & 16.00 & 16.00 &
  3037.80 & 4.78 \\
  \bottomrule
\end{tabular}
}
}
\end{table*}

\clearpage


\begin{figure*}[!t]
  \centering
  \includegraphics[width=\textwidth]{figures/case5.svg_compressed.pdf}
  \caption{Case study 4: batching evolution and delta-patch backfilling for a DeepWideSearch query (updating $B$). In Case study 4, we demonstrate the evolution of batching and delta-patch backfilling for sparse fields (updating $B$). The query requests the Forbes Global Billionaires top-10 list for each year from 2019 to 2024 (60 rows), including net worth and \textsc{Age}. Evolution refines batching by switching from $per\_atom$ to $by\_attr(Year)$, ensuring that each agent has a single year, thereby significantly increasing execution efficiency. However, this execution leaves \textsc{Age} cells missing. To address this, a delta-patch operator is introduced: a second MapReduce call fills in the missing \textsc{Age} values, followed by a keyed join.}
  \label{fig:case_ws_en_071}
\end{figure*}

\begin{table*}[!t]
\centering
\caption{The performance of case study 4. We report row-level and item-level metrics, together with wall-clock runtime and API cost.As shown in Table, the evolution results in significant improvements: Row F1 increases from 0.00\% to 58.33\%, and Item F1 rises from 51.52\% to 79.20\%. Additionally, execution time and cost decrease by 18.3\% and 22.5\%, respectively, demonstrating the effectiveness of delta-patch backfilling in improving performance while reducing both time and cost.}
\label{tab:case_ws_en_071}
\setlength{\tabcolsep}{5pt}
{\small
\resizebox{\textwidth}{!}{%
\begin{tabular}{lccccccccc}
  \toprule
  
  \textbf{Framework} & \textbf{Evo.} &
  \textbf{Item P(\%)} & \textbf{Item R(\%)} & \textbf{Item F1(\%)} &
  \textbf{Row P(\%)} & \textbf{Row R(\%)} & \textbf{Row F1(\%)} &
  \textbf{Delay(s)} & \textbf{Cost(\$)} \\
  \midrule
  \multirow{2}{*}{\textbf{A-MapReduce}} & \textbf{\checkmark} &
  \textbf{79.20} & \textbf{79.20} & \textbf{79.20} &
  \textbf{58.33} & \textbf{58.33} & \textbf{58.33} &
  \textbf{1917.00} & \textbf{1.21} \\
  & $\times$ &
  56.70 & 47.20 & 51.52 &
  0.00 & 0.00 & 0.00 &
  2341.20 & 1.56 \\
  \bottomrule
\end{tabular}
}
}
\end{table*}

\newpage

\begin{table*}[!t]
\centering
\begingroup
\scriptsize
\renewcommand{\arraystretch}{0.86}
\caption{Notation and Definition}
\label{tab:notation_method}
\setlength{\tabcolsep}{3pt}
\begin{tabular}{@{}>{\raggedright\arraybackslash}p{0.17\textwidth}
                >{\raggedright\arraybackslash}p{0.31\textwidth}
                >{\raggedright\arraybackslash}p{0.17\textwidth}
                >{\raggedright\arraybackslash}p{0.31\textwidth}@{}}
\toprule
\textbf{Notation} & \textbf{Definition} & \textbf{Notation} & \textbf{Definition} \\
\midrule
$q=(q^{\text{text}},S)$ & Wide-search query: a natural-language request plus an explicit schema. & $\mathcal{D}=\{\epsilon_i\}_{i=1}^{N_{\mathcal{D}}}$ & Set of past records; $N_{\mathcal{D}}$ is the set size. \\
$S=\{s_k\}_{k=1}^{K}$ & Schema attributes (columns); $K$ is the number of requested fields. & $\epsilon_i=(q_i,\Theta_i,\tau_i,u_i)$ & Record: query, decision, trace, and utility feedback. \\
$E=\{e_i\}_{i=1}^{N}$ & Latent target entities (rows) induced by $q$; $N$ is the number of entities. & $\mathcal{H}$ & Hint pool of reusable decision guidance (about $\Theta=(M,P,B)$). \\
$Y \in \mathcal{V}^{N\times K}$ & Structured table output; $\mathcal{V}$ is the value space of cells. & $h$ & A hint rule's content used to bias planning/decision sampling. \\
$\Theta$ & High-level execution decision sampled by the MAS (controls decomposition/scheduling/execution). & $w_h$ & Online hint score updated from utility feedback (higher implies more reliable guidance). \\
$p(\Theta\mid q)$ & Query-conditioned decision distribution to be refined by experience-based evolution. & $\mathcal{S}(h)$ & Hint provenance: supporting record indices in $\mathcal{D}$ (for retrieval/voting). \\
$Q(Y;q)$ & Quality evaluator of $Y$ w.r.t.\ the query $q$. & $\Delta_q$ & Query-conditioned experiential prior injected as an decision anchor during planning. \\
$C(\tau),\,D(\tau)$ & Efficiency terms derived from $\tau$: cost and delay/runtime, respectively. & $\delta$ & Utility threshold for splitting retrieved records into positive vs.\ negative sets. \\
$\lambda_c,\lambda_t$ & Weights that trade off quality vs.\ cost and delay in the utility definition. & $\mathcal{I}_q^{+},\mathcal{I}_q^{-}$ & Retrieved positive/negative record index sets for query $q$. \\
$u(q,\Theta)$ (or $u$) & Utility/feedback associated with executing $\Theta$ on $q$. & $\mathcal{H}_{\mathrm{cand}}(q)$ & Candidate hints whose provenance overlaps retrieved record indices. \\
$\Theta_q=(M_q,P_q,B_q)$ & MapReduce-style decision for query $q$: task matrix, template, and batching strategy. & $s(h)$ & Feedback-calibrated hint score used for selecting query-relevant hints. \\
$M_q \in \mathcal{V}^{N\times K_0}$ & Task matrix with $N$ entities and $K_0$ known/seed attributes ($K_0<K$). & $v(h;q)$ & Query-conditioned relevance term in $s(h)$. \\
$P_q$ & Template with placeholders aligned to $M_q$ for instantiating atomic retrieval tasks. & $\mathcal{H}_q$ & Top-$k_{\mathcal{H}}$ selected query-relevant hints used to build $\Delta_q$. \\
$B_q$ & Batching strategy that partitions $\mathcal{T}(q)$ into disjoint batches. & $\Phi(q;\mathcal{H}_q)$ & Optional projection that compresses $\mathcal{H}_q$ into concise guidance $h_s$. \\
$\mathcal{T}(q)$ & Atomic task set instantiated from $(M_q,P_q)$ in the mapping stage. & $h_s$ & Condensed guidance derived from selected hints for planning-time conditioning. \\
$\{Bat_k\}_{k=1}^{m}$ & Batches produced by $B_q$; $m$ is the number of batches. & $\epsilon_{\mathrm{pos}},\epsilon_{\mathrm{neg}}$ & Most similar positive/negative exemplar records included in $\Delta_q$. \\
$\mathcal{A}^{\text{manage}}$ & Manager agent: sample $\Theta_q$, dispatch batches, reduce/validate outputs. & $\Theta_q^{(0)}$ & Initial decision sample before conditioning on the experiential prior $\Delta_q$. \\
$\mathcal{A}^{\text{search}}_k$ & $k$-th search agent that executes batch $Bat_k$ in parallel. & $\epsilon_{\text{new}}$ & New record appended after executing the current query, e.g., $\epsilon_{\text{new}}=(q,\Theta_q,\tau,u)$. \\
$\hat{Y}^k,\ \hat{Y}$ & Partial table from batch $k$ and their union before final reduction/validation. & $\mathcal{D}^{\text{new}}$ & Updated record set after appending $\epsilon_{\text{new}}$. \\
$M_{q}^{\mathrm{rep}}$ & Repair task matrix (a small matrix over $\mathcal{E}_{\mathrm{miss}}$). & $i_{\text{new}}$ & Index of $\epsilon_{\text{new}}$ in $\mathcal{D}^{\text{new}}$. \\
$P_{q}^{\mathrm{rep}}$ & Repair template restricted to $S_{\mathrm{miss}}$ (for patch-only retrieval). & $g(u)$ & Mapping from utility to a bounded signed reward for updating hint scores. \\
$\Theta_{q}^{\mathrm{rep}}$ & Repair decision resampled for the patch round. & $F_{\psi}$ & Distillation operator that consolidates $\mathcal{D}$ into an updated hint pool. \\
$\hat{Y}_{\mathrm{patch}}$ & Patched table returned by the repair MapReduce round. & $\{\mathcal{I}_k\}_{k=1}^{K_c}$ & Task clusters over records used by the distillation operator. \\
$\{Obs_t\}_{t=1}^{T}$ & Sequential observations accumulated during lightweight planning. & $h_k$ & Representative hint distilled for cluster $\mathcal{I}_k$. \\
$\mathcal{M}=\{\mathcal{D},\mathcal{H},F_{\psi}\}$ & Experiential memory: record set, hint pool, and a distillation operator. & $\mathcal{H}^{\text{new}}$ & Updated hint pool output by the distillation operator. \\
\bottomrule
\end{tabular}
\endgroup
\end{table*}

\end{document}